\documentclass{article}
\usepackage[T1]{fontenc}
\usepackage{authblk}
\usepackage{geometry}
\usepackage{amsfonts}
\usepackage{graphicx,amssymb,mathrsfs,amsmath,color,fancyhdr}
\usepackage{subcaption}
\usepackage{amsthm}
\usepackage{manfnt}
\usepackage{cases}
\usepackage{mathbbol}
\usepackage{bm,bbm}
\usepackage{setspace}
\usepackage{appendix}
\usepackage[justification=centering]{caption}
\usepackage{hyperref}
\hypersetup{colorlinks = True,linkcolor = blue,anchorcolor =red,citecolor = blue,filecolor = red,urlcolor = blue, pdfauthor=author}
\newtheorem{theorem}{Theorem}
\newtheorem{definition}{Definition}

\newtheorem{proposition}{Proposition}

\title{The Effects of High-Frequency Anticipatory Trading:\\
Small Informed Trader vs. Round-Tripper}
\author[1]{Ziyi Xu}
\author[1]{Xue Cheng \thanks{chengxue@pku.edu.cn}}
\affil[1]{\footnotesize LMEQF, Department of Financial Mathematics, \authorcr School of Mathematical Sciences, \authorcr
Peking University, Beijing 100871, China.}
\date{}
\begin{document}
\maketitle
\begin{abstract}
In an extended Kyle's model, the interactions between a large informed trader and a high-frequency trader (HFT) who can anticipate the former's incoming order are studied. We find that, in equilibrium, HFT may play the role of Small-IT or Round-Tripper: both of them trade in the same direction as IT in advance, but when IT's order arrives, Small-IT continues to take liquidity away, while Round-Tripper supplies liquidity back. So Small-IT always harms IT, while Round-Tripper may benefit her. What's more, with an anticipatory HFT, normal-speed small uninformed traders suffer less and price discovery is accelerated. 
\end{abstract}


\newpage
\section{Introduction}

High-frequency traders (HFTs) are one of the most active participants in today's financial markets. According to Breckenfelder (2020) \cite{breckenfelder2020does}, high-frequency trading has represented about 50\% of
trading volume in US equity markets and 24\%-43\% in Europe.  Based on data from the NASDAQ stock market, Hirschey (2021) verifies that HFTs are indeed anticipatory traders who predict and trade ahead of other
investors' order flow. This widespread conduction of anticipatory trading by HFTs makes it a major concern for both investors and regulators (see SEC (2010) \cite{sec2010}, SEC (2014) \cite{sec2014} and SEC (2020) \cite{sec2020}).

The effects of anticipatory strategies are still debatable in research. 
A series of literature suggests that they are negative. For instance, Brunnermeier and Pedersen (2005) \cite{brunnermeier2005predatory} believes that predators who detect larger traders' need for liquidation will take away liquidity that might otherwise have gone to the latter; Korajczyk and Murphy (2019) \cite{korajczyk2019high} empirically finds that HFTs consume liquidity by submitting same-direction orders during institutional trades, which brings higher transaction costs for large traders; Hirschey (2021) \cite{hirschey2021high} estimates that, due to HFT's anticipatory trading, non-HFTs undertake 14\% additional price impact. 
On the other hand, some studies have different views: Bessembinder et al. (2016) \cite{bessembinder2016liquidity} shows both theoretically and empirically that in a sufficiently resilient market, anticipatory traders tend to provide liquidity and thus benefit the large trader; Murphy and Thirumalai (2017) \cite{murphy2017short}, Br{\o}gger \cite{brogger2021market} (2021) and Yan et al. (2022) \cite{yan2022sunshine} all find evidence that the predictable order flow attracts liquidity suppliers and obtains a lower price impact. 

Motivated by these, in this paper, through an extended Kyle's (1985) \cite{kyle1985continuous} model, we consider the interactions between a large informed trader (IT) and a high-frequency trader (HFT) who predicts, to a certain extent, IT's future order. HFT is faster than IT in the sense that she can not only trade with IT, but also lay out in advance by trading ahead of IT. When optimizing, she considers the anticipation to IT's order, the advantage of speed, and importantly, her aversion to inventory.  We are interested in HFT's trading strategies and their effects on IT and market quality. We find that HFT will play, though sharing some common features, two distinct roles, under different conditions. One role is displayed as betting on the asset value like an informed investor, which we call ``Small-IT''. The other appears as exploiting the market impact caused by the large order, which we call ``Round-Tripper''. Although their initial trading behavior is similar, they ultimately have completely different effects.

We show that, prior to the execution of IT's order, both Small-IT and Round-Tripper build up positions in the same direction as it. When the informed order is being executed, Small-IT will still trade along with it in the same direction, while Round-Tripper will trade against it. For example, IT plans to sell a large amount of an asset. Once this plan is anticipated by HFT, no matter which role she plays, she will sell ahead of IT, establishing short positions at a more favorable price. When IT's sell order comes, Small-IT still sells while Round-Tripper buys back. Small-IT infers the asset value from the prediction signal and profits from the accumulated inventories, while Round-Tripper holds fewer inventories and profits from the price impact generated by IT's order.

If HFT cares little about the inventory, she will act as Small-IT. Otherwise, HFT will make a round-trip if she predicts inaccurately. If HFT is extremely inventory averse, she will play the role of Round-Tripper in any case. Kirilenko et al. (2017) \cite{2017The} studies the E-mini S\&P 500 futures market during a few days around Flash Crash and categorizes fast traders into   ``opportunistic traders'' and ``high-frequency traders'': both of them trade at a high speed, but the former adjust holdings with larger fluctuations and lower frequency. This classification coincides with the classification of Small-IT and Round-Tripper. Actually, \cite{2017The} further finds that before liquidity demanders arrive, ``high-frequency traders'' aggressively remove the last few contracts at the best bid and ask, then provide liquidity at a new price level, offsetting their position, which implies that ``high-frequency traders'' are actually Round-Trippers.

The influences of Small-IT and Round-Tripper on IT are totally different. Compared to the case without an anticipatory HFT, IT is worse off with Small-IT as imagined, while Round-Tripper possibly benefits her. This happens when there are adequate high-speed noise orders in the market or Round-Tripper's signal is rough. 

These results may shed light on the reason why existing empirical studies on the impact of anticipatory trading have drawn seemingly inconsistent conclusions. For example, in the sample of Hirschey (2021) \cite{hirschey2021high}, HFTs are not required to maintain low inventory during execution. Therefore, they are more likely to play the role of Small-IT and unwind their positions after the predicted order is completed, which brings non-HFTs extra costs. 
Likewise, the behavior of HFT submitting same-direction orders during institutional trades in  Korajczyk and Murphy (2019) \cite{korajczyk2019high} is similar to the strategy of Small-IT.
In comparison, anticipatory traders in Bessembinder et al. (2016) \cite{bessembinder2016liquidity} supply liquidity back to the large trader and favor her in a resilient market. This kind of market can be regarded as one with active noise trading in our framework, where the anticipatory trader will act as Round-Tripper and benefit the predicted investor more often. Situations in \cite{murphy2017short}, \cite{brogger2021market} and \cite{yan2022sunshine} are roughly analogous.

In practice, there is a special type of anticipatory trading called ``front-running'' that is prohibited by Securities and
Exchange Commission (SEC). However, driven by huge profits, there are still traders trying to do it, which can be seen in recent SEC news. From the conclusion of this paper, in addition to imposing fines on such transactions, more restrictions on traders' intraday positions may make things better. If the limit on inventory increases, anticipatory traders will be guided to play the role of Round-Tripper, thereby providing liquidity and 
potentially benefiting other investors.


We also explore the influences of HFT's prediction accuracy. When IT is harmed, she will be less hurt if HFT receives a less accurate signal, which is consistent with the finding
in Sa{\u{g}}lam (2020) \cite{sauglam2020order}.
When IT is benefited by Round-Tripper, her profit may decrease as the signal gets noisier.
This seems counterintuitive, but the finding in Br{\o}gger \cite{brogger2021market} (2021) supports it: 
with anticipatory traders who supply liquidity, less predictable flows are associated with higher price impact. 


As for market quality, the existence of an anticipatory HFT reduces the loss of normal-speed noise traders. It is because the asset information is disclosed earlier through HFT's preemptive trading, which lowers the subsequent intensity of price impact and consequently cuts down noise traders' costs. It provides a possible explanation for the finding in Korajczyk and Murphy (2019) \cite{korajczyk2019high} that high-frequency trading brings lower transaction costs for small, uninformed trades. For the same reason, price is discovered earlier, which is in accordance with the finding in Brogaard et al. (2014)  \cite{brogaard2014discovery} that the direction of HFTs' liquidity demanding orders predicts short-horizon price movements. However, HFT does not produce any extra information, so the ultimate pricing error does not decrease.

Equilibria under extreme conditions are studied and some interesting and analytical results arise. First, HFT does not always take advantage of speed. If almost all noise traders are as slow as IT, HFT will only trade along with IT at the same time and in the same direction, as back-runners in Yang and Zhu (2020) \cite{yang2020back}. In other words, HFT becomes a normal-speed anticipatory trader. Second, when HFT's inventory aversion tends to infinity, she becomes a short-term Round-Tripper who leaves market with zero positions, i.e., she consumes liquidity first but provides equal liquidity back, as in Li (2018) \cite{li2018high}.

Anticipatory traders in this paper differ from those in \cite{li2018high} and \cite{yang2020back} in two key aspects. The first difference lies in the inventory management: (1) HFTs in \cite{li2018high} can be regarded as infinitely inventory averse, they clear all the positions; (2) back-runners in \cite{yang2020back} bear no inventory pressure, they accumulate positions; (3) HFTs in this paper can have different inventory aversion and the aversion decides their anticipatory strategies, specific conditions where HFT holds assets or not are obtained. The second difference lies in the trading speed: (1) anticipatory traders in \cite{li2018high} are high-frequency in the sense that they are able to trade before large traders; (2) back-runners in \cite{yang2020back} are normal-speed in the sense that they only trade at the same time as large traders; (3) both scenarios are considered in this paper, including high-frequency and normal-speed anticipatory trading.

Conclusions about high-frequency anticipatory trading and its effects are robust when both IT and HFT trade on multiple correlated assets. However, in the real world, HFT may not hold exactly the same assets as IT. In this case, on the assets that are traded by both IT and HFT, HFT may play two kinds of anticipatory roles, depending on the inventory aversion; on the assets that are only traded by HFT, she always plays the role of Small-IT. That is to say, the existence of a large trader to generate price impact and be the counterparty is a necessary condition for HFT to play the role of Round-Tripper.

The paper is organized as follows. Section \ref{related} illustrates some related literature.
In Section \ref{secmodel}, we set up the model. In Section \ref{secresults}, we discuss the equilibrium through both theoretical and numerical ways, also in certain limit situations. In Section \ref{secmulti}, extensions to multi-asset market are discussed. Section \ref{secconclusion} concludes and all the proofs are displayed in the \nameref{secappendix}.

\section{Related Literature}
\label{related}
This paper contributes to the theoretical literature on anticipatory trading. Li (2018) \cite{li2018high} models the speed competition
between fast traders who observe a common signal about incoming orders. Yang and Zhu (2020) \cite{yang2020back} models back-runners who use order anticipation strategies based on past order flows and concludes that
informed traders counteract this by randomizing their orders. The contribution of this paper relative to \cite{yang2020back} and \cite{li2018high} lies in the analyses regarding to inventory aversion and trading speed, as mentioned in last section.

When it comes to inventory management, a related work is Ro{\c{s}}u (2019) \cite{rocsu2019fast}, which studies traders who differ in their speed of processing information and receive a stream of signals. Fast traders use immediate information, while slow traders use lagged ones. As a result,
the former's order flow predicts the latter's. When the fast trader is averse to inventory enough, she no longer makes long-term bets but unloads her inventory to slower traders. This is proved in this paper as well. In our model, IT has firsthand information about the asset value and HFT infers part of it through a prediction signal.
 HFT employs this information before the informed order is executed, which also displays her ability to process information faster. This paper differs from \cite{rocsu2019fast} in several aspects. First, we focus on whether and when the predicted investor IT is harmed or benefited. Second, we consider the equilibria not only under different inventory aversion,  but also under different sizes of noise trading and signal accuracy, which enables us to get further conclusions. 

This paper also contributes to theoretical high-frequency trading literature. For this topic, see Hoffmann (2014) \cite{hoffmann2014dynamic}, Foucault et al. (2016) \cite{foucault2016news}, Baldauf and Mollner (2020) \cite{baldauf2020high}, and the survey of Menkveld (2016) \cite{menkveld2016economics}, as well as the works mentioned before.

This paper relates to predatory trading, too. Typical works include Brunnermeier and Pedersen (2005) \cite{brunnermeier2005predatory}, Carlin et al. (2007) \cite{carlin2007episodic}, Sch{\"o}neborn and Schied (2009) \cite{schoneborn2009liquidation}, and Bessembinder et al. (2016) \cite{bessembinder2016liquidity}, which discuss predatory trading in price impact models, where the impact coefficients are exogenously given. The main difference is that this paper follows Kyle's setting, where the information-transfer structure between the market and investors is modeled and the impact coefficients are endogenously decided. In this framework, some metrics of market quality, like price discovery, can be explored. 

\section{The Model and Equilibrium}
\label{secmodel}
We start by introducing the following two-stage dynamic market model for high-frequency trading, which is an extension of the classic Kyle's model \cite{kyle1985continuous}.

\textbf{Assets and participants.}
In this market, a risky asset is traded whose true value,
$v$, is normally distributed as
\begin{equation*}
    v\sim N(p_0,\sigma_v^2).
\end{equation*}
There is also a risk-free asset with zero interest rate, which provides inter-temporal value accumulations only.

Four types of participants are modeled: (1) \textit{dealers}, who observe the aggregate order flow and are assumed to be competitive and risk-neutral. The Bertrand competition forces them to make zero expected profit and hence set the transaction price as the expectation of $v$ conditional on their information; (2) a normal-speed large \textit{informed trader} (IT, for short), who privately knows $v$; (3) a strategic \textit{high-frequency trader} (HFT, for short), who is capable to get a signal about IT's future trading; (4) \textit{noise traders}, who trade randomly. We suppose that both IT and HFT are risk-neutral and seek to maximize their expected P\&L, while HFT may conduct inventory management as well.

\textbf{Timeline and trading structure.}  
We consider three time points \footnote{These time stamps are marked just for convenience. It is not required that the time lengths of $[0,1]$ and $[1,2]$ are equal. }, $t=0,\, 1,\, 2.$ At $t=0,$ IT sends a market order of quantity $i$, based on her private knowledge about $v$. However, for some reasons, e.g., the submission delay, as in \cite{li2018high}, \cite{baldauf2020high}, the order is not executed until $t=2$.

We assume that HFT detects IT's intention
immediately after the order $i$ is sent. There are multiple ways for HFT to implement such kind of anticipation. For example, as proved in Hirschey (2021) \cite{hirschey2021high}, the prediction of non-HFTs' order can be realized by constructing indicators from data of order flow and return. We will not get to the bottom of how HFT predicts the order specifically, which lies beyond the scope of this paper. Instead, HFT is supposed to receive a noisy signal about $i$:
\begin{equation*}
    \hat{i}=i+\varepsilon,
\end{equation*}
where $\varepsilon$ is independent of $i$ and follows $ N(0,\sigma_\varepsilon^2)$. 

Practically, the signal noise may come from (1) market regulations about information disclosure, which makes it difficult for HFT to filter useful news about large traders' trading; (2) the limitation of HFT's technology, which brings the prediction error.
 The standard deviation of $\varepsilon$ represents the accuracy of HFT's signal. The smaller the $\sigma_\varepsilon$, the higher the accuracy.


Since HFT rarely has a delay in sending orders, her orders are assumed to be executed at once: (1) during $(0,1)$, HFT sends the order $x_{1},$ which is executed at $t=1;$ (2) during $(1,2)$, HFT sends the order $x_{2},$ which is executed at $t=2.$ The noise orders during periods 1 and 2 are respectively denoted by $u_1$ and $u_2$, where
$$u_1\sim N(0,\sigma_1^2)\quad\text{and}\quad u_2\sim N(0,\sigma_2^2),\ \sigma_1,\sigma_2>0,$$are independent of each other and any other random variables.

To sum up, the total order flow $y_1$ and $y_2$ executed at $t=1$ and $t=2$ are
\begin{equation*}
    y_1=x_1+u_1 \quad\text{and}\quad y_2=i+x_2+u_2.
\end{equation*}

\textbf{Equilibrium.}
The profit of IT is:
\begin{equation*}
\pi^{\text{IT}}(i)=i(v-p_2).    
\end{equation*}
The profits of HFT in period 1 and 2 respectively are:
\begin{equation*}
\begin{aligned}
\pi_1^{\text{HFT}}(x_1)=x_1(v-p_1)\quad\text{and}\quad\pi_2^{\text{HFT}}(x_2)=x_2(v-p_2).
\end{aligned}
\end{equation*}
HFT additionally bears an inventory penalty and is represented as a quadratic function of position:
$$
-\gamma(x_1+x_2)^2,
$$
where $\gamma$ describes HFT's inventory aversion. If $\gamma$ grows, HFT will more consider the costs of ending position. 

The quadratic form of inventory cost is prevailing in existing literature on high-frequency trading (see, Ro{\c{s}}u (2019) \cite{rocsu2019fast} and Herrmann et al. (2020) \cite{herrmann2020inventory}) and optimal execution (see Cheng et al. (2017), Cardaliaguet and Lehalle (2018) \cite{cardaliaguet2018mean}, and Huang et al. (2019) \cite{huang2019mean}).

We can interpret it from an intuitive perspective:
 at some certain moment $T>2$, $v$ will be revealed publicly and everyone can clear her position at the price $v$. However, unlike long-term investor IT, due to certain regulatory requirements or limited risk-bearing capacity and capital constraints, as mentioned in \cite{2017The} and \cite{rocsu2019fast}, HFT may have to liquidate the position in advance. The liquidation will cause price impact and if the impact is assumed to be linear in trading volume, the transaction price will approximately be $v-\gamma (x_1+x_2)$. As a result, the objective function of HFT is:
 \begin{equation*}
    \begin{aligned}
&x_1(v-\gamma(x_1+x_2)-p_1)+x_2(v-\gamma(x_1+x_2)-p_2)\\
=&\pi_1^{\text{HFT}}(x_1)+\pi_2^{\text{HFT}}(x_2)-\gamma(x_1+x_2)^2.
    \end{aligned}
 \end{equation*}
Hence, the inventory aversion $\gamma$ is similar to the coefficient of price impact.

The reason why we only penalize the position at $t=2$ rather than positions at both $t=1$ and $2$, is that the total 
duration of the model can be very short, even a few seconds, so the penalty to the time-2 inventory is in line with the situation that the end-of-minute and end-of-day net inventories of HFTs are low, as verified in Kirilenko et al. (2017) \cite{2017The}.

We now give the definition of equilibrium. 
\begin{definition}
The equilibrium is defined as a collection of prices and strategies: $\{p_1, p_2,i,x_1,x_2\}$, such that the following weak-efficiency condition and two optimization conditions are satisfied.
\begin{enumerate}
    \item Given IT's strategy $i$ and HFT's strategies $x_1,x_2$, dealers set prices according to the weak-efficiency rule:
    \begin{equation*}
\begin{aligned}
&p_1=\mathbbm{E}\left(v|y_1\right),\\
&p_2=\mathbbm{E}\left(v|y_1,y_2\right).
\end{aligned}
    \end{equation*}
   \item Given dealers' quotes $p_1,p_2$ and HFT's strategies $x_1,x_2$, in the \textbf{pure-strategy equilibrium}, IT maximizes her expected profit over all measurable strategies $\Bar{i}(v)$:
    \begin{equation*}
i=\arg\max\ \mathbbm{E}\left(\left.\pi^{\text{IT}}(\Bar{i})\right|v\right).
    \end{equation*}
 In the \textbf{mixed-strategy equilibrium}, IT should be indifferent among all realizations of pure strategies, i.e., $\forall i=i(v),\mathbbm{E}\left(\left.\pi^{\text{IT}}(i)\right|v\right)$
    are the same.
    \item Given dealers' quotes $p_1,p_2$ and IT's strategy $i$, HFT maximizes her expected profit with inventory penalty over all measurable strategies $\Bar{x}_2(\hat{i},y_1)$ in period 2 and her total expected profit with inventory penalty over all measurable strategies $\Bar{x}_1(\hat{i})$ in period 1:
    \begin{equation*}
    \left\{\begin{aligned}
    &   x_2=\arg\max\ \mathbbm{E}\left(\pi_2^{\text{HFT}}(\Bar{x}_2)-\gamma(x_1+\Bar{x}_2)^2\Big|\hat{i},y_1\right);\\
&x_1=\arg\max\ \mathbbm{E}\left(\pi_1^{\text{HFT}}(\Bar{x}_1)+\pi_2^{\text{HFT}}(x_2)-\gamma(\Bar{x}_1+x_2)^2\Big|\hat{i}\right).
    \end{aligned}
\right.
    \end{equation*}
\end{enumerate}
\end{definition}



Within the normal-distribution framework, we conjecture a linear structure of the equilibrium, which is consistent with Kyle (1985) \cite{kyle1985continuous},  Bernhardt and Miao (2004) \cite{bernhardt2004informed} and Bernhardt and Taud (2008) \cite{bernhardt2008front}:
\begin{equation}
\label{conjecture}
\begin{aligned}
&p_1=p_0+\lambda_1y_1,\\
&p_2=p_0+\lambda_{21}y_1+\lambda_{22}y_2,\\
&i=\alpha (v-p_0)+z,\\
&x_1=\beta_1\hat{i},\\
&x_2=\beta_{21}\hat{i}+\beta_{22}u_1+\beta_{23}x_1,
\end{aligned}
\end{equation}
where $z\sim N(0,\sigma_z^2)$ is the endogenous noise added by IT. When considering the mixed-strategy equilibrium, $\sigma_z>0$; while in the pure-strategy equilibrium, $\sigma_z=0$. IT's mixed strategy follows Huddart, Hughes and Levine (2001) \cite{2001Public} and Yang and Zhu (2020) \cite{yang2020back}. 

Note that, HFT's time-2 strategy is
\begin{equation*}
x_2=\beta_{21}\hat{i}+\beta_{22}y_1+(\beta_{23}-\beta_{22})x_1.
\end{equation*}
The first part represents HFT's reliance on signal $\hat{i}$, the second part represents the adjustment according to the information already exposed through $y_1$, and the third part represents the adjustment to the existing inventory $x_1$. 

HFT's strategies $x_1$ and $x_2$ can also be written as
\begin{equation*}
\begin{aligned}
&x_1=\beta_1\hat{i},\\
&x_2=(\beta_{21}+\beta_{23}\beta_1)\hat{i}+\beta_{22}u_1,
\end{aligned}
\end{equation*}
where the coefficients $\beta_1$ and $\beta_{21}+\beta_{23}\beta_1$ represent 
whether or not she follows the signal to trade. Hence, these two coefficients stand for the trading directions of $x_1$ and $x_2$. When the coefficient is positive (negative), we say that HFT tends to trade in the same (opposite) direction as IT. Based on it, two kinds of strategies are defined:
\begin{itemize}
    \item \textit{Small-IT}: HFT trades in the same direction as IT in both periods;
    \item \textit{Round-Tripper}: HFT trades in the same direction as IT in period 1 but in the opposite direction in period 2.
\end{itemize}

\begin{figure}[!htbp]
    \centering
    \includegraphics[height=0.12\textheight]{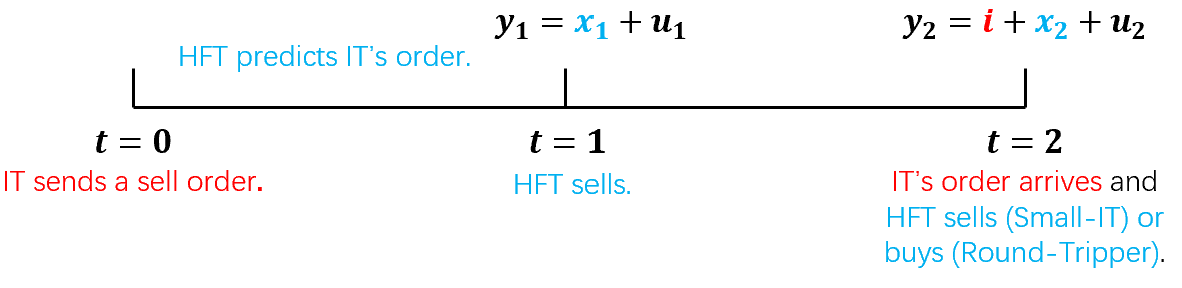}
    \caption{Timeline of the model when IT sells.}
    \label{figtimeline}
\end{figure}

A specific example is provided here. At $t=0,$ IT decides to sends a sell order and this intention is predicted by HFT, as shown in Figure \ref{figtimeline}. No matter HFT acts as Small-IT or Round-Tripper, she will preempt IT by first selling at $t=1,$ building up a short position when the market price has not been pushed down. When IT's sell order arrives at $t=2,$ Small-IT still sells and her short position increases, while Round-Tripper buys back from IT and her short position decreases. If IT decides to buy, the situations are similar: HFT will preempt it by first buying at $t=1.$ When IT's buy order arrives at $t=2,$ Small-IT still buys but Round-Tripper sells to IT.

Small-IT accumulates positions and makes profits from holding assets. While for Round-Tripper, the total profit can be divided into two parts, one is from the market impact and the other is from the holding of the asset: 
\begin{equation}
\label{impactholding}
\pi_1^{\text{HFT}}(x_1)+\pi_2^{\text{HFT}}(x_2)=\underbrace{(-x_2)(p_2-p_1)}_{\text{Impact profit}}+\underbrace{(x_1+x_2)(v-p_1)}_{\text{Holding profit}}.
\end{equation}

At last, we introduce three parameters which will be useful to characterize the equilibrium:
\begin{equation*}
\begin{aligned}
&\theta_1=\sigma_1^2/\sigma_2^2>0,\ \text{the relative size of market noise over two periods;}\\
&\theta_{\varepsilon}=\sigma_\varepsilon^2/\sigma_2^2\geq0,\ \text{HFT's signal accuracy;}\\
&\Gamma=\frac{\gamma}{\sigma_v/\sigma_2}\geq0,\ \text{the dimensionless form of HFT's inventory aversion.}
\end{aligned}
\end{equation*}

\section{Main Results}
\label{secresults}
First, we study the optimal strategies of  dealers, HFT and IT respectively.

\textbf{Dealers' quotes.}
As stated in the former section, risk-neutral and competitive dealers set the transaction prices as the expectations of $v$ conditioned on the order flow information.

At $t=1$, when HFT builds up the position $x_1$, the transaction price is
\begin{equation*}
p_{1}=\mathbbm{E}(v|y_1)=p_0+\lambda_1y_1,
\end{equation*}
where, given by the linear conjecture and projection theorem,
\begin{equation}
\label{lam1}
    \lambda_1=\frac{\sigma_{v}}{\sigma_{y_1}}\rho_{(v,y_1)}=\frac{\alpha\beta_1\sigma_v^2}{\beta_1^2(\alpha^2\sigma_v^2+\sigma_z^2+\sigma_{\varepsilon}^2)+\sigma_1^2}.
\end{equation}

At $t=2$, when IT's order $i$ and HFT's order $x_2$ are being executed, the transaction price is
\begin{equation*}
    p_2=\mathbbm{E}(v|y_1,y_2)=p_0+\lambda_{21}y_1+\lambda_{22}y_2,
\end{equation*}
where, still by the projection theorem,
\begin{equation}
\label{lam21}
\lambda_{21}=\frac{\sigma_{v}}{\sigma_{y_1}}  \frac{\rho_{(v,y_1)}-\rho_{(y_1,y_2)} \rho_{(v,y_2)}}{1-\rho^{2}_{(y_1,y_2)}}=\frac{\alpha\sigma_v^2\{-\sigma_1^2\beta_{22}[1+\beta_{21}+\beta_1(\beta_{23}-\beta_{22})]+\sigma_2^2\beta_1-\sigma_\varepsilon^2\beta_1(\beta_{21}+\beta_{23}\beta_1)\}}{\Sigma},
\end{equation}
\begin{equation}
\label{lam22}
\lambda_{22}=\frac{\sigma_{v}}{\sigma_{y_2}} \frac{\rho_{(v,y_2)}-\rho_{(v,y_1)} \rho_{(y_1,y_2)}}{1-\rho_{(y_1,y_2)}^{2}}=\frac{\alpha\sigma_v^2\{\sigma_1^2[1+\beta_{21}+\beta_1(\beta_{23}-\beta_{22})]+\sigma_{\varepsilon}^2\beta_1^2\}}{\Sigma},
\end{equation}
and
\begin{equation*}
\Sigma=(\alpha^2\sigma_v^2+\sigma_z^2)\{\sigma_1^2[1+\beta_{21}+\beta_1(\beta_{23}-\beta_{22})]^2+(\sigma_2^2+\sigma_\varepsilon^2)\beta_1^2\}+\sigma_1^2\sigma_\varepsilon^2[\beta_{21}+\beta_1(\beta_{23}-\beta_{22})]^2+\sigma_2^2(\beta_1^2\sigma_\varepsilon^2+\sigma_1^2).
\end{equation*}

In the analyses below, we will also be interested in the total impact of the informed order $i$ on the time-2 price $p_2$. Under linear structures, this impact is still linear in $i$, and we denote its coefficient as $\lambda_2$. Since HFT partly detects the intention of IT and trades along/against $i$, $\lambda_2$ turns out to be a combination of different-time impacts:
\begin{equation}
\label{lambda2}
\lambda_{2}=\lambda_{21}\beta_1+\lambda_{22}(1+\beta_{21}+\beta_{23}\beta_1).
\end{equation}
The part $\lambda_{21}\beta_1$ is from HFT's time-1 trading $x_1$ and the part $\lambda_{22}(1+\beta_{21}+\beta_{23}\beta_1)$ is from $i+x_2$.

\textbf{HFT's strategies.}
Given dealers' quotes $p_1,p_2$ and IT's strategy $i$, HFT's objective function in period 2 is
\begin{equation*}
\begin{aligned}
&\mathbbm{E}\left(\pi_2^{\text{HFT}}-\gamma(x_1+x_2)^2\Big|\hat{i},y_1\right)\\
=&-(\lambda_{22}+\gamma)x_2^2+x_2[\left(]\frac{\alpha\sigma_v^2(1-\lambda_{22}\alpha)-\lambda_{22}\sigma_z^2}{\alpha^2\sigma_v^2+\sigma_z^2+\sigma_\varepsilon^2}\hat{i}-\lambda_{21}u_1-(\lambda_{21}+2\gamma)x_1\right)-\gamma x_1^2.
\end{aligned}
\end{equation*}
It is maximized at $x_2=\beta_{21}\hat{i}+\beta_{22}u_1+\beta_{23}x_1$ when the second order condition (SOC)
\begin{equation}
\label{SOC1}
\lambda_{22}+\gamma>0
\end{equation}holds, where
\begin{equation}
\label{beta21}
\beta_{21}=\frac{\alpha\sigma_v^2(1-\lambda_{22}\alpha)-\lambda_{22}\sigma_z^2}{2(\lambda_{22}+\gamma)(\alpha^2\sigma_v^2+\sigma_z^2+\sigma_\varepsilon^2)},
\end{equation}
\begin{equation}
\label{beta22}
\beta_{22}=\frac{-\lambda_{21}}{2(\lambda_{22}+\gamma)},
\end{equation}
\begin{equation}
\label{beta23}
\beta_{23}=\frac{-(\lambda_{21}+2\gamma)}{2(\lambda_{22}+\gamma)}.
\end{equation}
HFT's objective function in period 1 is
\begin{equation*}
\begin{aligned}
&\mathbbm{E}\left(\pi_1^{\text{HFT}}+\pi_2^{\text{HFT}}-\gamma(x_1+x_2)^2\Big|\hat{i}\right)\\
=&-\big[\lambda_1+\gamma-(\lambda_{22}+\gamma)\beta_{23}^2\big]x_1^2+x_1\hat{i}\left(\frac{\alpha\sigma_v^2}{\alpha^2\sigma_v^2+\sigma_z^2+\sigma_{\varepsilon}^2}+2(\lambda_{22}+\gamma)\beta_{21}\beta_{23}\right)+(\lambda_{22}+\gamma)(\beta_{21}^2\hat{i}^2+\beta_{22}^2\sigma_1^2).
\end{aligned}
\end{equation*}
It is maximized at $x_1=\beta_1\hat{i}$ when the SOC 
\begin{equation}
\label{SOC2}
\lambda_1+\gamma-(\lambda_{22}+\gamma)\beta_{23}^2>0
\end{equation}
holds, where
\begin{equation}
\label{beta1}
\beta_1=\frac{\frac{\alpha\sigma_v^2}{\alpha^2\sigma_v^2+\sigma_z^2+\sigma_\varepsilon^2}+2(\lambda_{22}+\gamma)\beta_{21}\beta_{23}}{2\left[\lambda_1+\gamma-(\lambda_{22}+\gamma)\beta_{23}^2\right]}.
\end{equation}

\textbf{IT's strategy.} Given dealers' quotes $p_1,p_2$ and HFT's strategies $x_1,x_2$, IT's expected profit is
\begin{equation*}
\begin{aligned}
\mathbbm{E}\left(\left.\pi^{\text{IT}}\right|v\right)
=&-\lambda_2 i^2+i(v-p_0).
\end{aligned}
\end{equation*}
In the \textit{pure-strategy equilibrium}, the expected profit is maximized at $i=\alpha(v-p_0)$ when the SOC 
\begin{equation}
\label{SOC3}
\lambda_2>0
\end{equation}
holds, where 
\begin{equation}
    \label{alpha}
\alpha=\frac{1}{2\lambda_{2}}.
\end{equation}
In the \textit{mixed-strategy equilibrium}, IT is indifferent to all $i=i(v),$ so the coefficients of $i^2$ and $i$ in the expected profit should both be zero, which is impossible. Consequently, it is always better for IT to take the pure strategy. We only need to consider pure-strategy equilibrium in the following, that is, $\sigma_z=0$. And we call it ``equilibrium'' directly for simplicity.

\textbf{Equilibrium.} Given asset's volatility $\sigma_v$ and time-2 market noise $\sigma_2$, define \begin{equation*}
\begin{aligned}
 &\Lambda_{1}=\frac{\lambda_1}{\sigma_v/\sigma_2},\Lambda_{21}=\frac{\lambda_{21}}{\sigma_v/\sigma_2},\Lambda_{22}=\frac{\lambda_{22}}{\sigma_v/\sigma_2},A=\frac{\alpha}{\sigma_2/\sigma_v}.\\
\end{aligned}
\end{equation*}
From the former discussions, prices and strategies are linear functions decided by intensities $\lambda_1,\lambda_{21},\lambda_{22},\alpha,\\
\beta_1,\beta_{21},\beta_{22},\beta_{23}.$ So the equilibrium can be represented in the intensity form:
\begin{equation}
\label{intensityform}
\{\Lambda_1,\Lambda_{21},\Lambda_{22},A,\beta_1,\beta_{21},\beta_{22},\beta_{23}\}.
\end{equation}
Substitute the new-defined intensities into equations \eqref{lam1},\eqref{lam21},\eqref{lam22},\eqref{beta21},\eqref{beta22},\eqref{beta23},\eqref{beta1},\eqref{alpha} and SOCs \eqref{SOC1},\eqref{SOC2},\eqref{SOC3}, we find that, the equilibrium \eqref{intensityform} is decided by $\{\theta_1,\theta_\varepsilon,\Gamma\}.$
 In the theorem below, the equilibrium conditions are further simplified.
\begin{theorem}
\label{mainthm}
The equilibrium is characterized by $\{\Lambda_{22},A,\beta_1\}$ through system \eqref{systemmain}. In equilibrium,
\begin{equation*}
\begin{aligned}
&\Lambda_1=\frac{A\beta_1}{\beta_1^2 (A^2 +\theta_\varepsilon)+\theta_1},\\
&\Lambda_{21}=\frac{A^3 \Lambda_{22} [2 (\beta_1-1) \Gamma-\Lambda_{22}]+A^2 \Gamma+2 A \theta_\varepsilon \Lambda_{22} [(\beta_1-1) \Gamma-\Lambda_{22}]+\theta_\varepsilon (\Gamma+\Lambda_{22})}{A\beta_1 (A^2+\theta_\varepsilon) (2 \Gamma+\Lambda_{22})},\\
&\beta_{21}=\frac{A (1-A\Lambda_{22})}{2 (A^2+\theta_\varepsilon) (\Gamma+\Lambda_{22})},\\
&\beta_{22}=-\frac{\Lambda_{21}}{2 (\Gamma+\Lambda_{22})},\\
&\beta_{23}=-\frac{2 \Gamma+\Lambda_{21}}{2 (\Gamma+\Lambda_{22})}.\\
\end{aligned}
\end{equation*}
Thus, the expected profit of IT is
\begin{equation*}
\mathbbm{E}(\pi^{\text{IT}})=\frac{\sigma_2\sigma_v }{2}A;
\end{equation*}
the expected profit and expected inventory penalty of HFT are
\begin{equation*}
\begin{aligned}
\mathbbm{E}(\pi_1^{\text{HFT}}+\pi_2^{\text{HFT}})=&\sigma_2\sigma_v \{A (\beta_{1} \beta_{23}+\beta_{21}) [1-A (\beta_{1} \beta_{23} \Lambda_{22}+\beta_{1} \Lambda_{21}+\beta_{21} \Lambda_{22}+\Lambda_{22})]+A \beta_{1} (1-A \beta_{1} \Lambda_{1})\\
-&\beta_{1}^2 \theta_\varepsilon \Lambda_{1}-\theta_\varepsilon (\beta_{1} \beta_{23}+\beta_{21}) [\beta_{1} (\beta_{23} \Lambda_{22}+\Lambda_{21})+\beta_{21} \Lambda_{22}]-\beta_{22} \theta_1(\beta_{22} \Lambda_{22}+\Lambda_{21})\},\\
\mathbbm{E}[-\gamma(x_1+x_2)^2]=&-\sigma_2\sigma_v\Gamma [\beta_{22}^2\theta_1+(A^2+\theta_\varepsilon) (\beta_{1} \beta_{23}+\beta_{1}+\beta_{21})^2];
\end{aligned}
\end{equation*}
the price discovery variables are
\begin{equation*}
\begin{aligned}
\mathbbm{E}(v-p_1)^2&=\sigma_v^2[(A \beta_{1} \Lambda_{1}-1)^2+\beta_{1}^2  \Lambda_{1}^2\theta_\varepsilon+\Lambda_{1}^2\theta_1],\\
\mathbbm{E}(v-p_2)^2&=\frac{\sigma_v^2}{2};
\end{aligned}
\end{equation*}
and the aggregate loss of noise traders is
\begin{equation*}
\sigma_2\sigma_v(\Lambda_1\theta_1+\Lambda_{22}).
\end{equation*}
\end{theorem}

When the signal inaccuracy $\theta_\varepsilon=0$ and the inventory aversion $\Gamma$ is relatively small, system \eqref{systemmain} has no solution. In fact, under this condition, HFT receives a noiseless signal and bears little inventory pressure. If IT traded, HFT would be perfectly aware of $v$ by learning the informed order and also fully engaged in trading. IT would lose the informational monopoly through trading, which is not optimal for her. However, if she did not trade, the price impact coefficients would all be zero, which means trading more is better. In both cases, we are faced with apparent contradictions.
Nevertheless, note that HFT perfectly learns about $v$, we come up with a equilibrium with duopolistic informed investors to overcome it. The conclusions are summarized in Proposition \ref{propexistence}:
\begin{proposition}
\label{propexistence}
Under conjectured strategy \eqref{conjecture}, equilibrium exists for all $\theta_\varepsilon>0$. When $\theta_\varepsilon=0,$ given $\theta_1>0,$ there exists a critical $\Tilde{\Gamma}=\Tilde{\Gamma}(\theta_1),$  when $\Gamma\in[0,\Tilde{\Gamma}),$ the equilibrium does not exist, while an equilibrium with duopolistic informed investors exists, where HFT takes the following strategy:
\begin{equation*}
\begin{aligned}
&x_1=\beta_1(v-p_0),\\
&x_2=\beta_{21}(v-p_0)+\beta_{22}u_1+\beta_{23}x_1,
\end{aligned}
\end{equation*}
strategies of dealers and IT remain. The equilibrium is characterized by system \eqref{systemduopolistic}. Both IT and HFT trade in the same direction as $v-p_0,$ so HFT acts as Small-IT.
\end{proposition}

\subsection{Equilibrium in the general dynamic model{}}
\label{secequilibriumgeneral}
In this section, we consider numerical solutions of system \eqref{systemmain}, as suggested by \cite{cox2006using}. We do a large number of experiments and investigate how HFT's inventory aversion $\Gamma$ and signal accuracy $\theta_\varepsilon$ affect the equilibrium, in different markets distinguished by the relative size of noise trading $\theta_1$. 
According to model assumptions, noise trading in period 2 comes from both high-speed and normal-speed traders, but noise traders in period 1 are all high-speed. So it is sensible to assume that $\sigma_1\leq\sigma_2,$ i.e., $\theta_1\in(0,1]$. To illustrate the results more clearly, we first present some conclusions about investors' strategies and profits.

\begin{figure}[!htbp]
    \centering
\includegraphics[height=0.19\textheight]{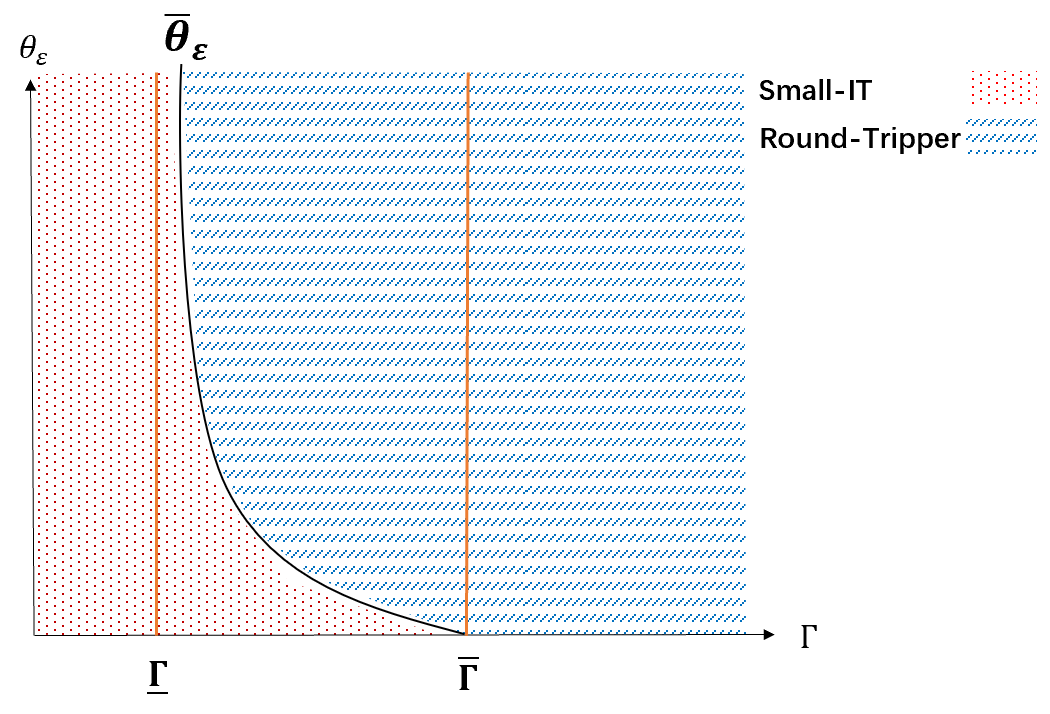}
\caption{HFT's strategies.}
\label{figconclusion1}
\end{figure}

\textbf{HFT's strategies.} For any $\theta_1 \in(0,1]$, there exists a $\underline{\Gamma}=\underline{\Gamma}(\theta_1)$ and a $\overline{\Gamma}=\overline{\Gamma}(\theta_1),$ with $\frac{\partial\underline{\Gamma}(\theta_1)}{\theta_1}<0,\frac{\partial\overline{\Gamma}(\theta_1)}{\theta_1}<0,$ to distinguish HFT's role:  (1) when $\Gamma\in[0,\underline{\Gamma}],$ HFT acts as Small-IT; (2) when $\Gamma\in[\overline{\Gamma},+\infty],$ HFT acts as Round-Tripper; (3) when $\Gamma\in(\underline{\Gamma},\overline{\Gamma}),$ there exists a $\overline{\theta}_\varepsilon=\overline{\theta}_\varepsilon(\theta_1,\Gamma),$ with $\frac{\partial\overline{\theta}_\varepsilon(\theta_1,\Gamma)}{\theta_1}<0,\frac{\partial\overline{\theta}_\varepsilon(\theta_1,\Gamma)}{\Gamma}<0,$ if $\theta_\varepsilon>\overline{\theta}_\varepsilon,$ HFT will act as Round-Tripper. Otherwise, she acts as Small-IT. These cases are displayed in Figure \ref{figconclusion1}.

\begin{figure}[!htbp]
    \centering
\includegraphics[height=0.2\textheight]{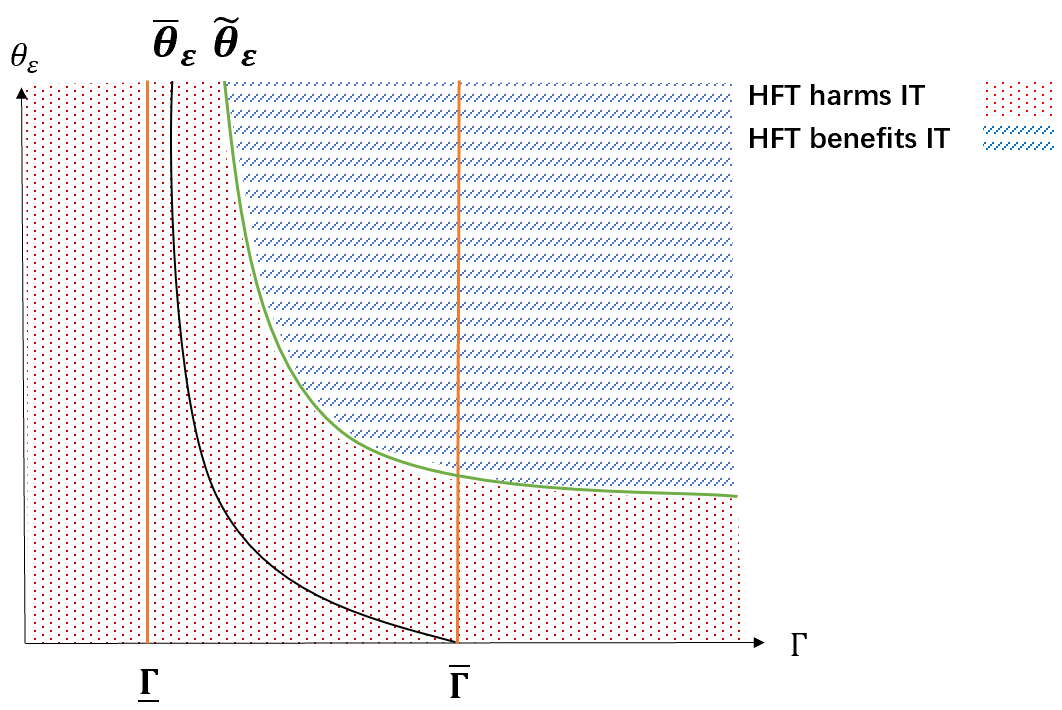}
\caption{HFT's influences on IT.}
\label{figconclusion2}
\end{figure}

\textbf{HFT's influences on IT.} Compared with the case without HFT, IT  is always harmed by Small-IT, but could make more profits with Round-Tripper. Given $\theta_1\in(0,1]$ and $\Gamma>\underline{\Gamma}(\theta_1)$, there exists a $\Tilde{\theta}_\varepsilon=\Tilde{\theta}_\varepsilon(\theta_1,\Gamma)\geq\overline{\theta}_\varepsilon(\theta_1,\Gamma),$ with $\frac{\partial\Tilde{\theta}_\varepsilon(\theta_1,\Gamma)}{\partial\theta_1}<0,\frac{\partial\Tilde{\theta}_\varepsilon(\theta_1,\Gamma)}{\partial\Gamma}<0,$ if $\theta_\varepsilon>\Tilde{\theta}_\varepsilon$, IT will be benefited by Round-Tripper, as shown in Figure \ref{figconclusion2}. 

\begin{figure}[!htbp]
    \centering
\includegraphics[height=0.2\textheight]{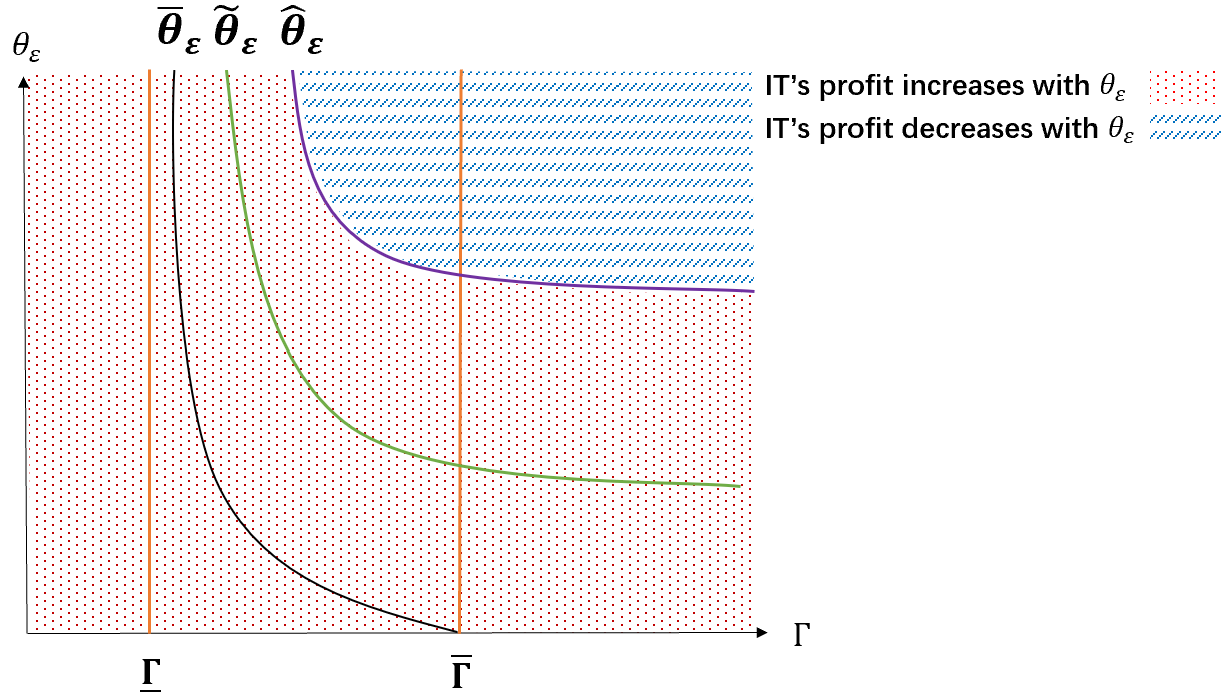}
\caption{$\theta_\varepsilon$'s influences on IT's profit.}
\label{figconclusion3}
\end{figure}

\noindent\textbf{$\theta_\varepsilon$'s influences on IT's profit.} When IT is harmed by HFT, increasing $\theta_\varepsilon$ always makes IT less harmed; when IT is benefited by HFT, there exists a $\hat{\theta}_\varepsilon=\hat{\theta}_\varepsilon(\theta_1,\Gamma)\geq\Tilde{\theta}_\varepsilon(\theta_1,\Gamma),$ with $\frac{\partial\hat{\theta}_\varepsilon(\theta_1,\Gamma)}{\partial\theta_1}<0,\frac{\partial\hat{\theta}_\varepsilon(\theta_1,\Gamma)}{\partial\Gamma}<0,$ if $\theta_\varepsilon>\hat{\theta}_\varepsilon,$ IT's profit will decrease with it, as shown in Figure \ref{figconclusion3}. 

As mentioned before, given $\theta_1$, the equilibrium is determined by two parameters $\theta_\varepsilon$ and $\Gamma$. Therefore, in the following parts, we will provide detailed explanations of the above conclusions, in the form of 3D graphs. These more detailed analyses also lead us to some interesting conclusions under extreme market conditions. 

The 3D graphs only present cases with
\begin{equation*}  \theta_\varepsilon\in[10^{-4},9],\ \text{i.e., }\frac{\sigma_\varepsilon}{\sigma_2}\in[10^{-2},3],
\end{equation*}
because when $\theta_\varepsilon<10^{-4}$ or $\theta_\varepsilon>9$, the results seem to have no essential changes.

\subsubsection{Strategies and profits of HFT and IT}
To show the results more clearly, we divide them into three situations with respect to HFT's different inventory-averse levels.
For IT, the profit has a linear relationship with the action, i.e., $\mathbbm{E}(\pi^{\text{IT}})=\frac{\sigma_2\sigma_v}{2}A,$ so we only show the results for profit. 

\begin{figure}[ht]
    \centering
\subcaptionbox{$\theta_1=10^{-4}.$}{
    \includegraphics[width = 0.28\textwidth]{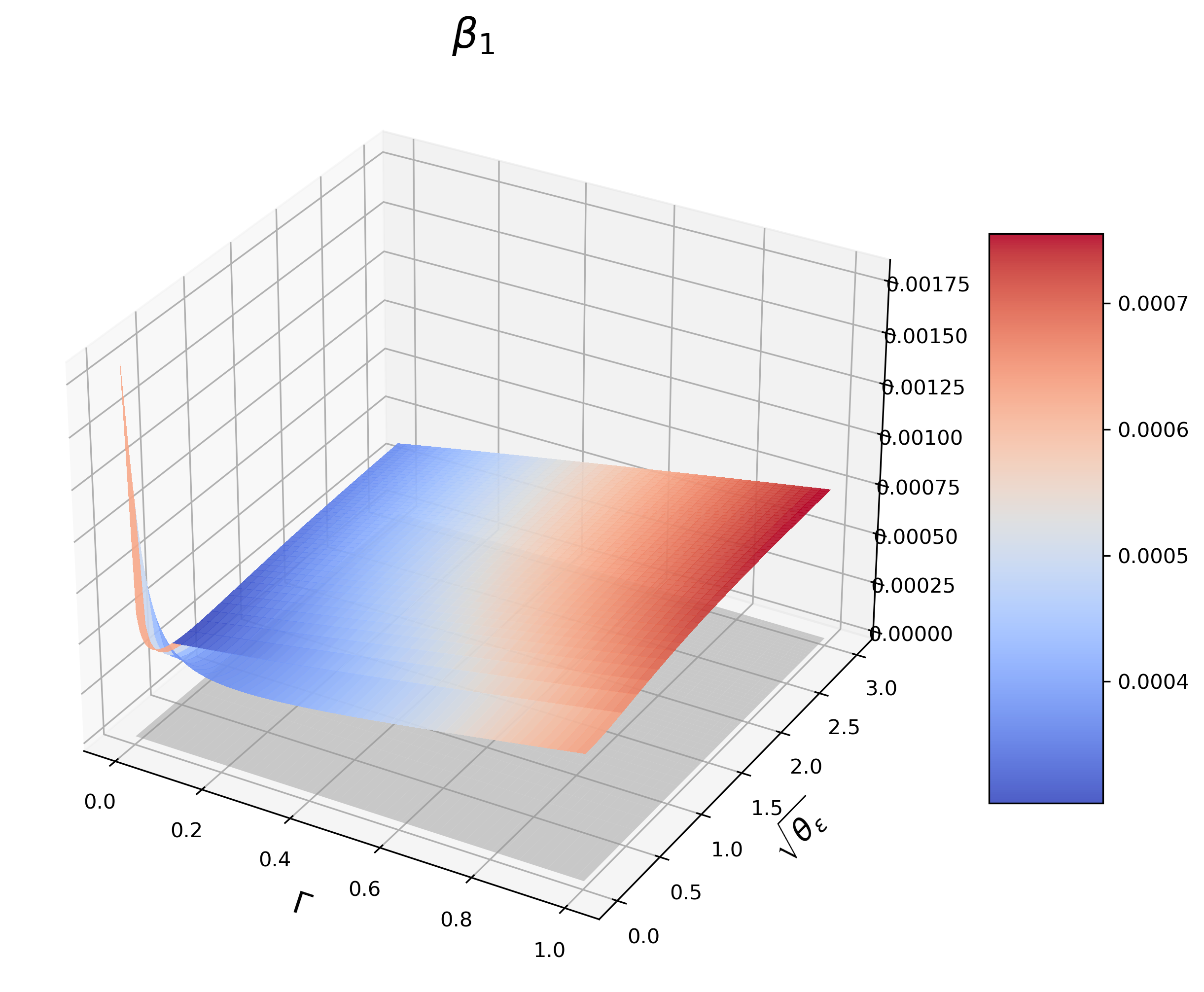}
    }
\subcaptionbox{$\theta_1=0.1.$}{
    \includegraphics[width = 0.28\textwidth]{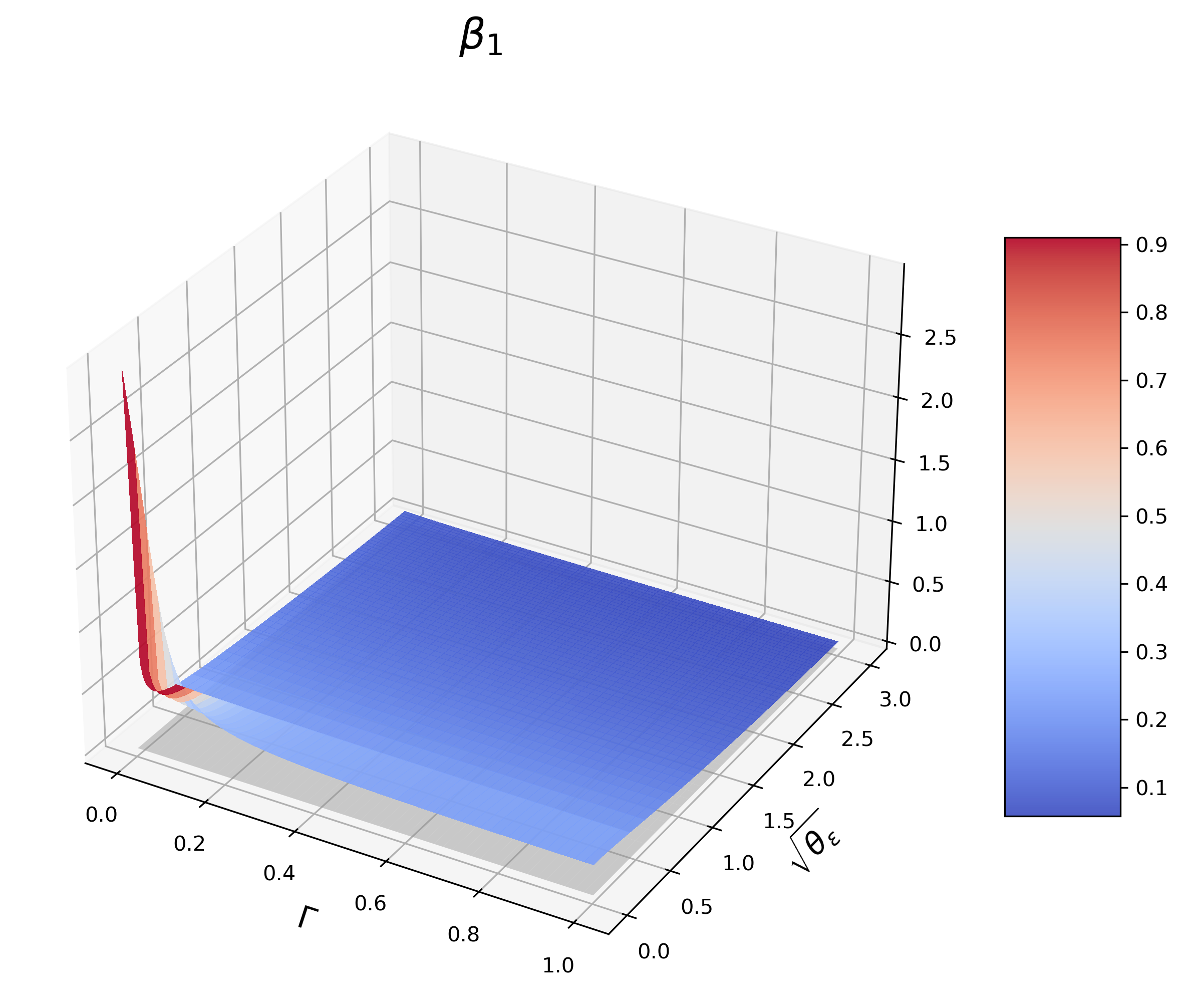}
    }
\subcaptionbox{$\theta_1=1.$}{
    \includegraphics[width = 0.28\textwidth]{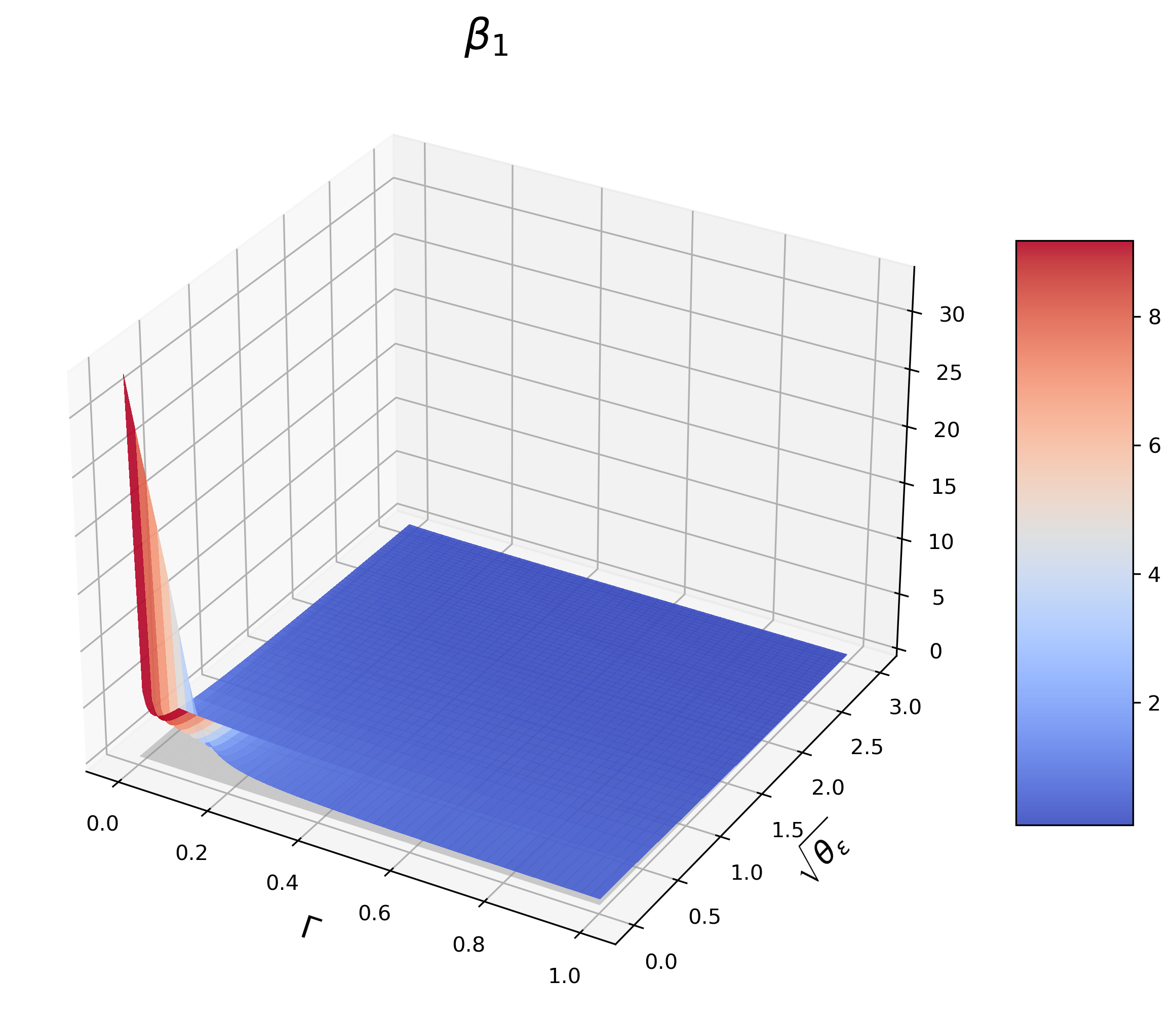}
    }
\caption{Direction of $x_1, \Gamma\in[0,1]$. The grey plane refers to $\beta_1=0$, \\
which is the dividing plane of $x_1$'s direction.}
\label{fig1direction1}
\end{figure}

\textbf{(1) HFT is light inventory averse $\boldsymbol{(\Gamma\in[0,1]).}$} Figure \ref{fig1direction1} presents the direction of HFT's order $x_1$: in period 1, HFT always builds up positions in the same direction as IT. 

\begin{figure}[htbp]
    \centering
\subcaptionbox{$\theta_1=10^{-4}.$}{
    \includegraphics[width = 0.28\textwidth]{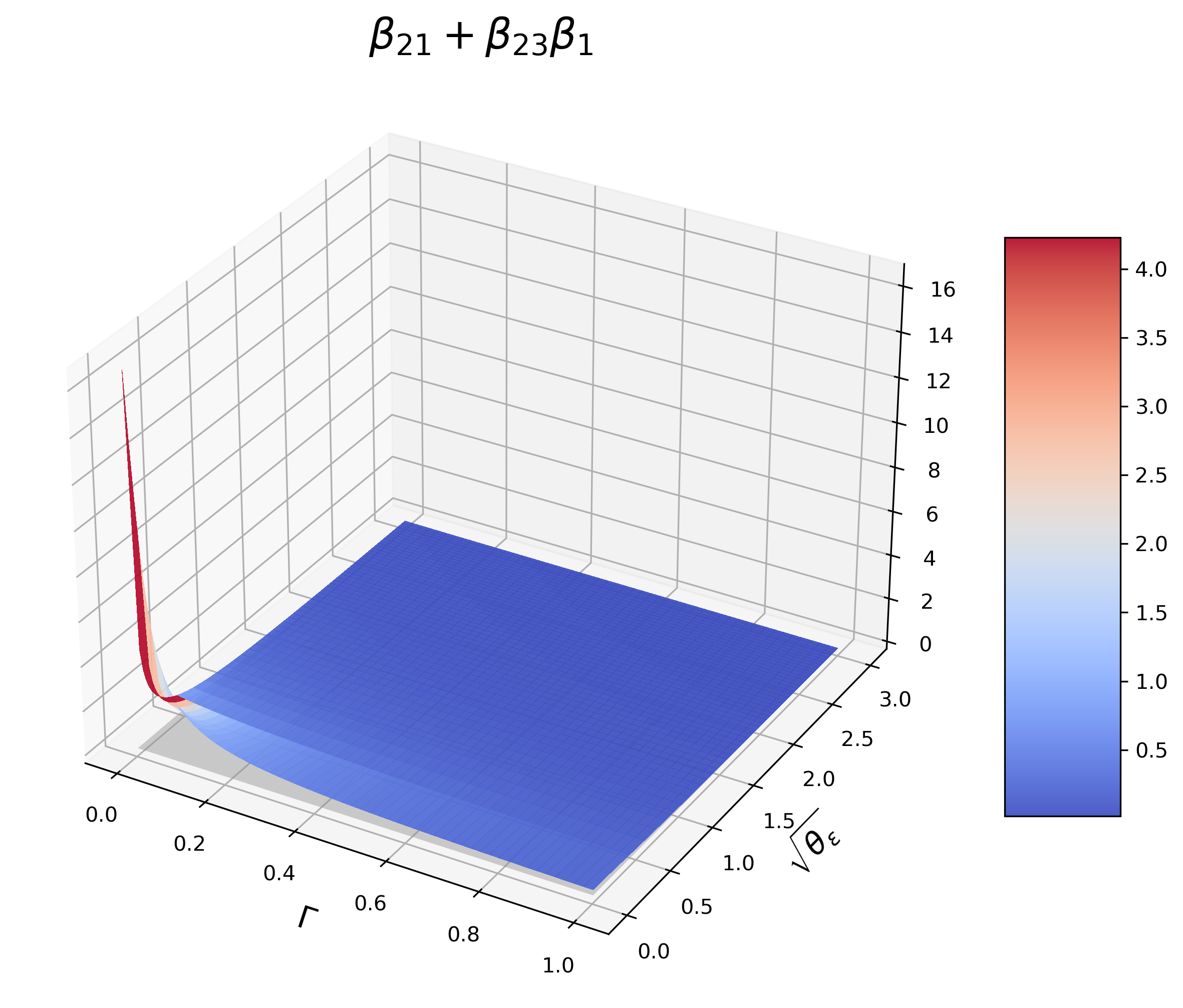}
    }
\subcaptionbox{$\theta_1=0.1.$}{
    \includegraphics[width = 0.28\textwidth]{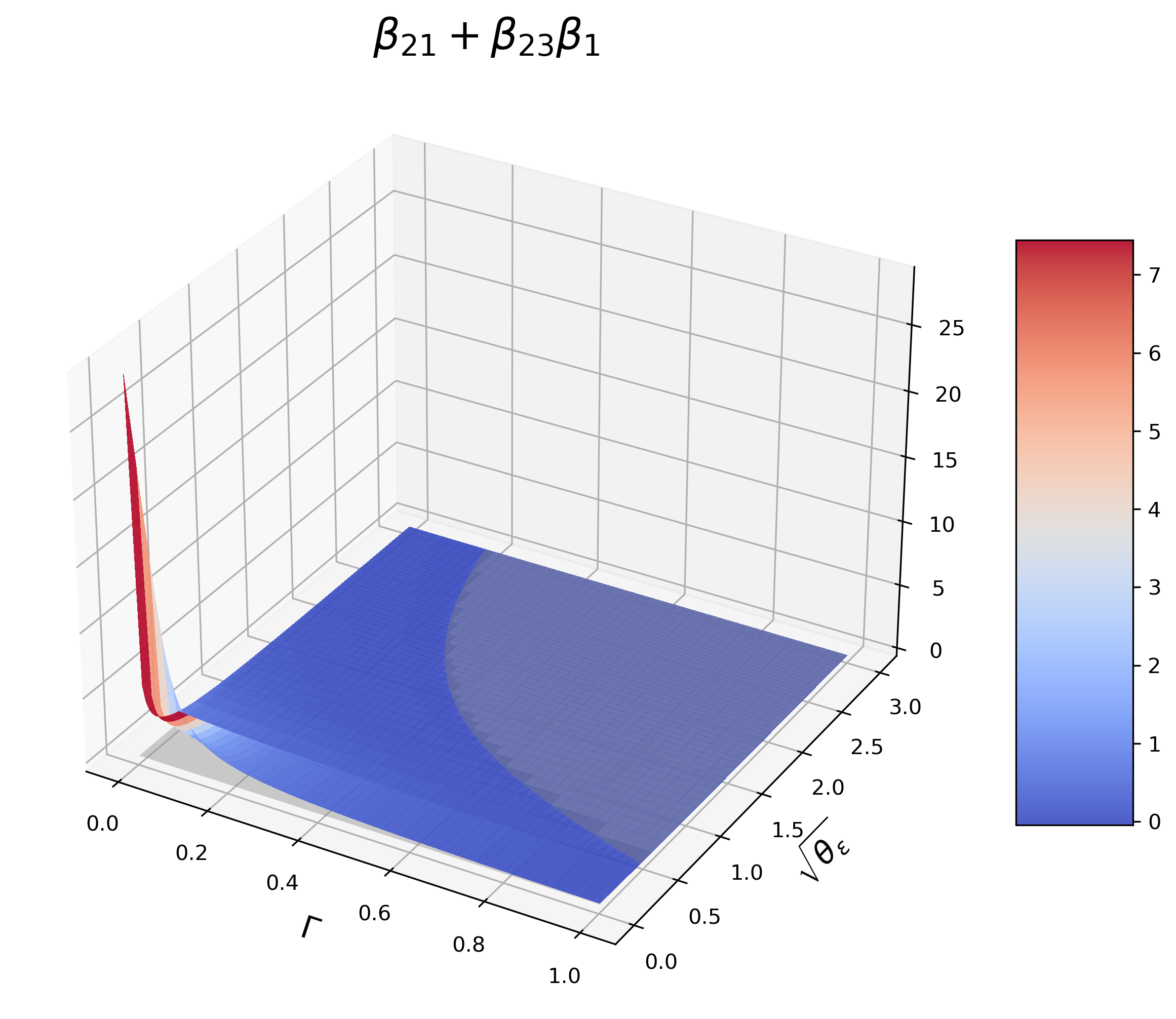}
    }
\subcaptionbox{$\theta_1=1.$}{
    \includegraphics[width = 0.28\textwidth]{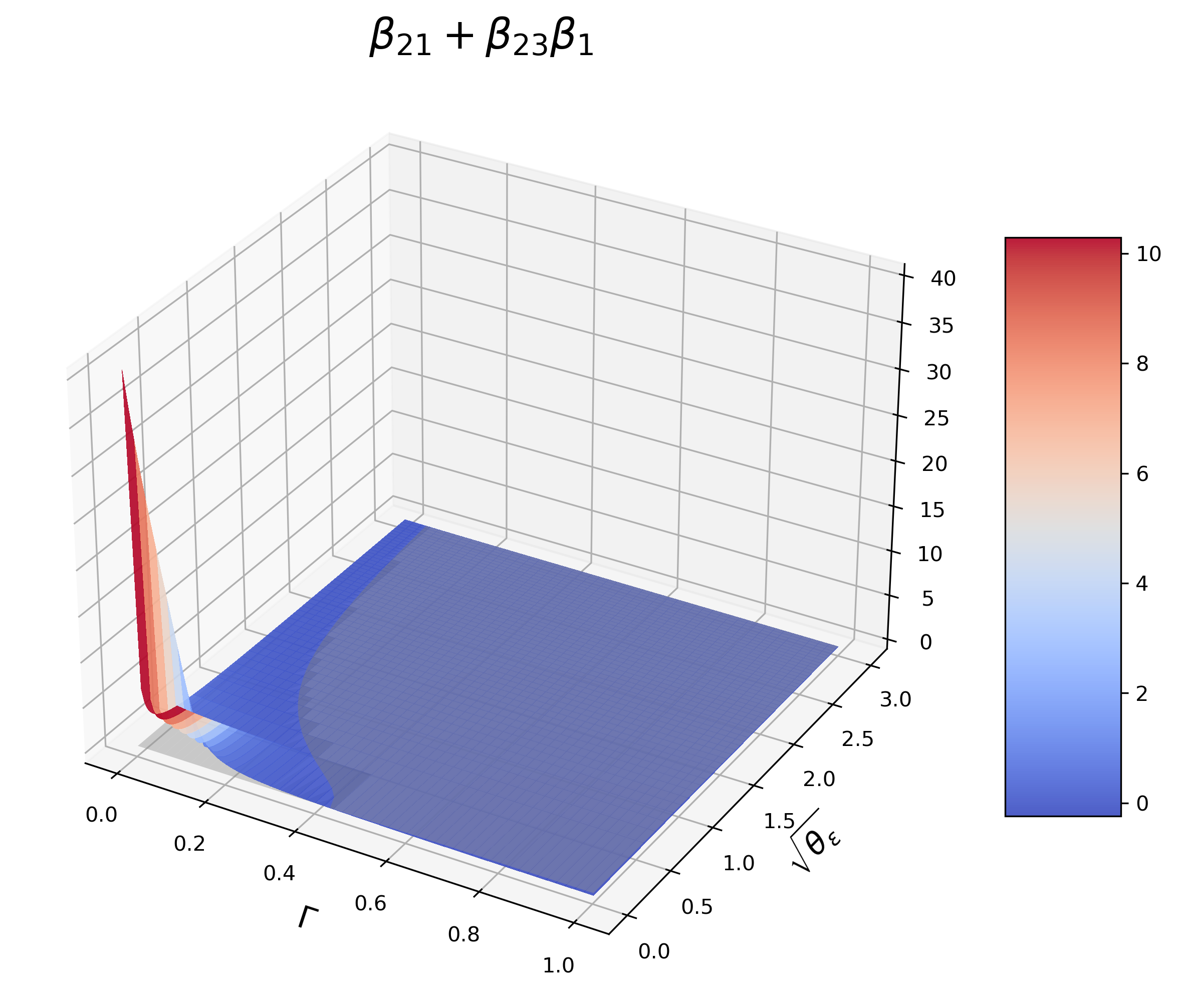}
    }
\caption{Direction of $x_2, \Gamma\in[0,1]$. The grey plane refers to $\beta_{21}+\beta_{23}\beta_1=0$, \\
which is the dividing plane of $x_2$'s direction.}
\label{fig1direction2}
\end{figure}

The direction of $x_2$ is shown in Figure \ref{fig1direction2}. When the colored surface falls below the grey plane, HFT provides liquidity for IT in period 2. As can be seen from (b) and (c), if the inventory penalty $\Gamma$ is mild, HFT acts as Small-IT. Otherwise, HFT acts as Round-Tripper when the signal is relatively vague. It also holds for $\theta_1=10^{-4}$, which will be illustrated later. The signal vagueness makes HFT quite uncertain about $v$, so in period 2, she closes some positions to reduce inventory cost.  From (c), we further see that when $\Gamma$ grows to a certain extent, HFT acts as Round-Tripper even if the signal is perfect.

Comparing (b) and (c) in Figure \ref{fig1direction2}, when $\theta_1$ is larger, HFT trades reversely for smaller $\Gamma$ and $\theta_\varepsilon$. Since larger the $\theta_1,$ less the impact $x_1$ can cause, and larger the position she is able to establish, which puts her under greater inventory pressure.

\begin{figure}[!htbp]
    \centering
\subcaptionbox{$\theta_1=10^{-4}.$}{
\includegraphics[width = 0.28\textwidth]{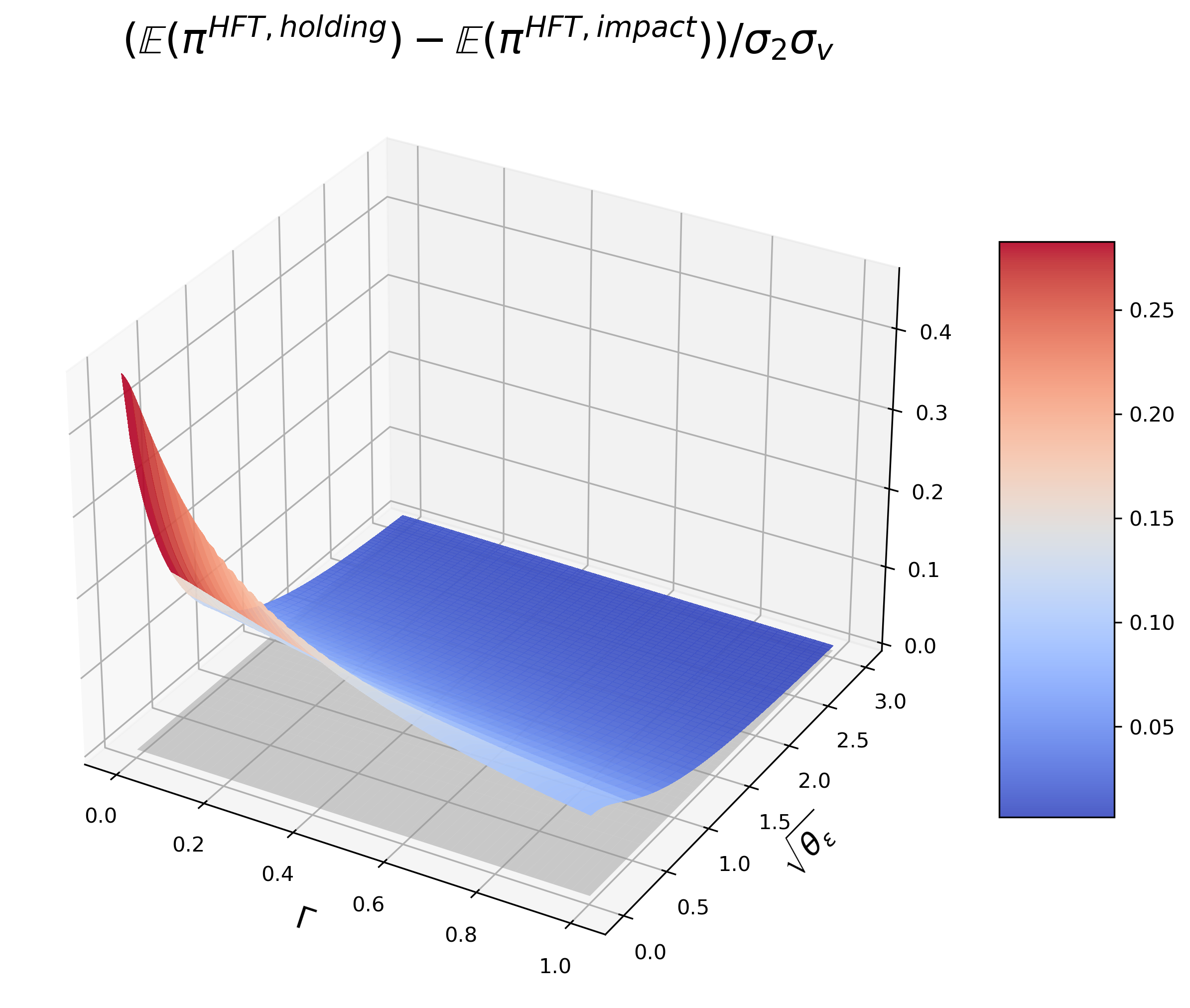}
    }
\subcaptionbox{$\theta_1=0.1.$}{
\includegraphics[width = 0.28\textwidth]{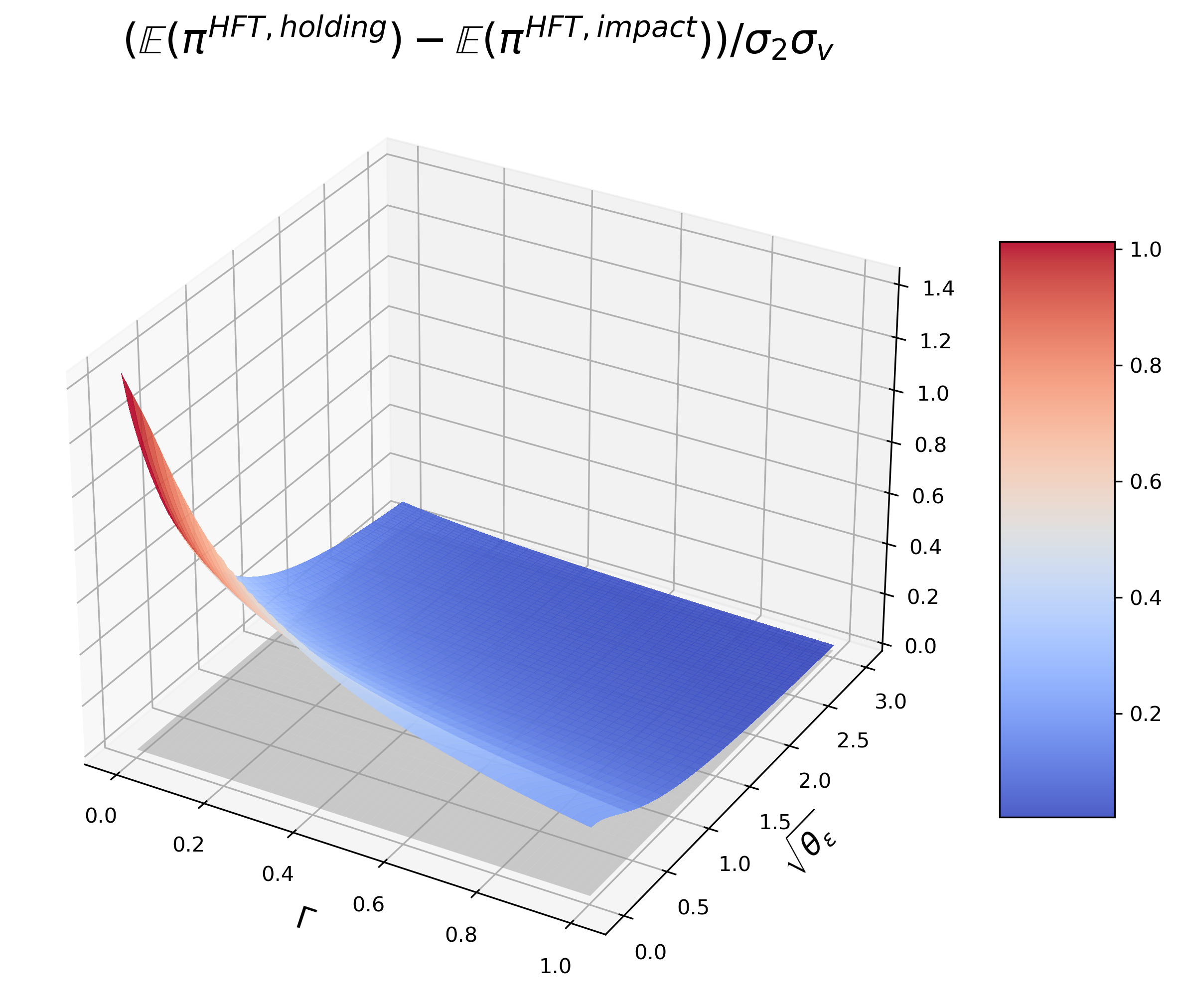}
    }
\subcaptionbox{$\theta_1=1.$}{
\includegraphics[width = 0.28\textwidth]{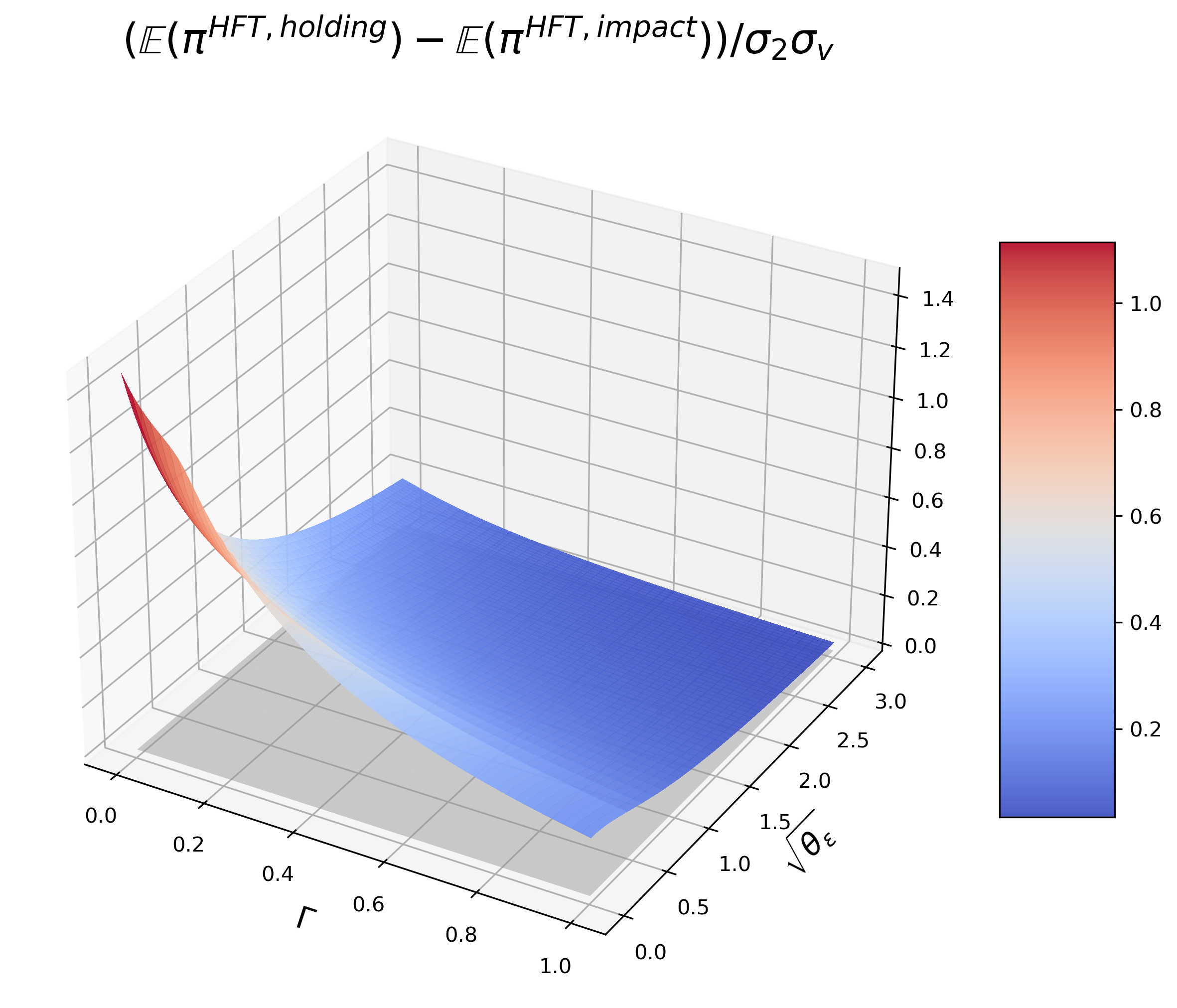}
    }
\caption{Comparation of HFT's holding and impact profit, $\Gamma\in[0,1]$.\\ The grey plane refers to $\mathbbm{E}(\pi^{\text{HFT,holding}})=\mathbbm{E}(\pi^{\text{HFT,impact}})$.}
\label{fig1piHFTmajor}
\end{figure}

Figure \ref{fig1piHFTmajor} compares HFT's holding profit and impact profit in \eqref{impactholding}, which indicates that the former has the upper hand. When acting as Small-IT or Round-Tripper with less inventory aversion, HFT infers information about $v$ and bets on it, the profits are mainly from investments in the risky asset. 

\begin{figure}[!htbp]
    \centering
\subcaptionbox{$\theta_1=10^{-4}.$}{
\includegraphics[width = 0.28\textwidth]{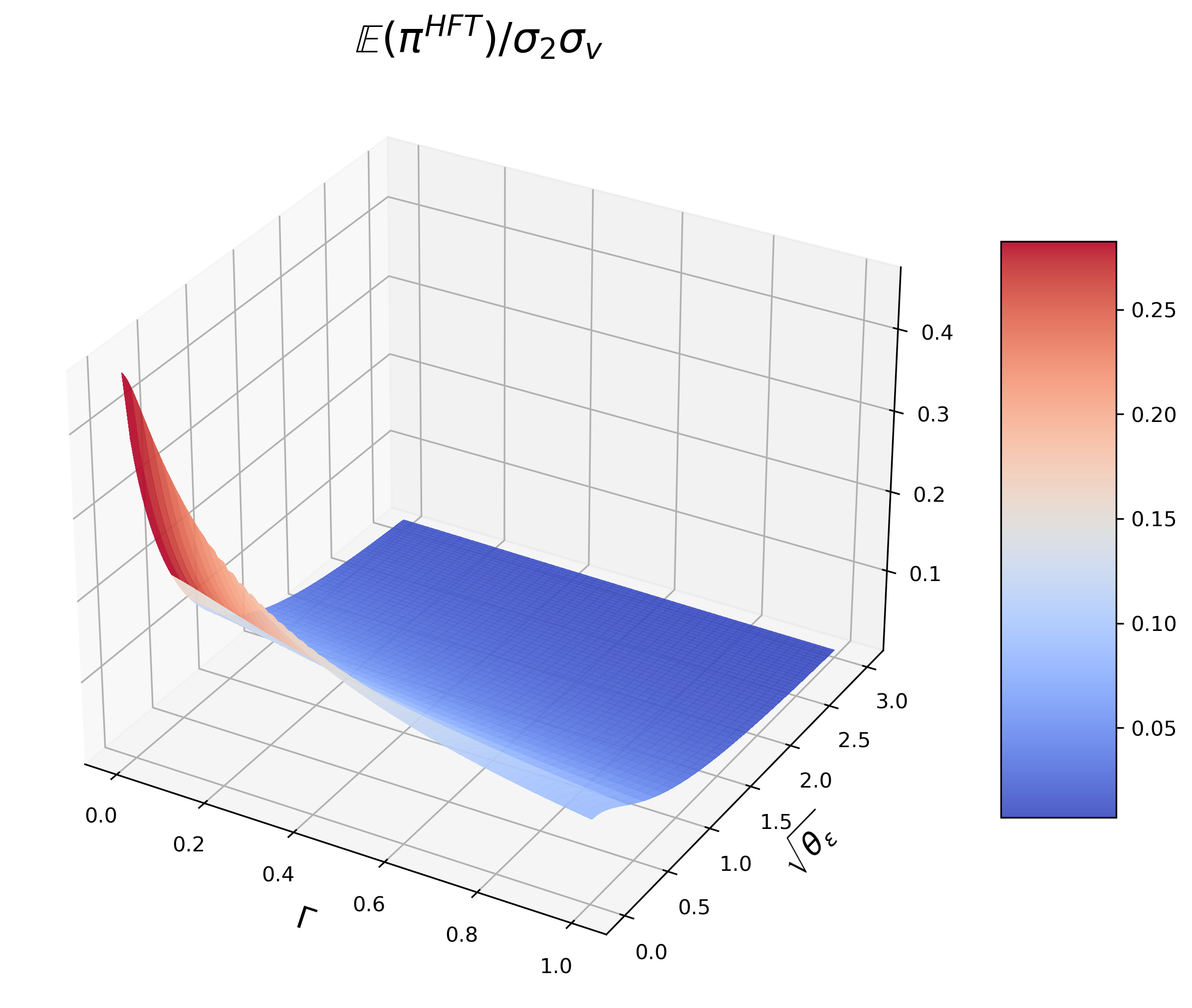}
    }
\subcaptionbox{$\theta_1=0.1.$}{
\includegraphics[width = 0.28\textwidth]{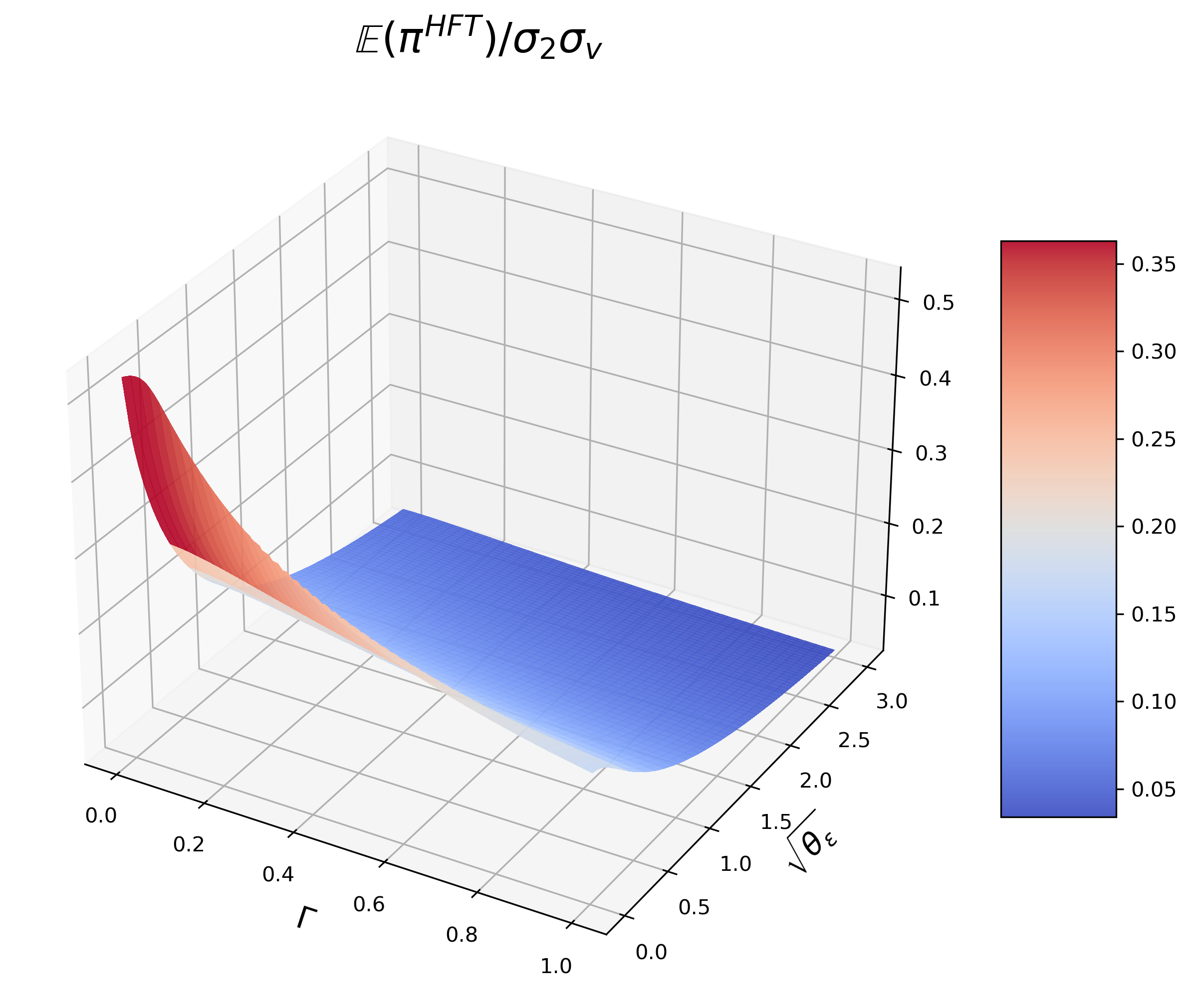}
    }
\subcaptionbox{$\theta_1=1.$}{
\includegraphics[width = 0.28\textwidth]{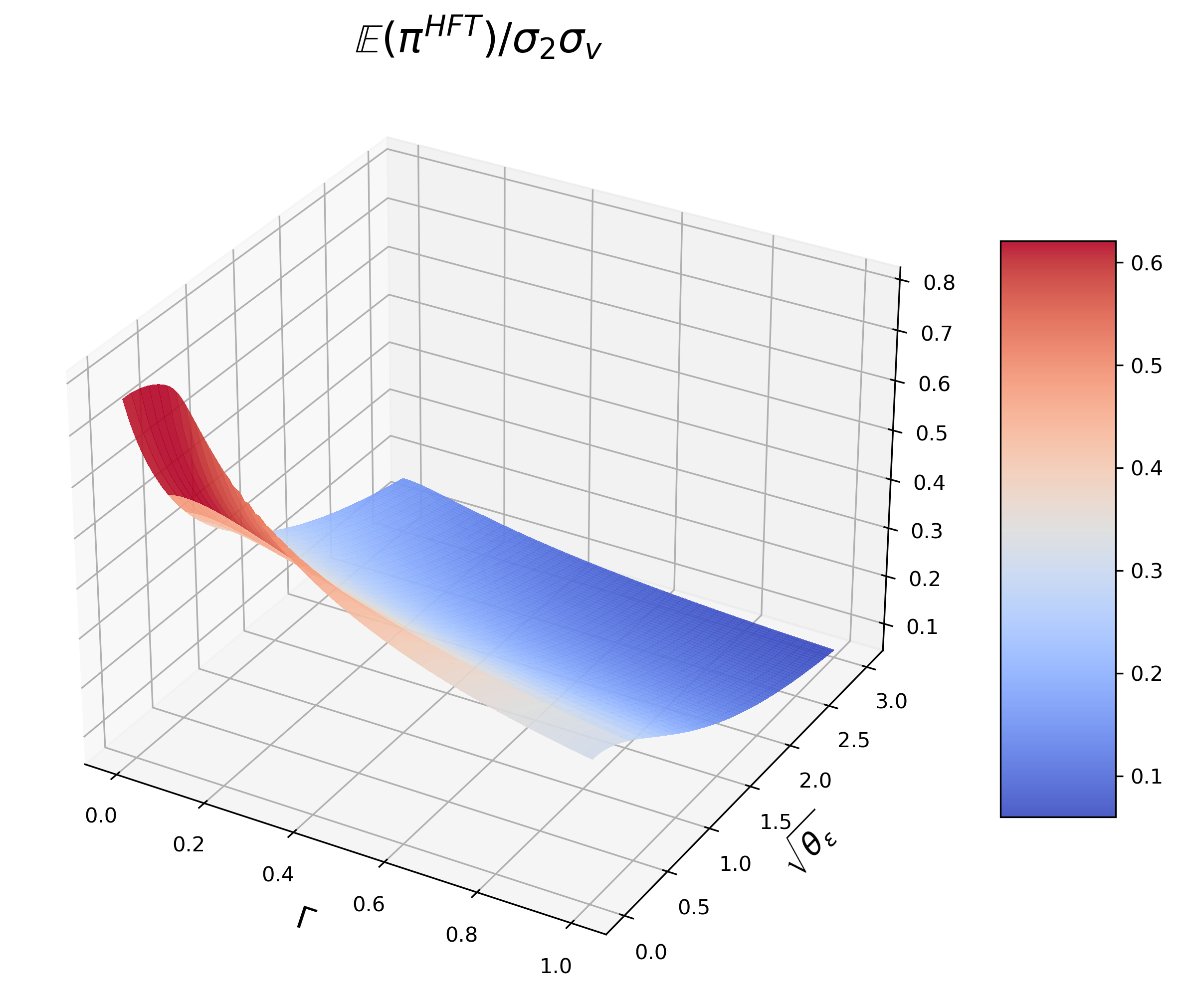}
    }
\caption{HFT's profit, $\Gamma\in[0,1]$.}
\label{fig1piHFT}
\end{figure}

HFT's total profit is given in Figure \ref{fig1piHFT}. In general, it decreases with $\theta_\varepsilon$ and $\Gamma$. A rough signal disables her to distinguish $v$ from $\hat{i}$ precisely, while the large inventory aversion prevents her from holding a large position, either of the above makes her trade more conservatively and receive fewer profits.

Surprisingly, from (c) of Figure \ref{fig1piHFT}, HFT's profit increases with her inventory aversion $\Gamma$ at the beginning, when $\theta_1$ is large and $\theta_\varepsilon$ is small. 
Since for minor $\Gamma$s, HFT's loss on signal inaccuracy drops rapidly as $\Gamma$ grows, which exceeds the decrease of investment income. When $\theta_1$ is small or $\theta_\varepsilon$ is large, the reduction in loss on signal noise is insignificant, as a result, HFT's profit decreases.

\begin{figure}[!htbp]
    \centering
\subcaptionbox{$\theta_1=10^{-4}.$}{
    \includegraphics[width = 0.28\textwidth]{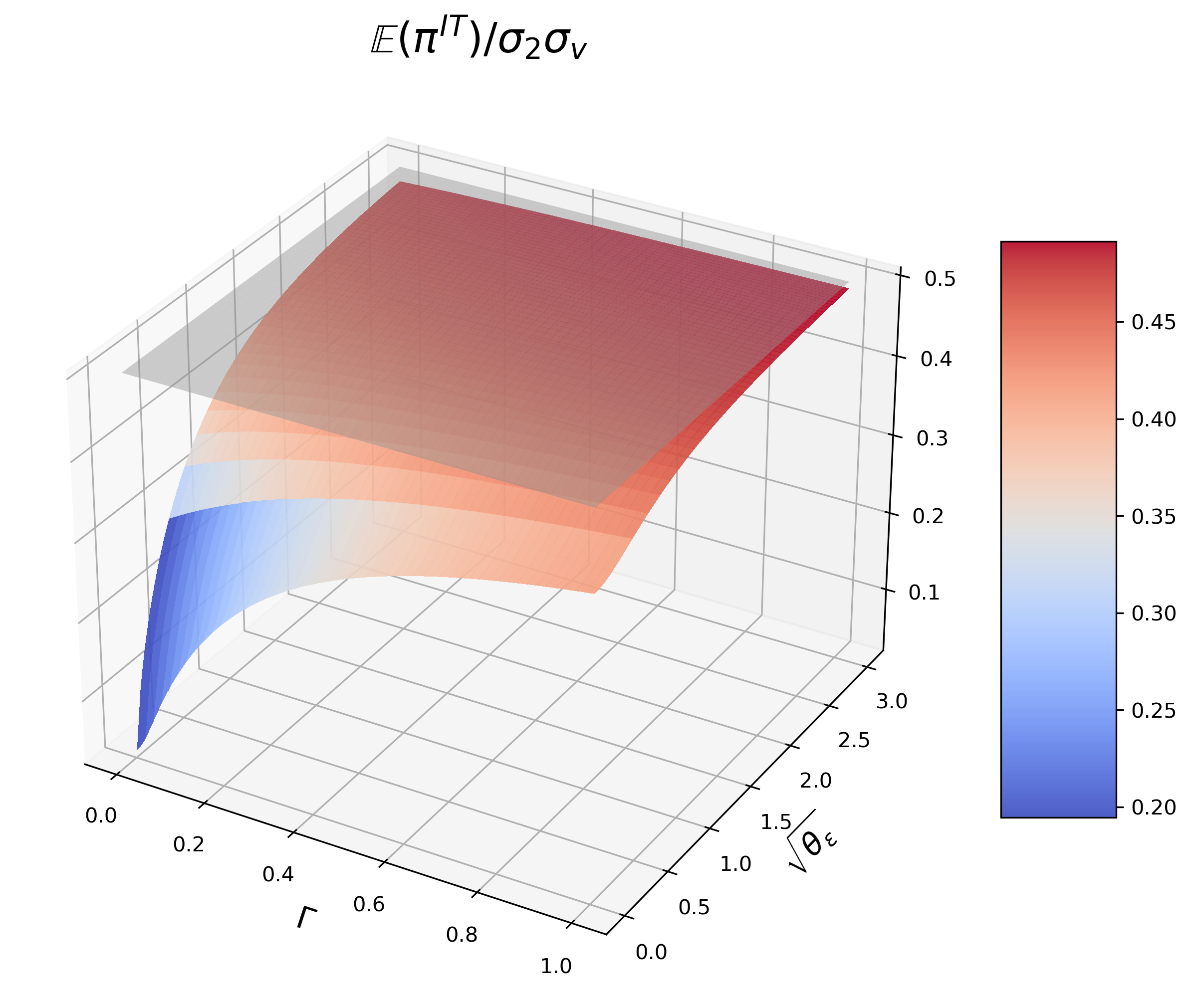}
    }
\subcaptionbox{$\theta_1=0.1.$}{
    \includegraphics[width = 0.28\textwidth]{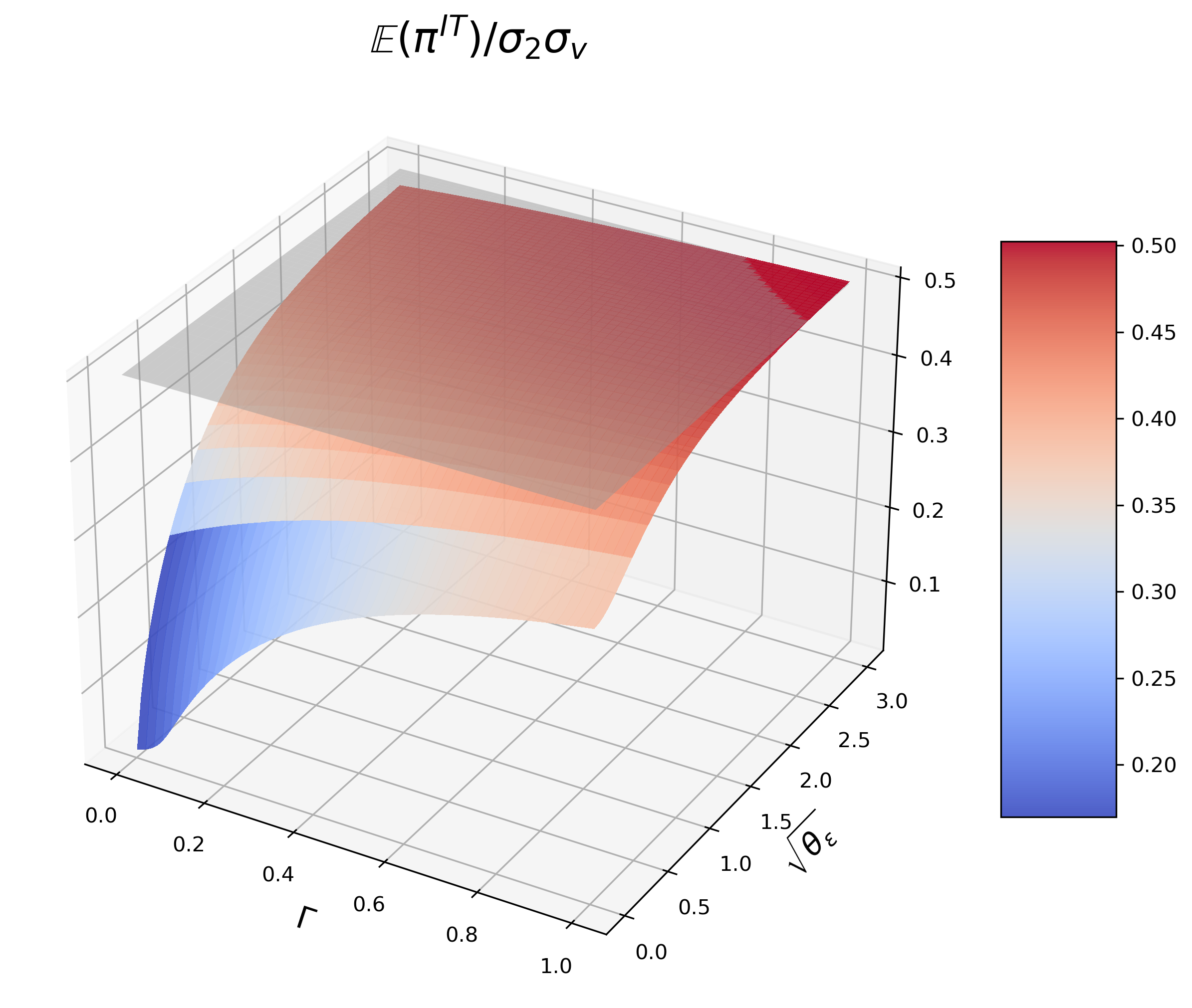}
    }
\subcaptionbox{$\theta_1=1.$}{
    \includegraphics[width = 0.28\textwidth]{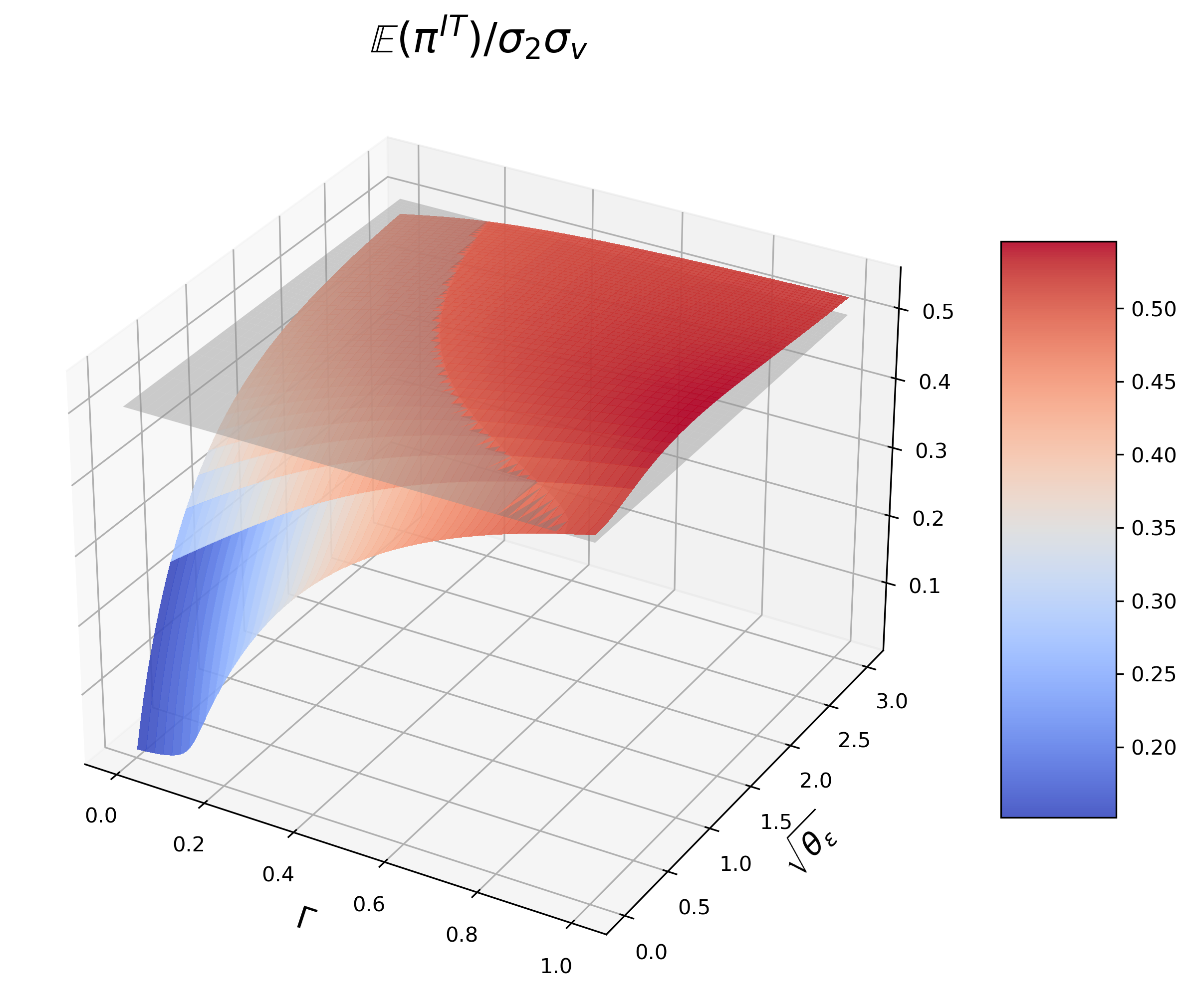}
    }
\caption{IT's profit, $\Gamma\in[0,1]$. The grey plane refers to $\mathbbm{E}(\pi^{\text{IT}})=\frac{\sigma_2\sigma_v}{2},$ \\
which is IT's profit without HFT.}
\label{fig1piIT}
\end{figure}

IT's profit is shown in Figure \ref{fig1piIT}.
Combining it and Figure \ref{fig1direction2}, we find that IT is always worse off with Small-IT. It is because the latter increases her price impact in both periods. When $\theta_1\geq0.1,$ HFT benefits IT as $\Gamma$ and $\theta_\varepsilon$ grow to a certain extent (see (b) and (c) of Figure \ref{fig1piIT}). In this case, HFT is Round-Tripper, the time-2 transaction cost she shares for IT exceeds the loss caused by her time-1 trades. When $\theta_1\in(0,0.1),$ it is also true but needs larger $\Gamma$ and $\theta_\varepsilon.$ 

Comparing IT's profits with different $\theta_1,$ we find that the phenomenon of HFT benefiting IT is more likely to appear for larger $\theta_1$. On the one hand, a larger $\theta_1$ leads HFT to take round-trip in more cases. On the other hand, more time-1 noise trading decreases $x_1$'s adverse impact. Both of the above are advantageous to IT.
Given $\theta_1$ and $\theta_\varepsilon,$ HFT's final position decreases with $\Gamma$, either from less same-direction trading or from more opposite-direction trading, which implies a greater likelihood of HFT benefiting IT. To sum up, there is a critical value $\Tilde{\theta}_\varepsilon$, with which HFT benefits IT, and it decreases with $\theta_1$ and $\Gamma$.

Given $\theta_1$ and $\Gamma$, in the area where HFT harms IT, IT's profit always increases with $\theta_\varepsilon$, since the signal noise protects IT from being precisely detected.

\begin{figure}[!htbp]
    \centering
\subcaptionbox{$\theta_1=10^{-4}.$}{
    \includegraphics[width = 0.28\textwidth]{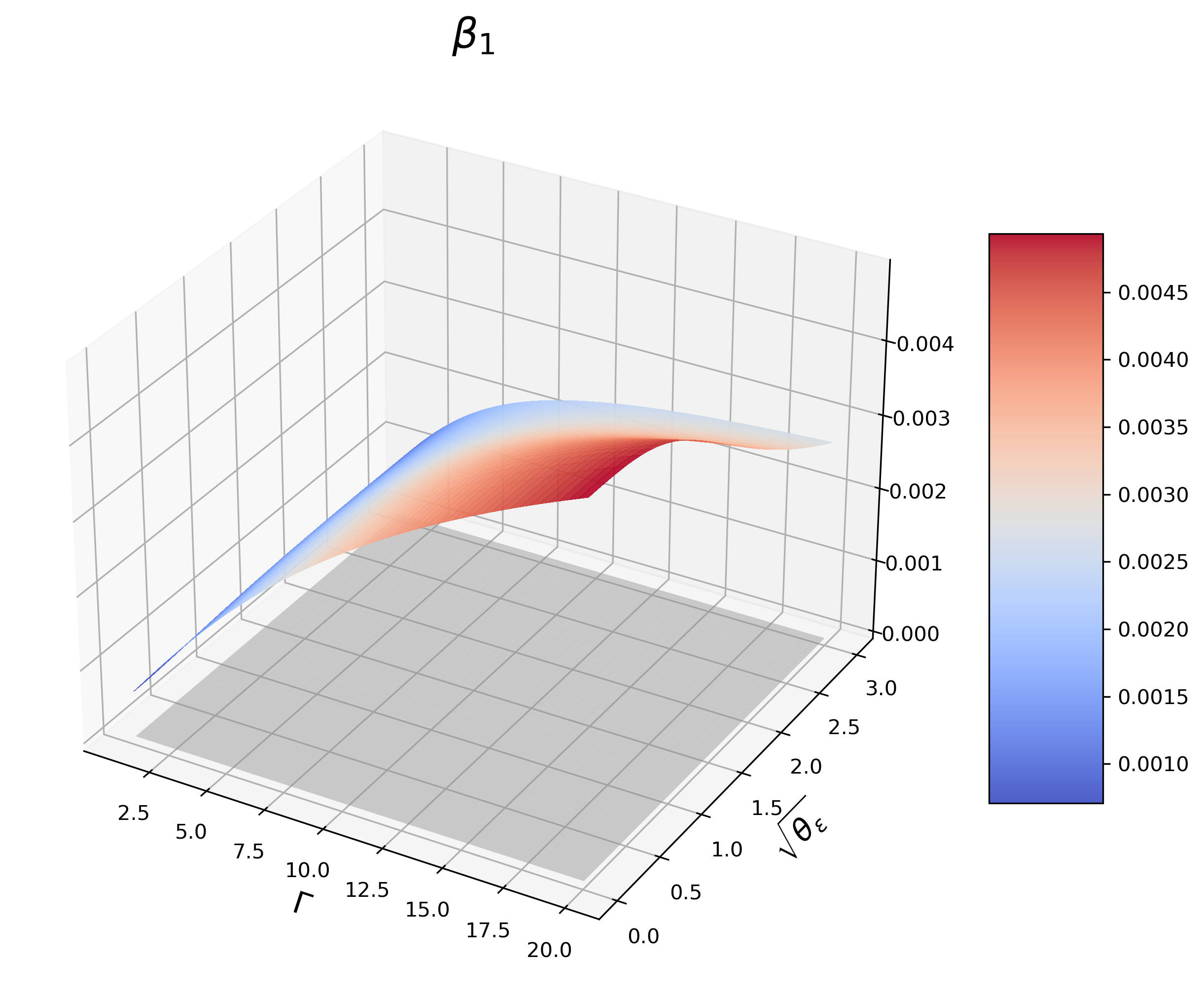}
    }
\subcaptionbox{$\theta_1=0.1.$}{
    \includegraphics[width = 0.28\textwidth]{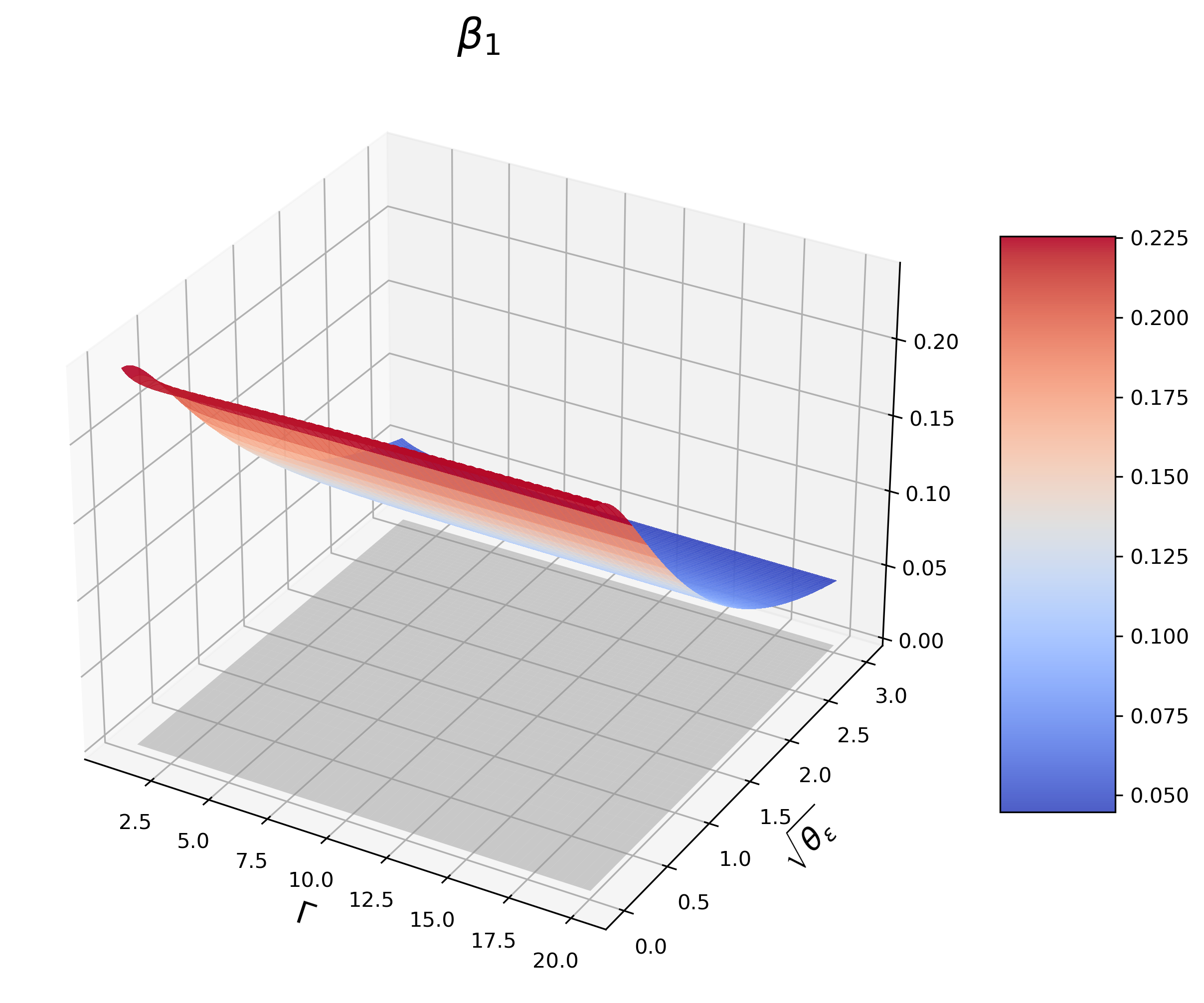}
    }
\subcaptionbox{$\theta_1=1.$}{
    \includegraphics[width = 0.28\textwidth]{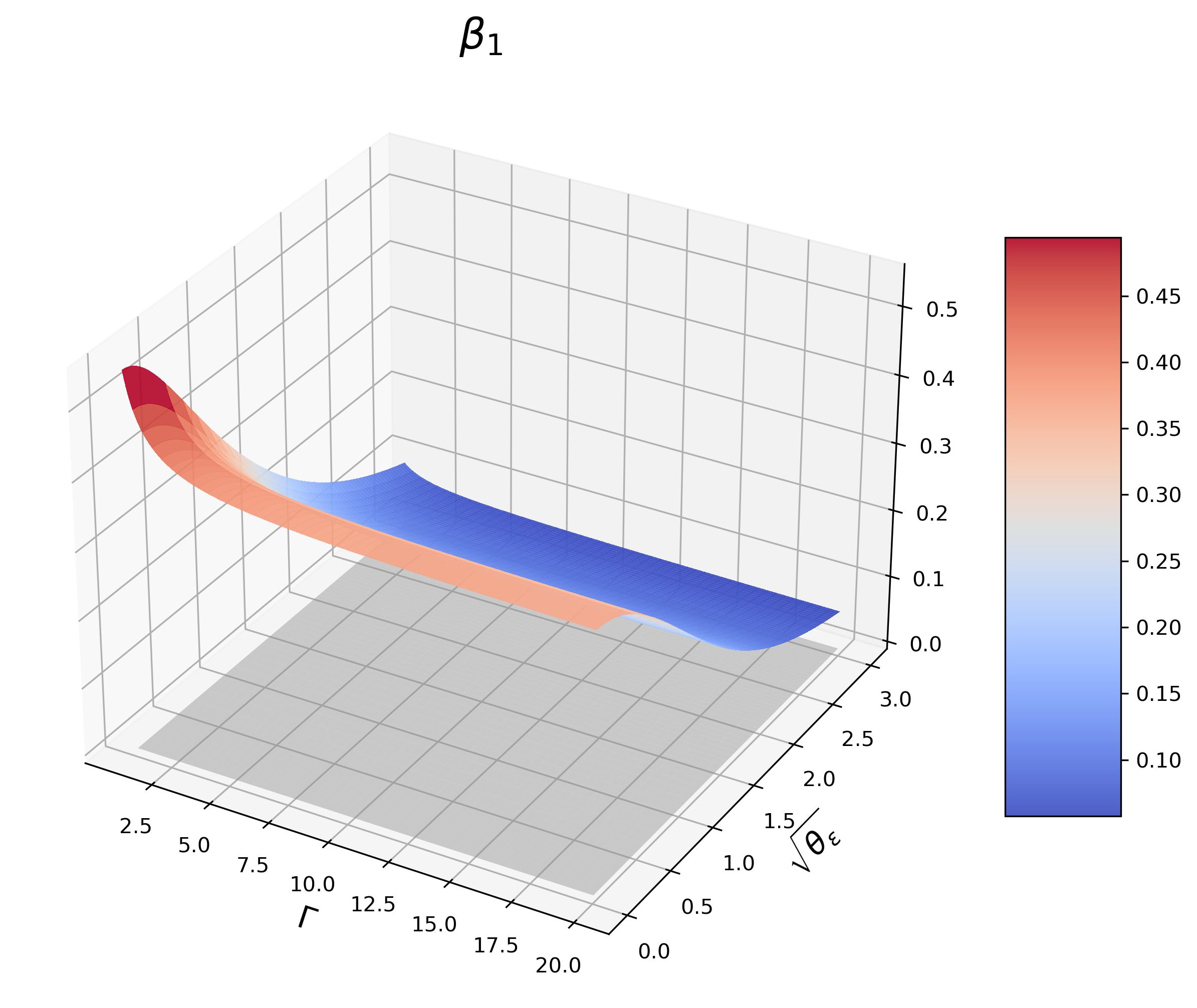}
    }
\caption{Direction of $x_1, \Gamma\in[1,20]$. The grey plane refers to $\beta_1=0$, \\
which is the dividing plane of $x_1$'s direction.}
\label{fig2direction1}
\end{figure}

\begin{figure}[!htbp]
    \centering
\subcaptionbox{$\theta_1=10^{-4}.$}{
    \includegraphics[width = 0.28\textwidth]{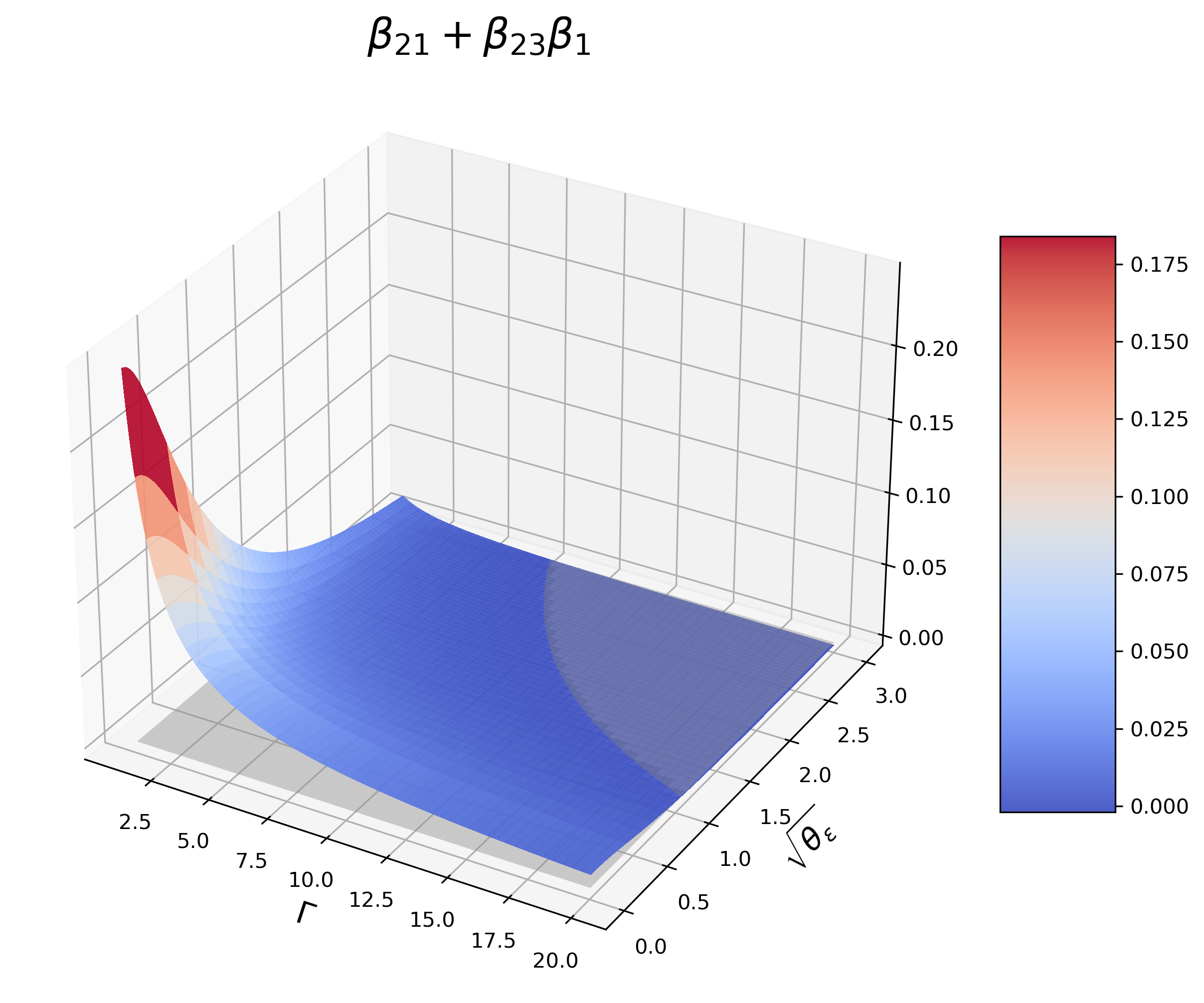}
    }
\subcaptionbox{$\theta_1=0.1.$}{
    \includegraphics[width = 0.28\textwidth]{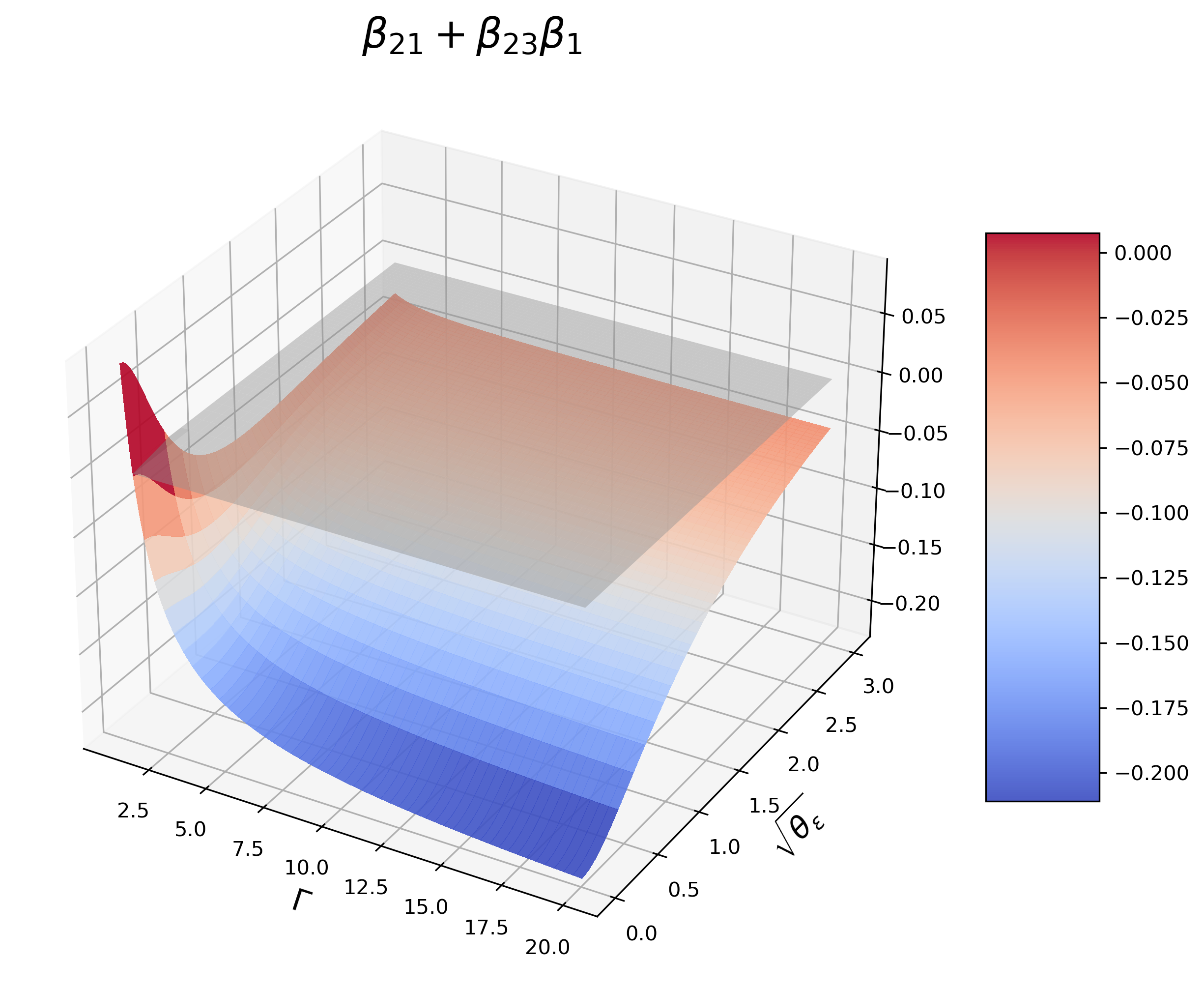}
    }
\subcaptionbox{$\theta_1=1.$}{
    \includegraphics[width = 0.28\textwidth]{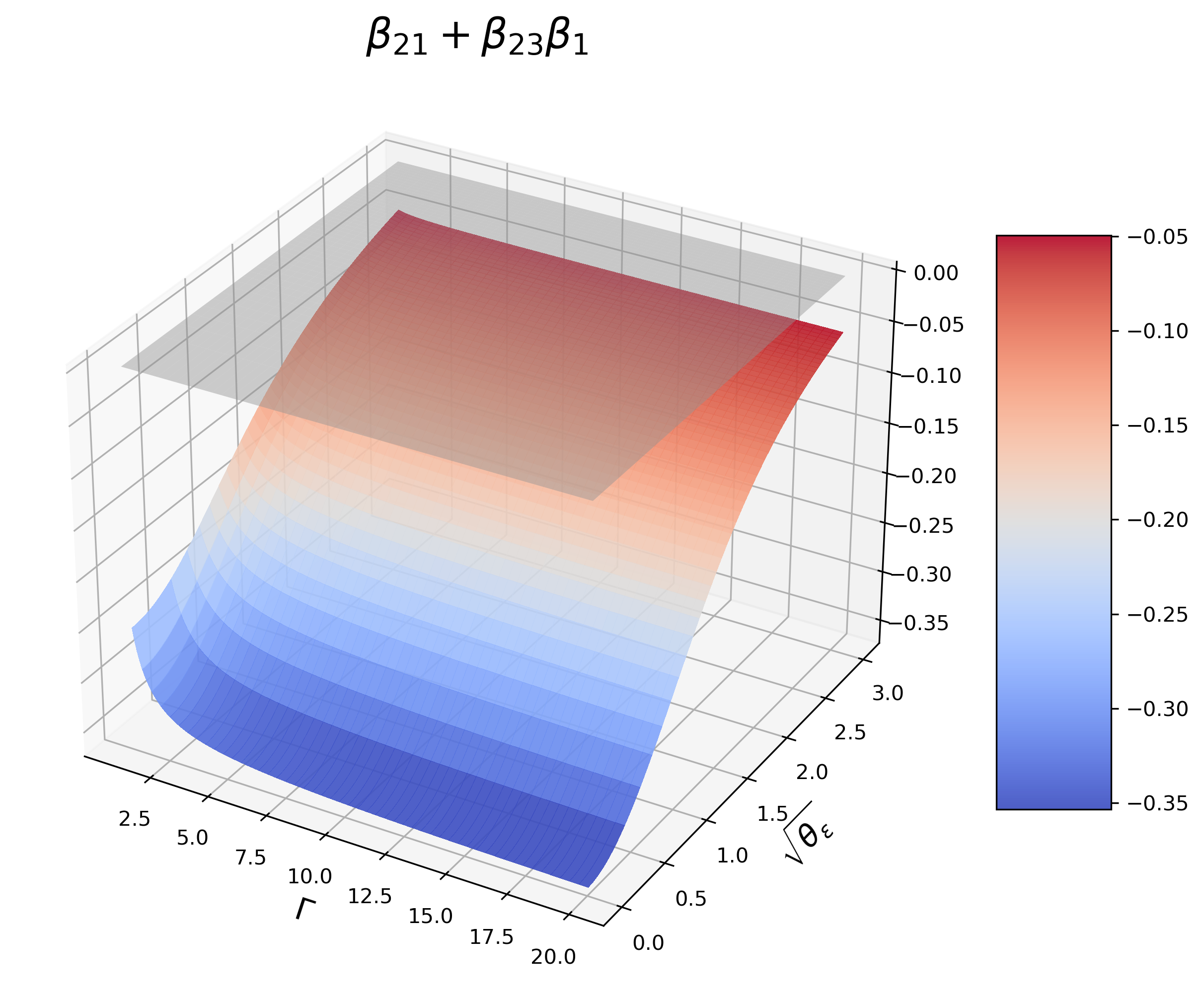}
    }
\caption{Direction of $x_2, \Gamma\in[1,20]$. The grey plane refers to $\beta_{21}+\beta_{23}\beta_1=0$, \\
which is the dividing plane of $x_2$'s direction.}
\label{fig2direction2}
\end{figure}

\textbf{(2) HFT is medium inventory averse $\boldsymbol{(\Gamma\in[1,20]).}$} Relying on the speed advantage, HFT still trades in the same direction as IT at $t=1$ (see Figure \ref{fig2direction1}). Comparing (a) of Figure \ref{fig1direction2} and Figure \ref{fig2direction2}, when HFT is more inventory averse, she is more likely to provide liquidity back at $t=2$. 

\begin{figure}[!htbp]
    \centering
\subcaptionbox{$\theta_1=10^{-4}.$}{
\includegraphics[width = 0.28\textwidth]{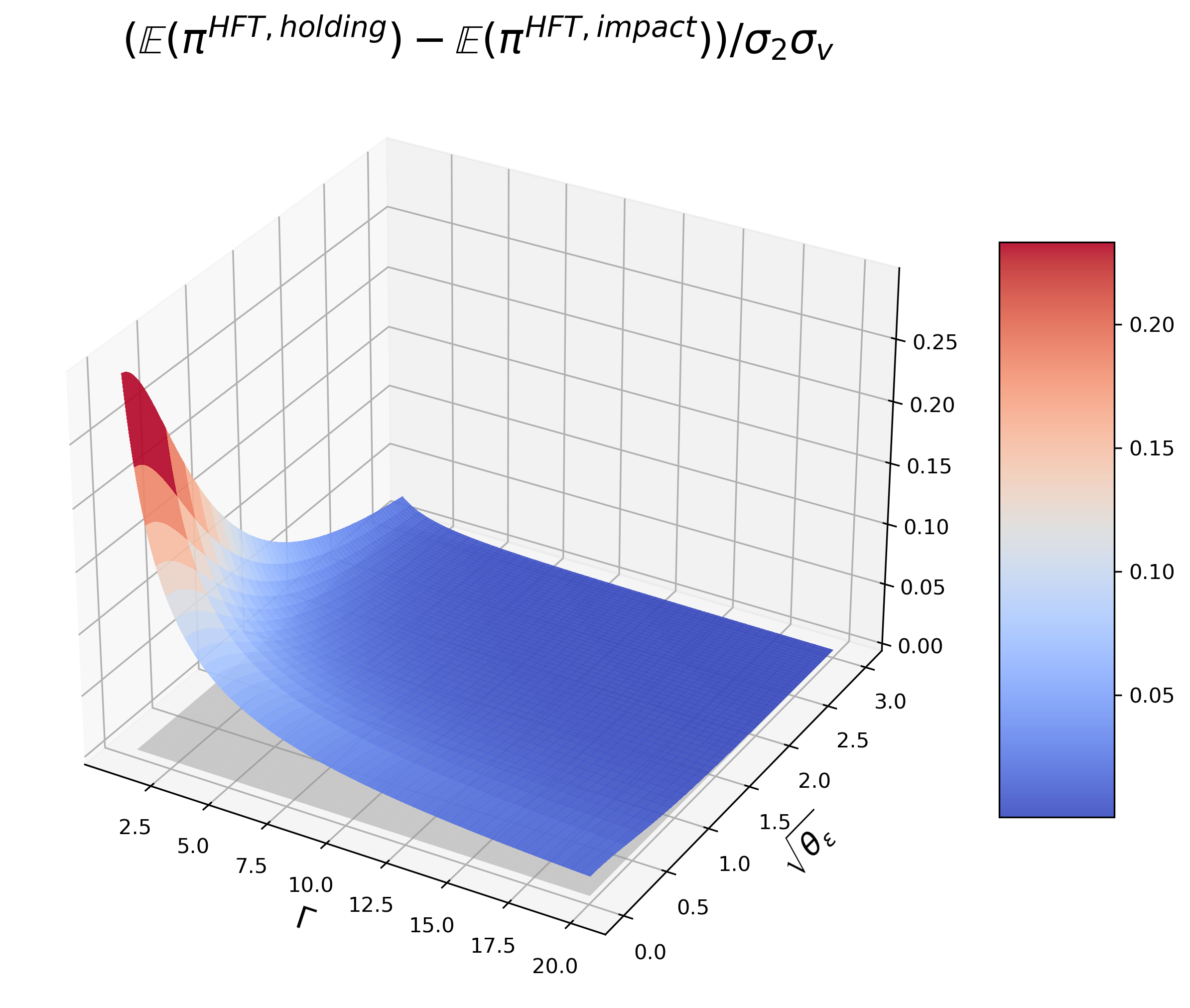}
    }
\subcaptionbox{$\theta_1=0.1.$}{
\includegraphics[width = 0.28\textwidth]{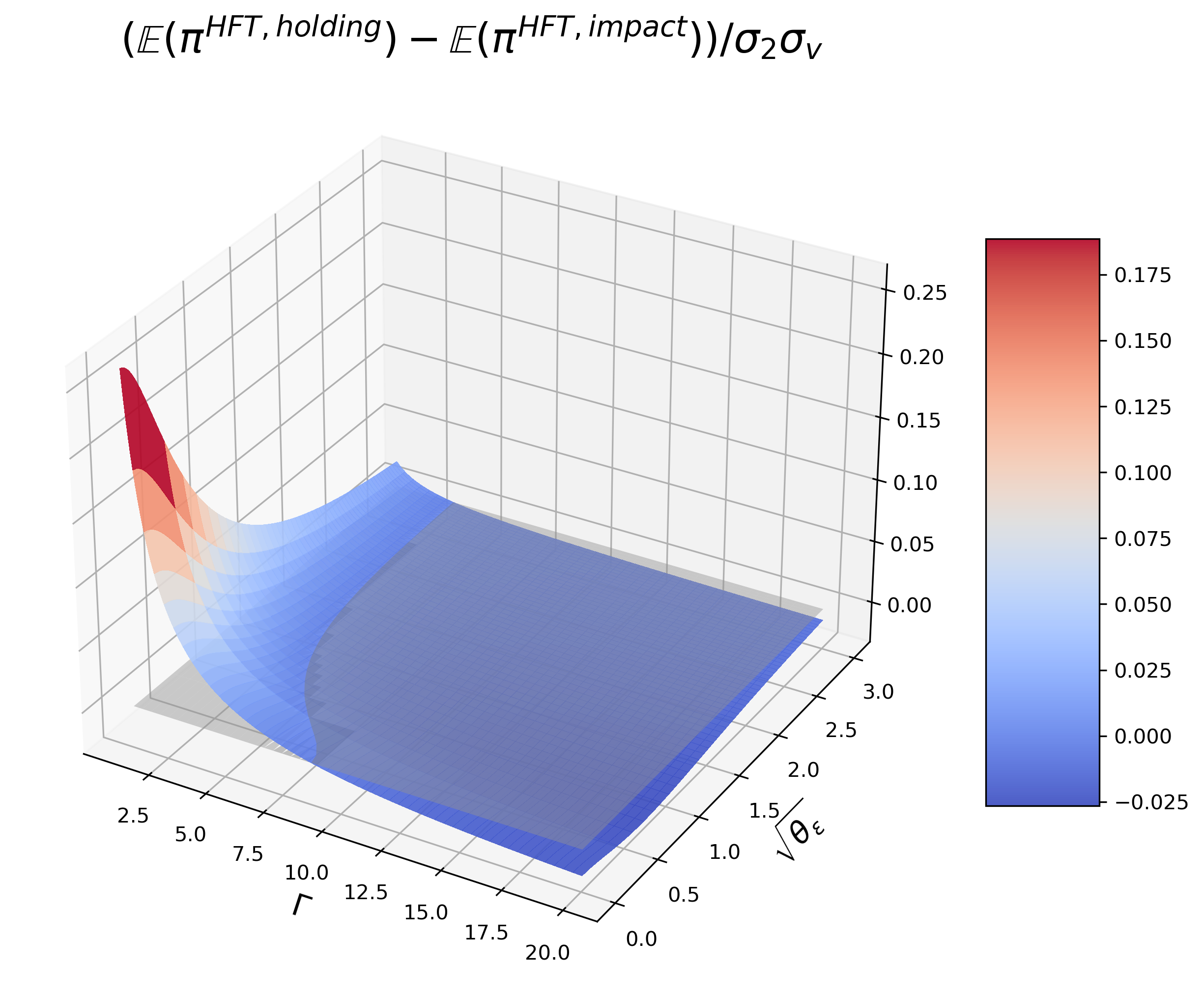}
    }
\subcaptionbox{$\theta_1=1.$}{
\includegraphics[width = 0.28\textwidth]{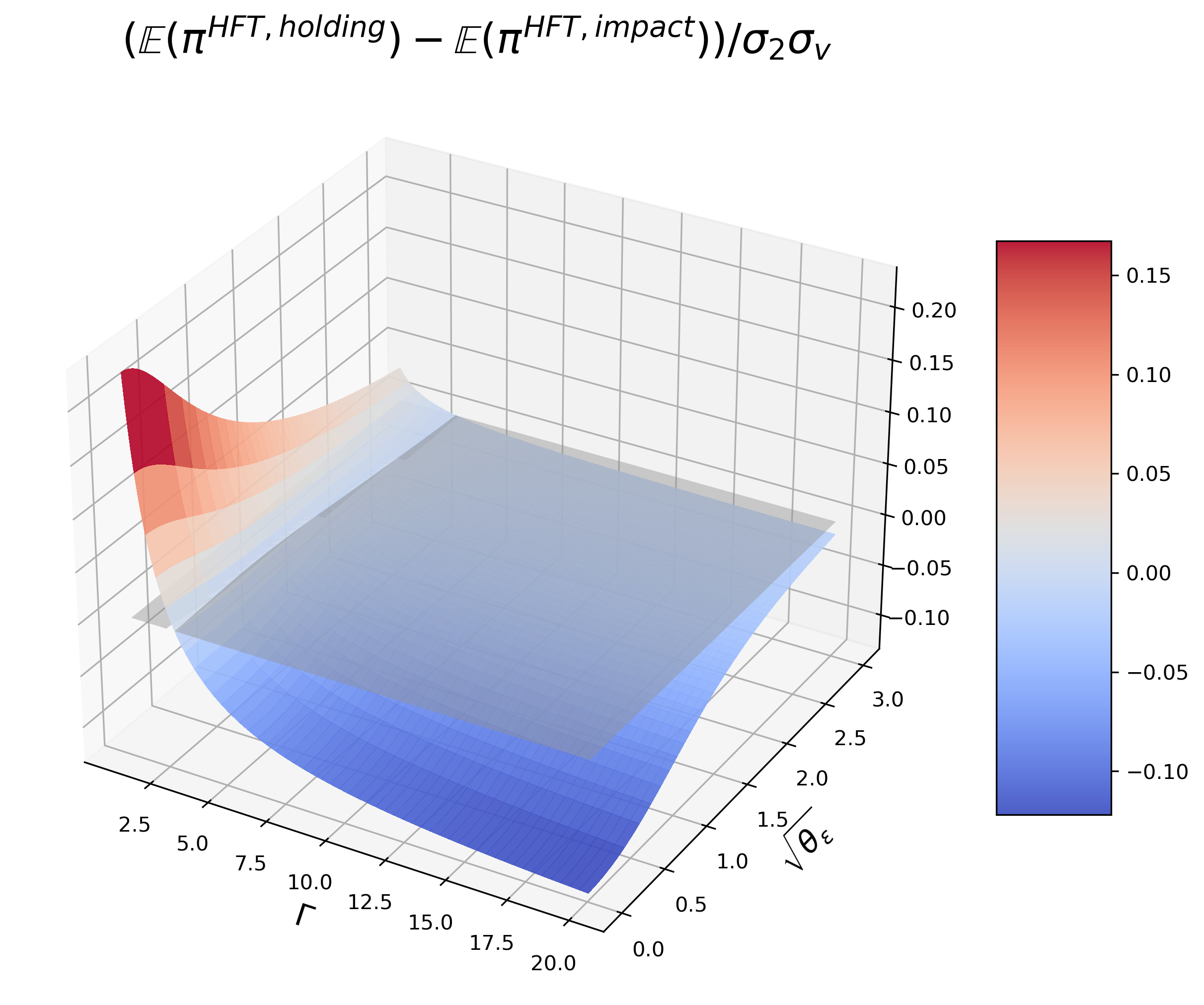}
    }
\caption{Comparation of HFT's holding and impact profit, $\Gamma\in[1,20]$.\\ The grey plane refers to $\mathbbm{E}(\pi^{\text{HFT,holding}})=\mathbbm{E}(\pi^{\text{HFT,impact}})$.}
\label{fig2piHFTmajor}
\end{figure}

\begin{figure}[!htbp]
    \centering
\subcaptionbox{$\theta_1=10^{-4}.$}{
    \includegraphics[width = 0.28\textwidth]{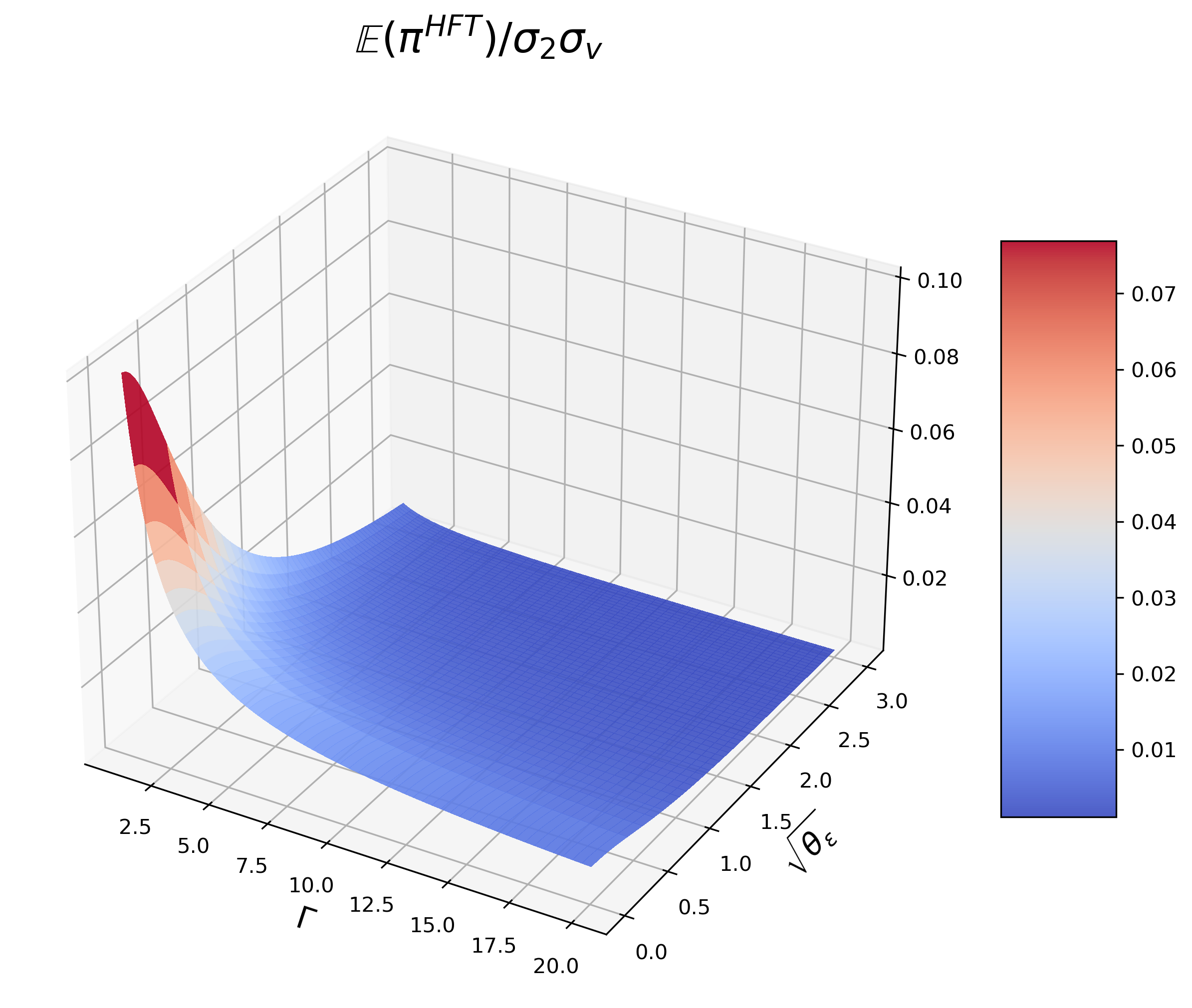}
    }
\subcaptionbox{$\theta_1=0.1.$}{
    \includegraphics[width = 0.28\textwidth]{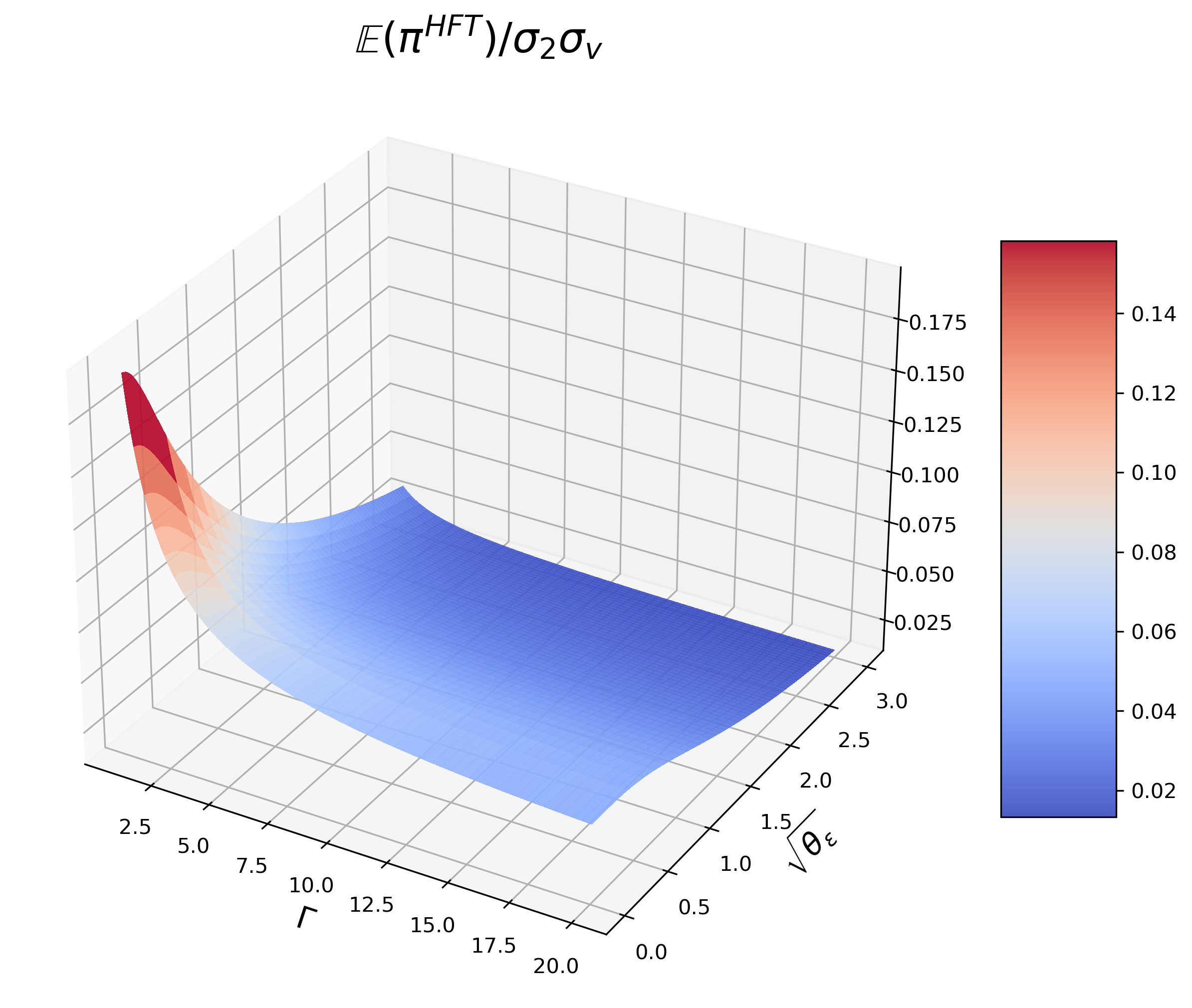}
    }
\subcaptionbox{$\theta_1=1.$}{
    \includegraphics[width = 0.28\textwidth]{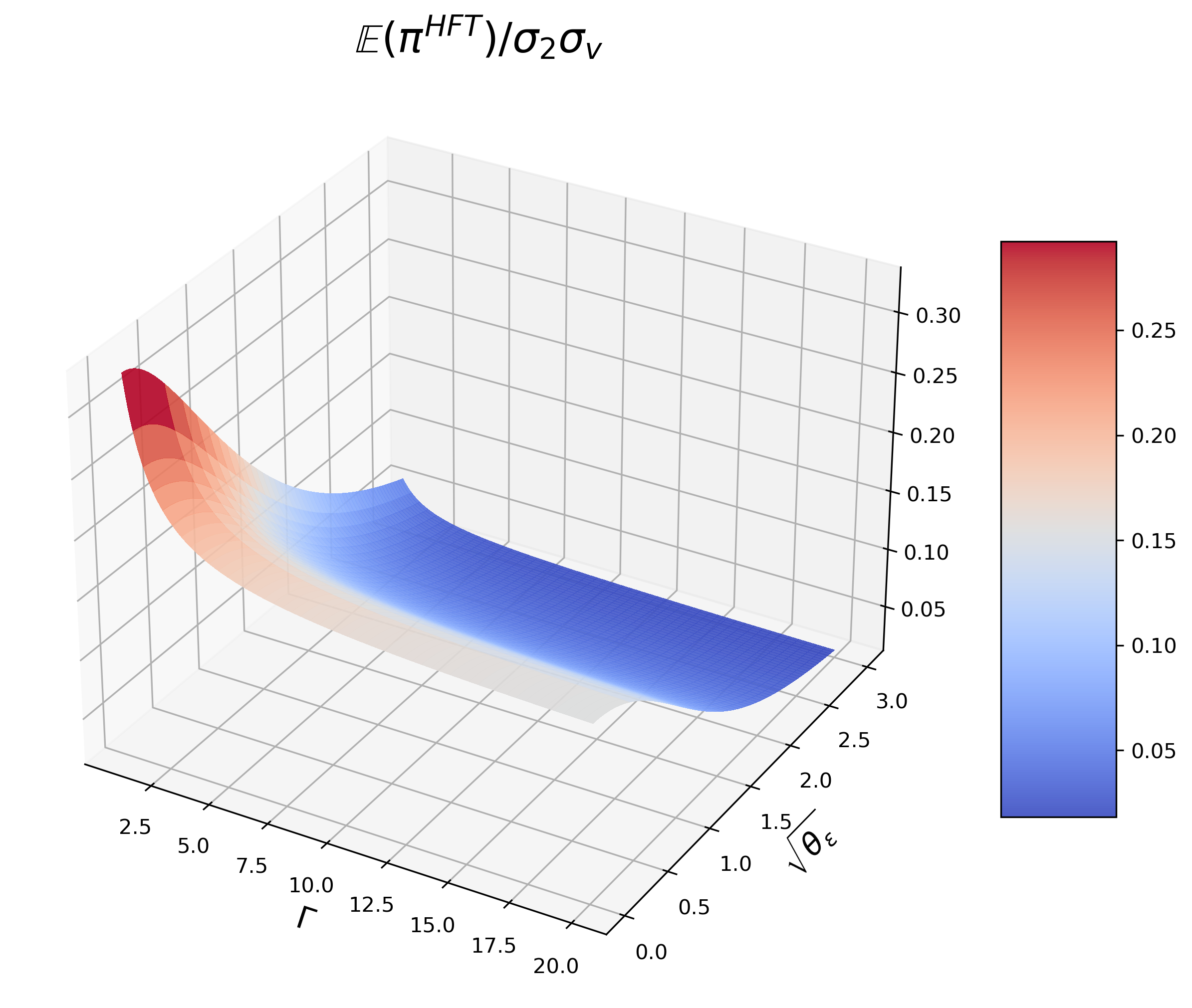}
    }
\caption{HFT's profit, $\Gamma\in[1,20]$.}
\label{fig2piHFT}
\end{figure}

The origin of HFT's profit is further analyzed through Figure \ref{fig2piHFTmajor}. In (b) and (c), Round-Trippers with high inventory aversion no longer bet on $v$, their profits mainly come from the price impact caused by IT's order. It also holds for (a), which needs very large $\Gamma$s. HFT's total profit is given in Figure \ref{fig2piHFT}, it still decreases with $\Gamma$ and $\theta_\varepsilon$.

\begin{figure}[!htbp]
    \centering
\subcaptionbox{$\theta_1=10^{-4}.$}{
    \includegraphics[width = 0.28\textwidth]{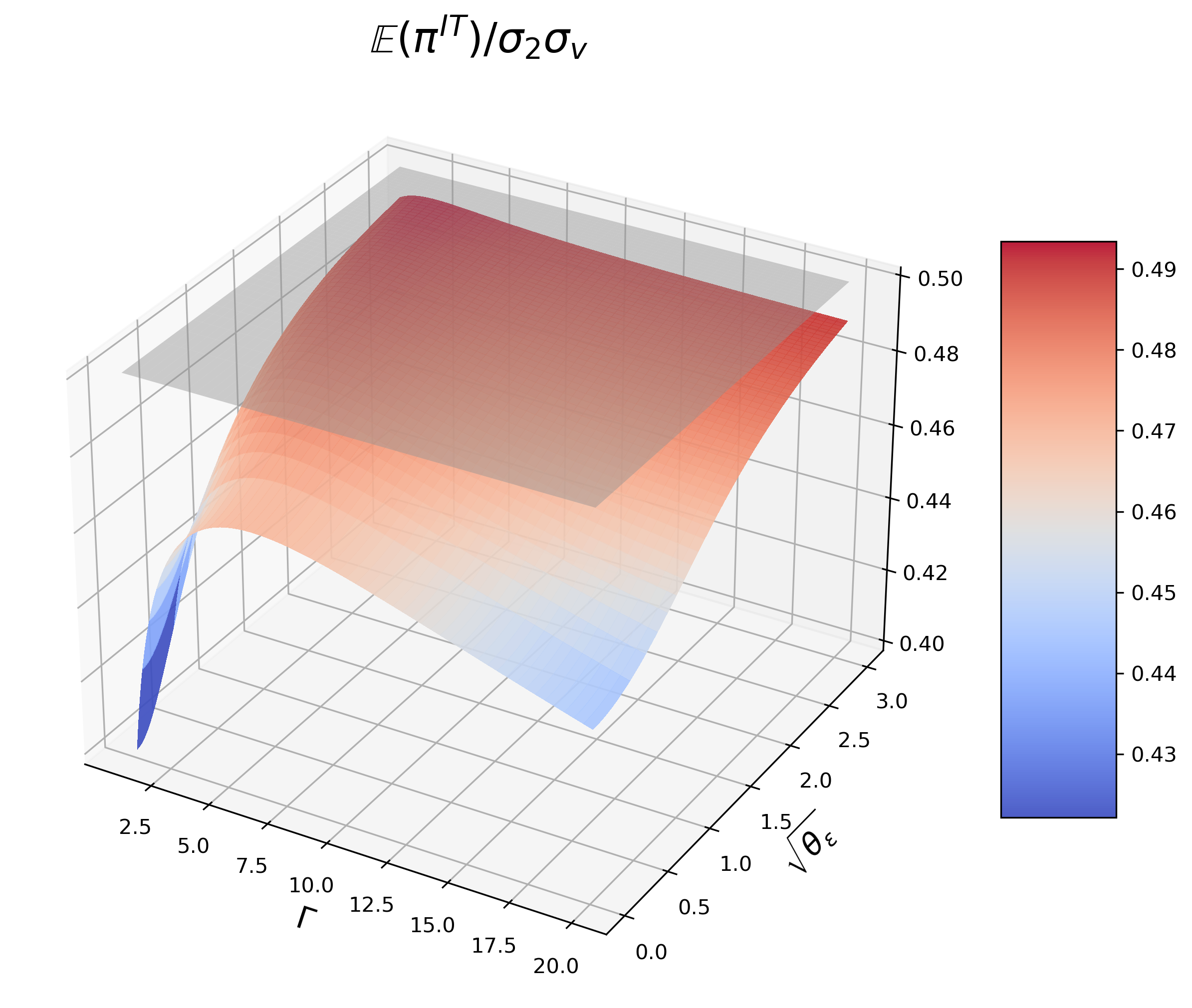}
    }
\subcaptionbox{$\theta_1=0.1.$}{
    \includegraphics[width = 0.28\textwidth]{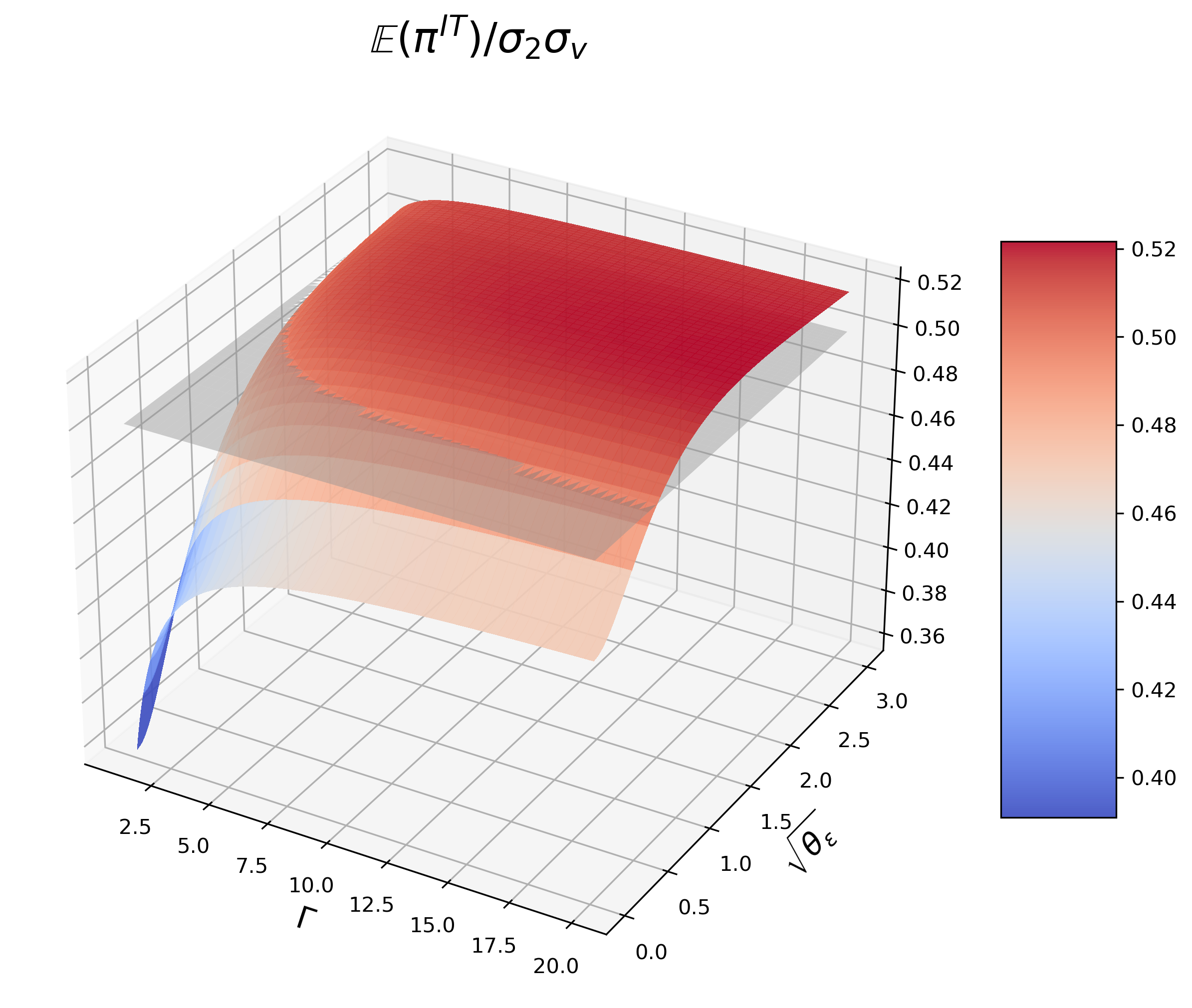}
    }
\subcaptionbox{$\theta_1=1.$}{
    \includegraphics[width = 0.28\textwidth]{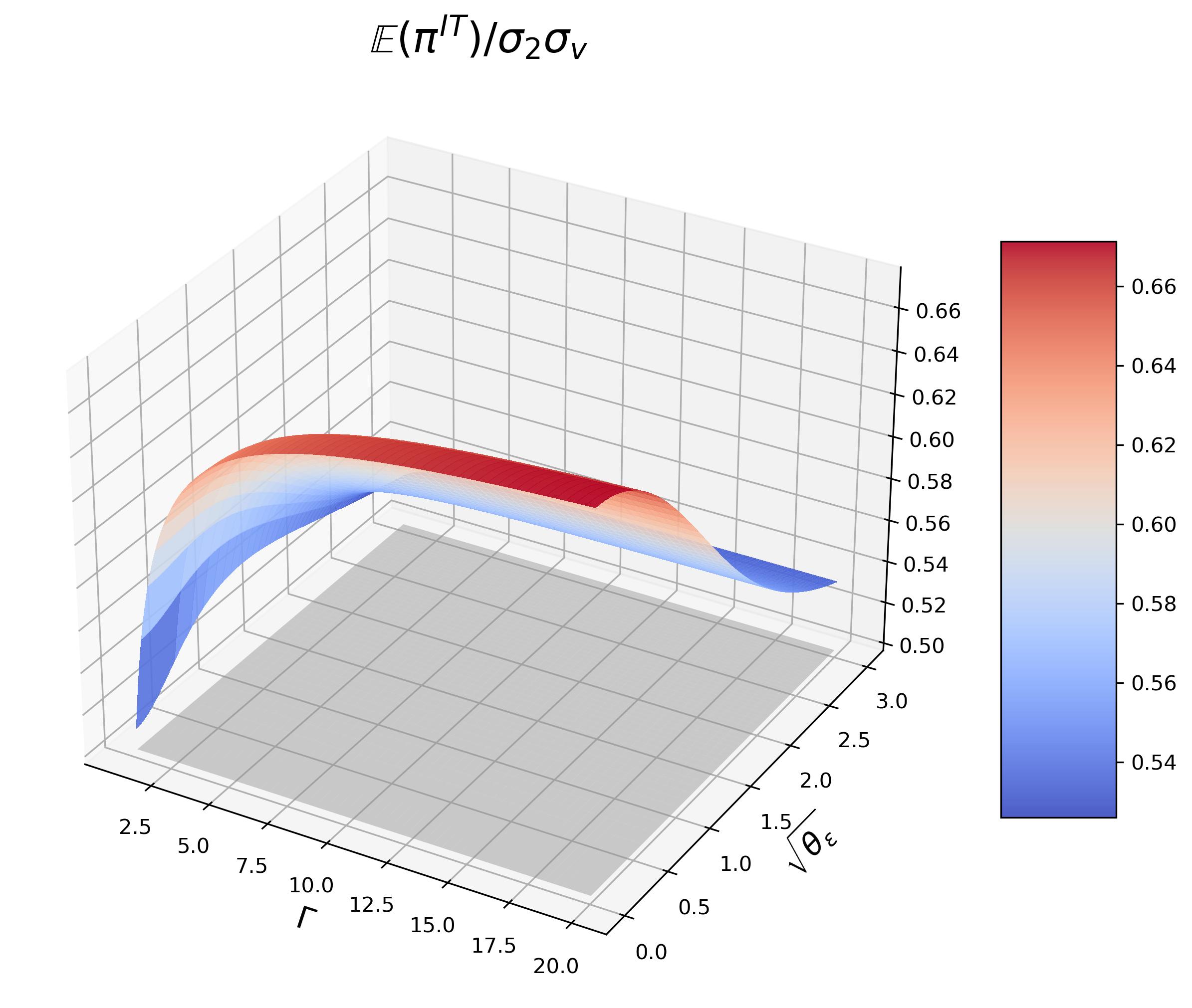}
    }
\caption{IT's profit, $\Gamma\in[1,20]$. The grey plane refers to $\mathbbm{E}(\pi^{\text{IT}})=\frac{\sigma_2\sigma_v}{2},$ \\
which is IT's profit without HFT.}
\label{fig2piIT}
\end{figure}

As shown in (b) and (c) of Figure \ref{fig2piIT}, IT can still be benefited by Round-Tripper. Surprisingly, in this case, as indicated by (c), her profit may decrease when HFT gets a noisier signal. The reason is: given $\theta_1$ and $\Gamma$, when the signal noise increases to a certain extent, HFT is unwilling to establish large positions at $t=1$ (see (c) of Figure \ref{fig2direction1}), and hence there is no need for her to liquidate large positions at $t=2$ (see (c) of Figure \ref{fig2direction2}), which makes her share less impact for IT and IT's profit starts to decrease. Actually, this holds for any $\theta_1\in(0,1],$ but is more likely to happen for larger $\theta_1$: if market noise has already provided enough shelter for IT, more signal noise may backfire. However, although IT's profit is declining, it stays higher than that without Round-Tripper.

Another noteworthy phenomenon is that given $\theta_1\in(0,0.07266)$, when HFT's signal is not overly noisy, IT's profit could decrease with $\Gamma$, which is not quite in line with common sense because a larger $\Gamma$ should deter HFT from establishing a large position to anticipate IT's trading.

\begin{figure}[!htbp]
    \centering
\subcaptionbox{IT's profit.}{
    \includegraphics[width = 0.28\textwidth]{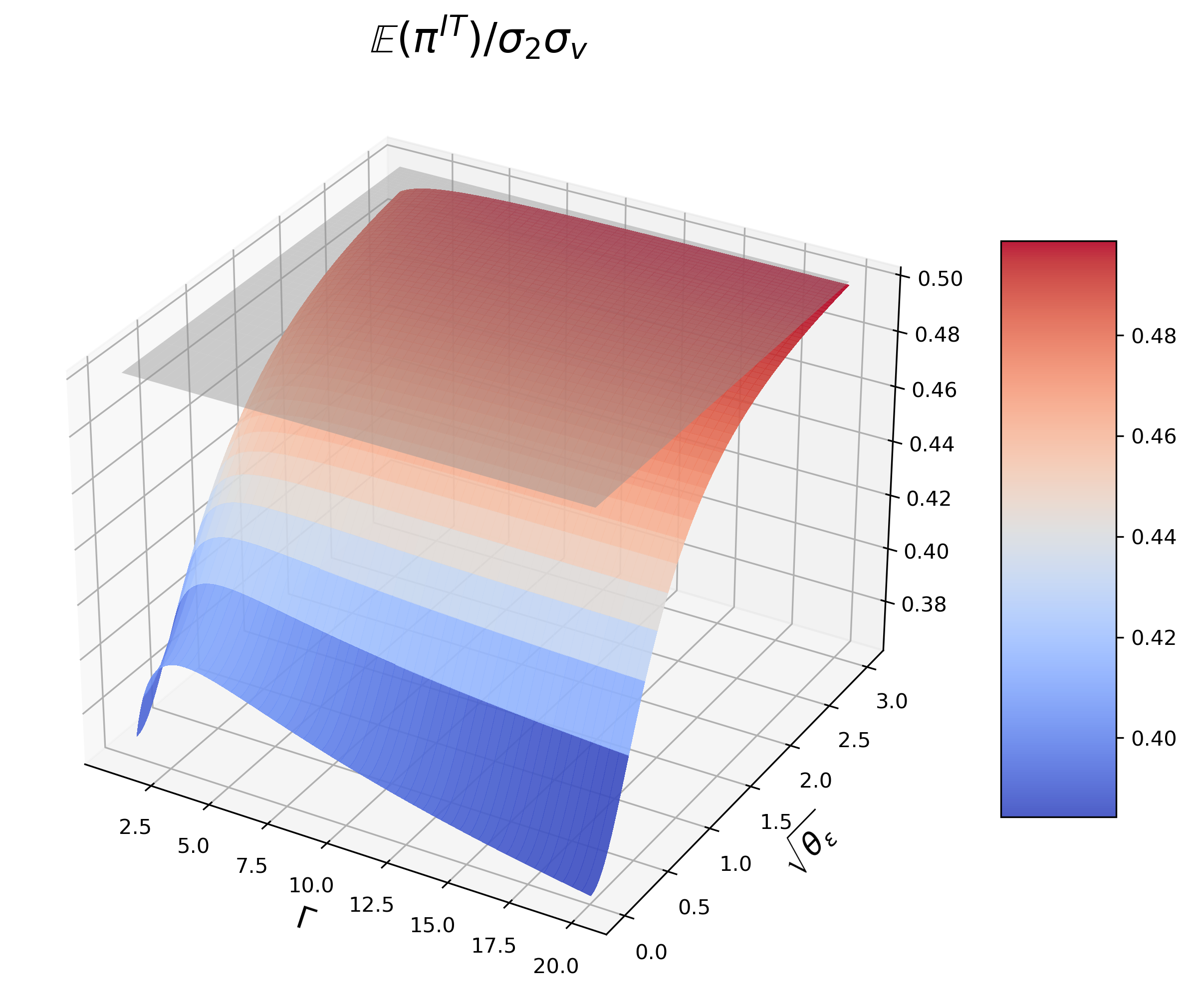}
    }
\subcaptionbox{HFT's directions, $\theta_\varepsilon=10^{-4}.$}{
    \includegraphics[width = 0.28\textwidth]{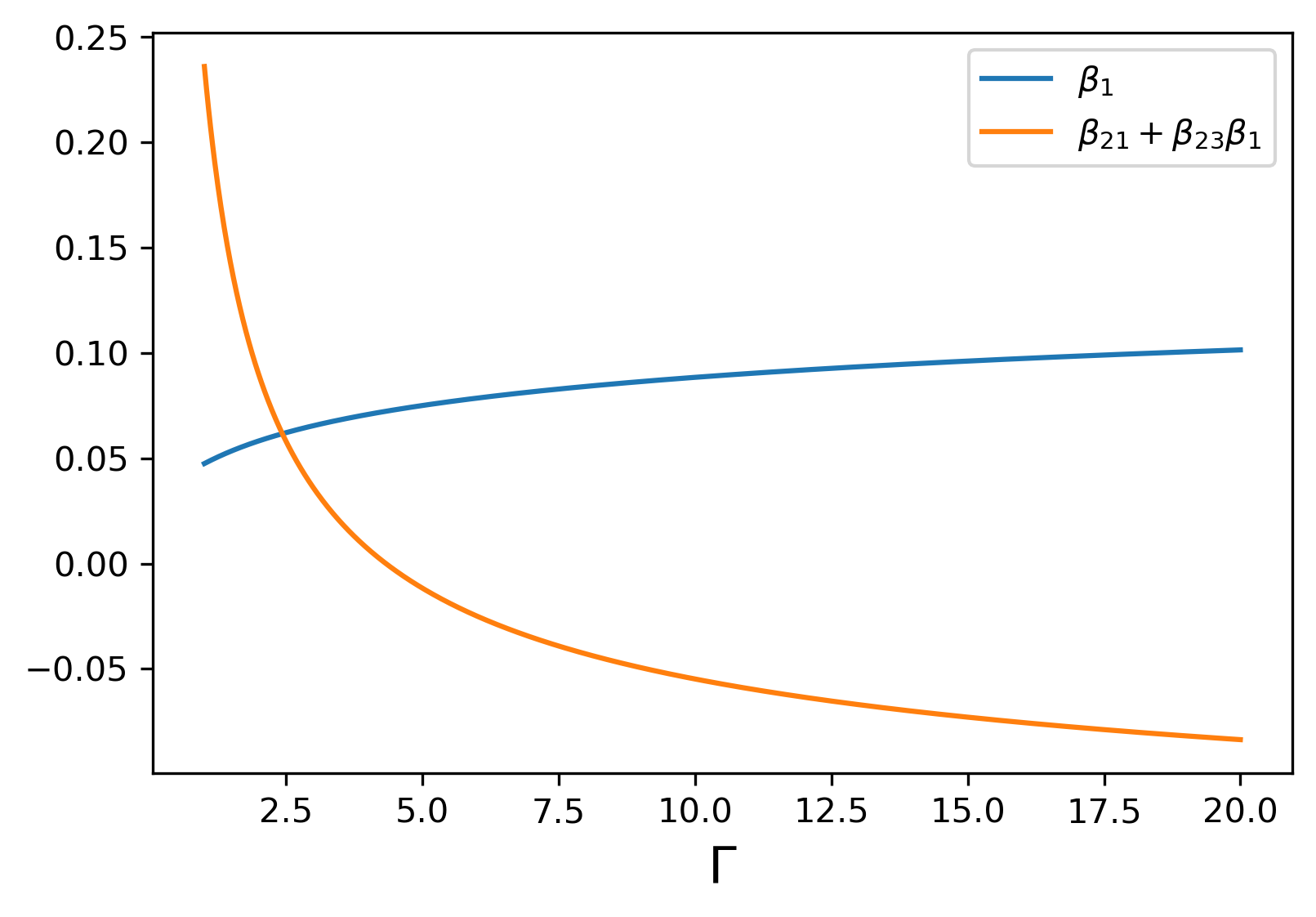}
    }
\subcaptionbox{HFT's directions, $\theta_\varepsilon=4.$}{
    \includegraphics[width = 0.28\textwidth]{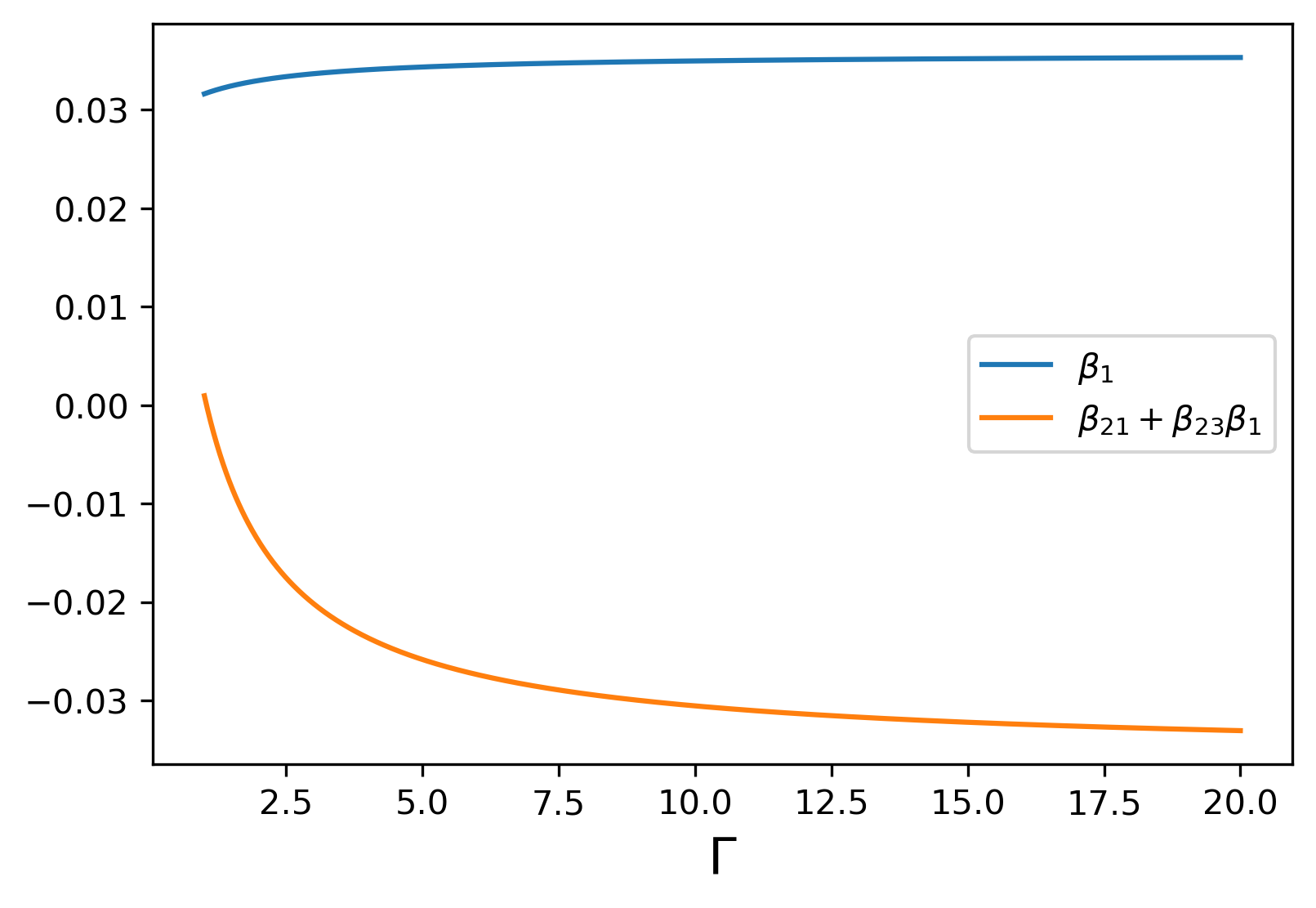}
    }
\caption{$\theta_1=0.01$, IT's profit and HFT's trading directions.}
\label{fignoteworthy}
\end{figure}

In order to explain this pattern, we take $\theta_1=0.01$ as an example (in Figure \ref{fignoteworthy}). When $\theta_\varepsilon\leq1.0531,$ IT's profit indeed follows the increase-decrease pattern, as shown in (a). From (b) and (c), when $\Gamma$ is relatively small, $x_2$ declines sharply in both figures, which causes less impact, so IT's profit increases. When $\Gamma$ is relatively large, things are different in (b) and (c). For (c), since $\theta_\varepsilon=4,$ HFT is unsure about $v$, the increase of $x_1$ is always slower than the decrease of $x_2$, IT's profit keeps increasing. However, in (b), both $x_1$ and $x_2$ change steadily.
Since there is little noise trading during period 1, the increase of $x_1$ brings larger adverse impact for IT, and IT's profit decreases.

We may wonder what kind of role HFT is playing when IT's profit attains its maximum. In most cases, the maximum is attained when HFT acts as Round-Tripper. But we find surprisingly that when the time-1 market noise is scarce and HFT's signal is relatively accurate, for example, (1) $\theta_1=10^{-4},\theta_\varepsilon\leq9$; (2) $\theta_1=0.01,\theta_\varepsilon\leq0.2,$ IT's profit is maximal when HFT is Small-IT. In other words, when HFT harms IT, in some extreme cases, she may harm IT more when she is Round-Tripper than she is Small-IT. At first glance, it contradicts the finding of Ro{\c{s}}u (2019) \cite{rocsu2019fast} that investors are better off with a sufficiently inventory-averse fast trader. This is because \cite{rocsu2019fast} compares a specific Small-IT ($\Gamma=0$) with a specific Round-Tripper ($\Gamma$ is sufficiently large), but we compare Small-IT who harms IT least with all Round-Trippers.

\begin{figure}[!htbp]
    \centering
    \subcaptionbox{HFT's profit.}{
    \includegraphics[width = 0.28\textwidth]{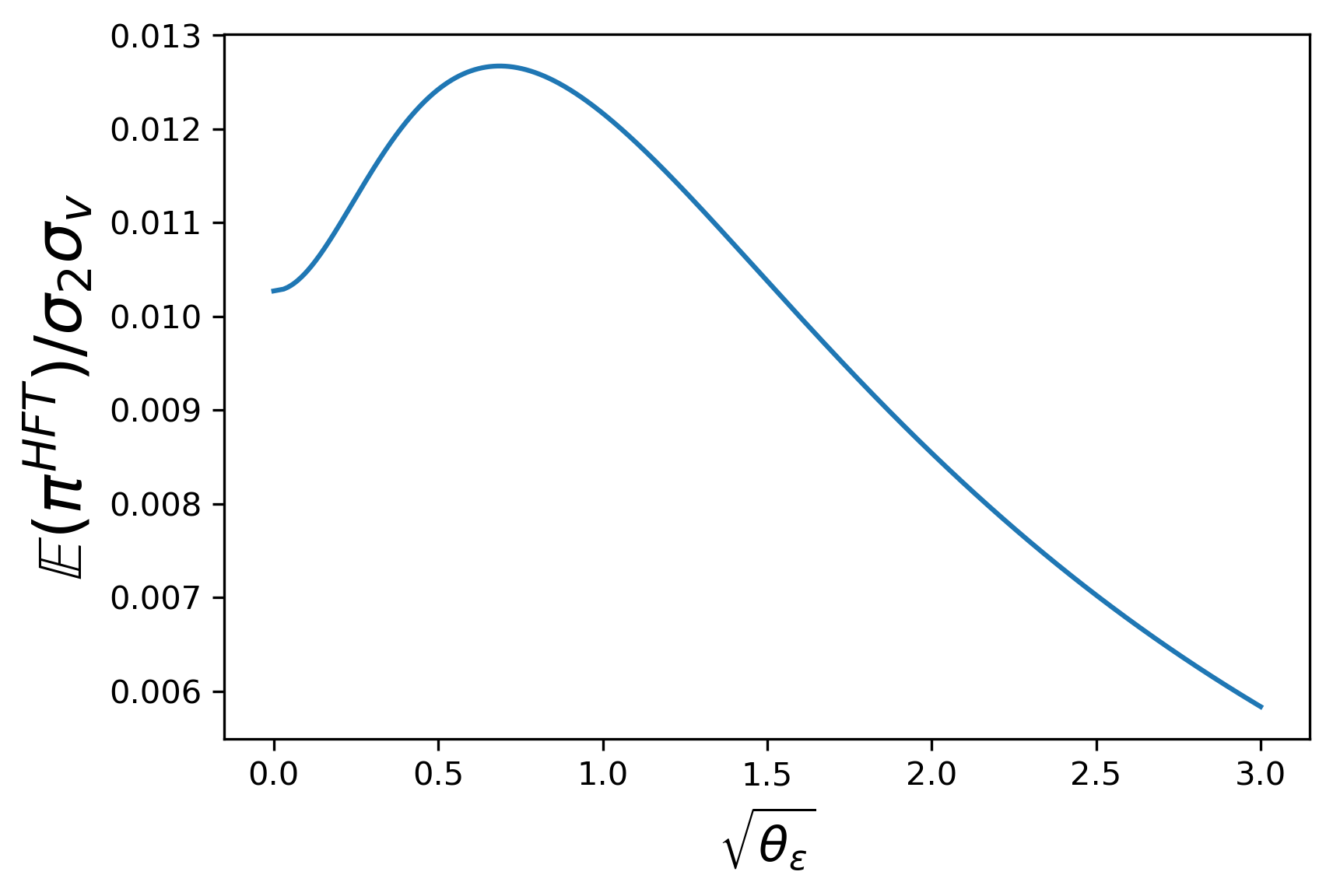}
    }
\subcaptionbox{HFT's impact profit.}{
\includegraphics[width = 0.28\textwidth]{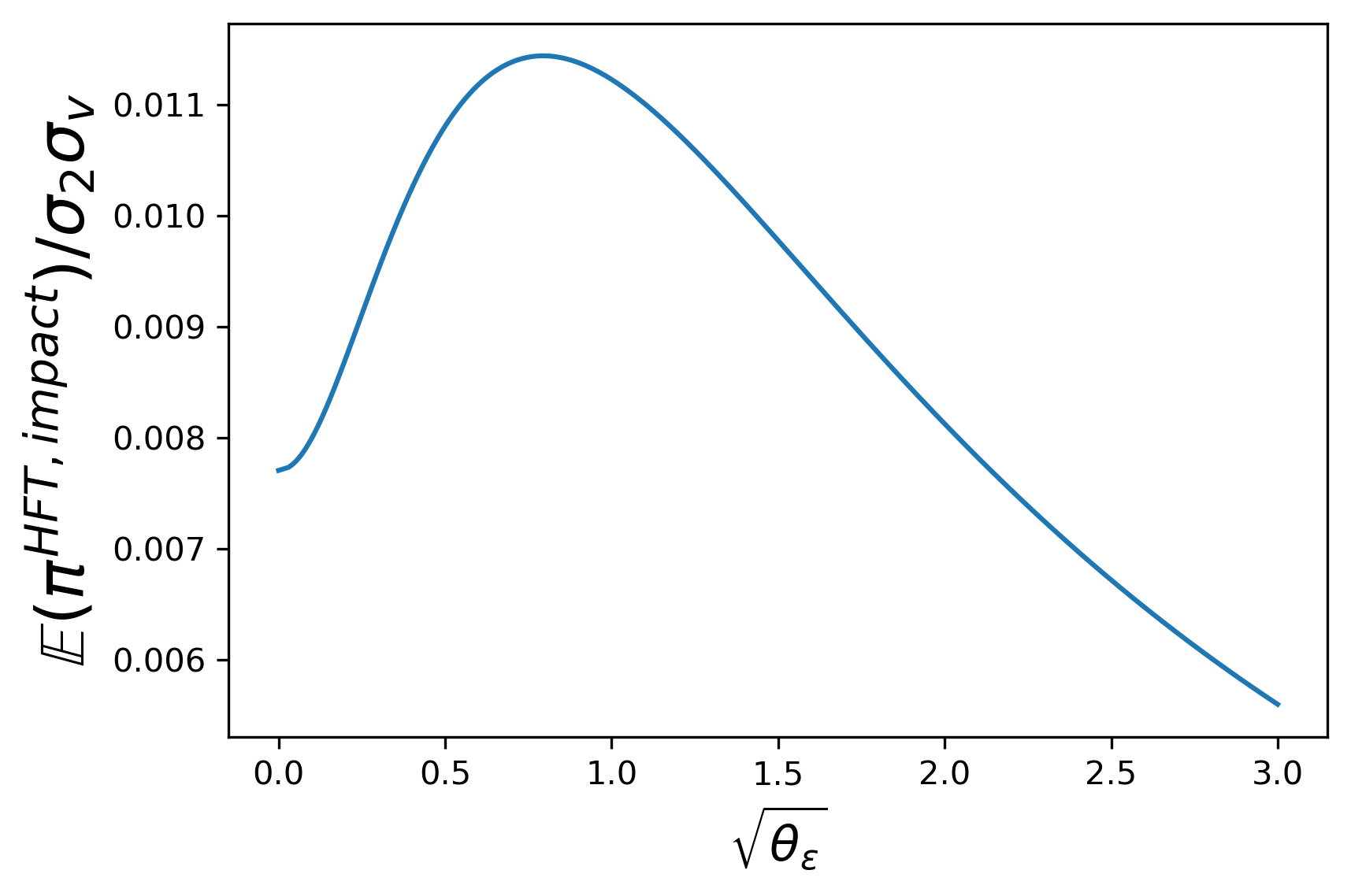}
    }
  \subcaptionbox{HFT's holding profit.}{
\includegraphics[width = 0.28\textwidth]{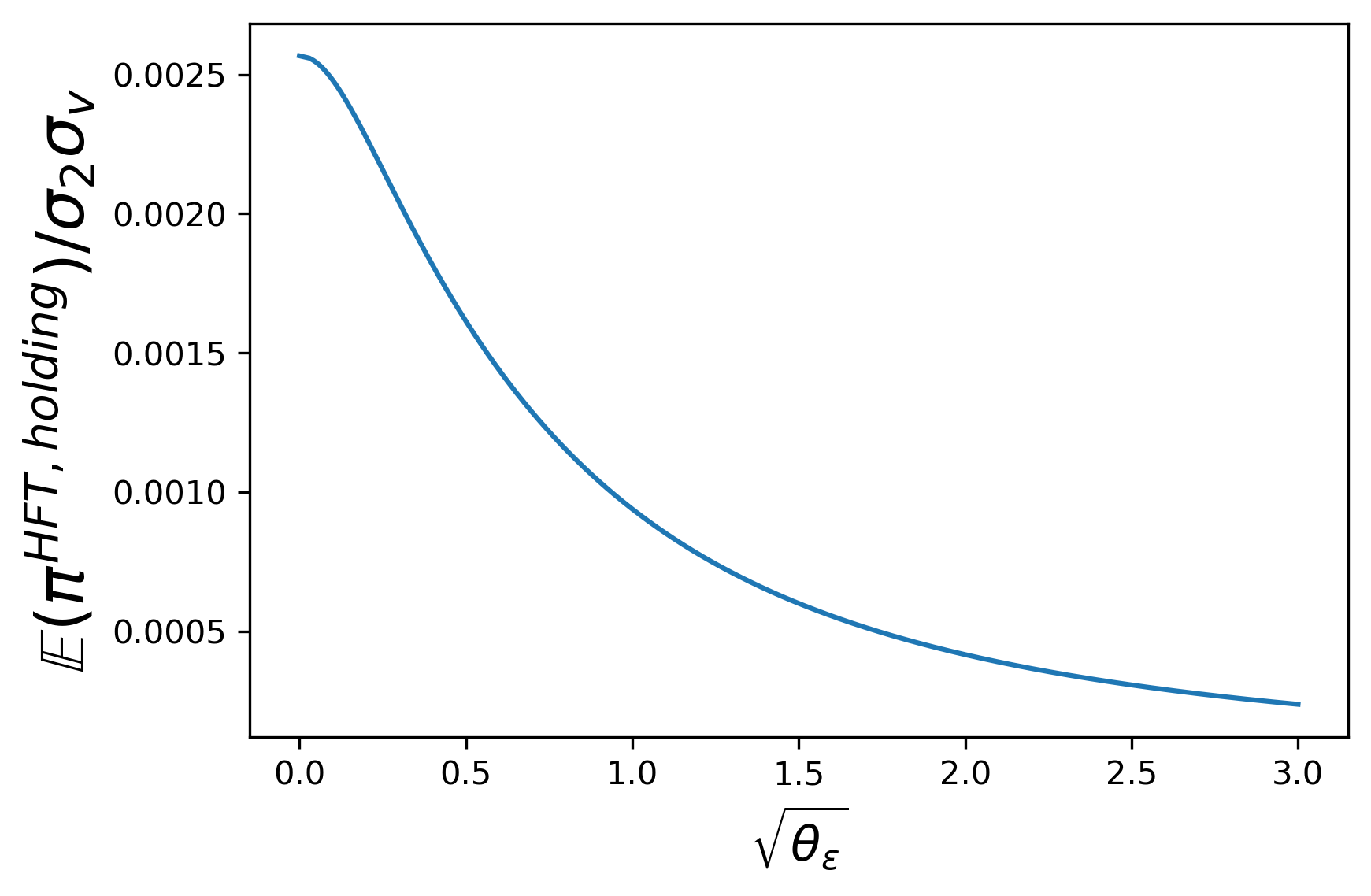}
    }
\caption{$\theta_1=0.01,\Gamma=100.$}
\label{figtheta1001Gam100}
\end{figure}

\textbf{(3) HFT is extremely inventory averse $\boldsymbol{(\Gamma>20).}$}  For extremely large $\Gamma$, most results remain the same as former cases, except for HFT's profit: when there is little high-speed noise trading, HFT may make more profits with a less accurate signal, as shown in (a) of Figure \ref{figtheta1001Gam100}. It is that HFT's profit is mainly decided by the impact caused by IT and her intensity. When $\theta_1$ is relatively small, HFT trades more conservatively but IT may trade more aggressively as $\theta_\varepsilon$ grows. 
When both $\theta_1$ and $\theta_\varepsilon$ are small, the latter dominates the former, so HFT's profit increases; when $\theta_1$ or $\theta_\varepsilon$ is large, the opposite is true. 

The reason why we do not see HFT's total profit increases with $\theta_\varepsilon$ when $\Gamma\in[0,20]$ is that, $\Gamma$ is not
yet large enough to allow the increase of impact profit to exceed the decrease of holding profit.

\subsubsection{Loss of noise traders}
\begin{figure}[ht]
    \centering
\subcaptionbox{$\theta_1=10^{-4}.$}{
    \includegraphics[width = 0.29\textwidth]{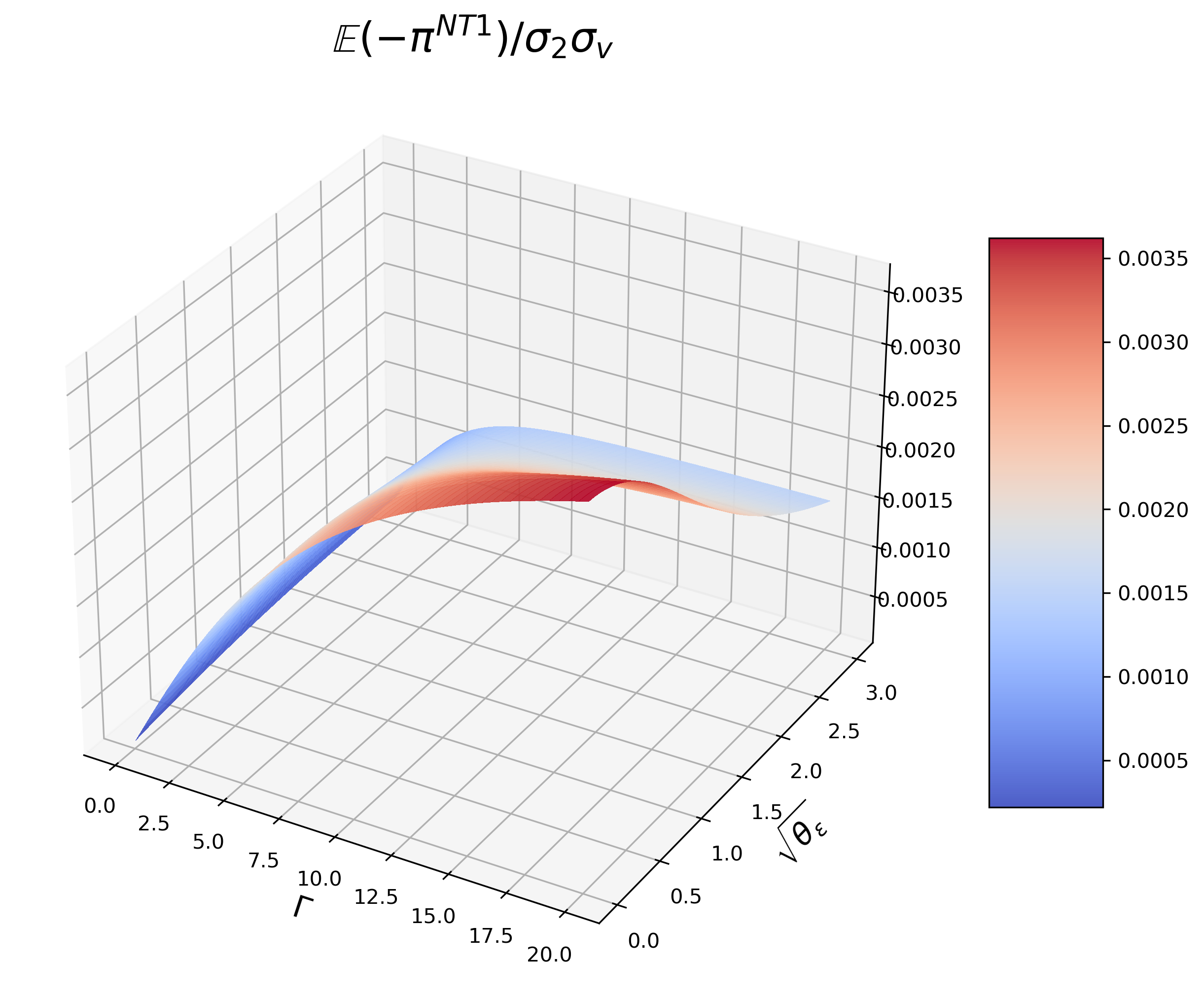}
    }
\subcaptionbox{$\theta_1=0.1.$}{
    \includegraphics[width = 0.29\textwidth]{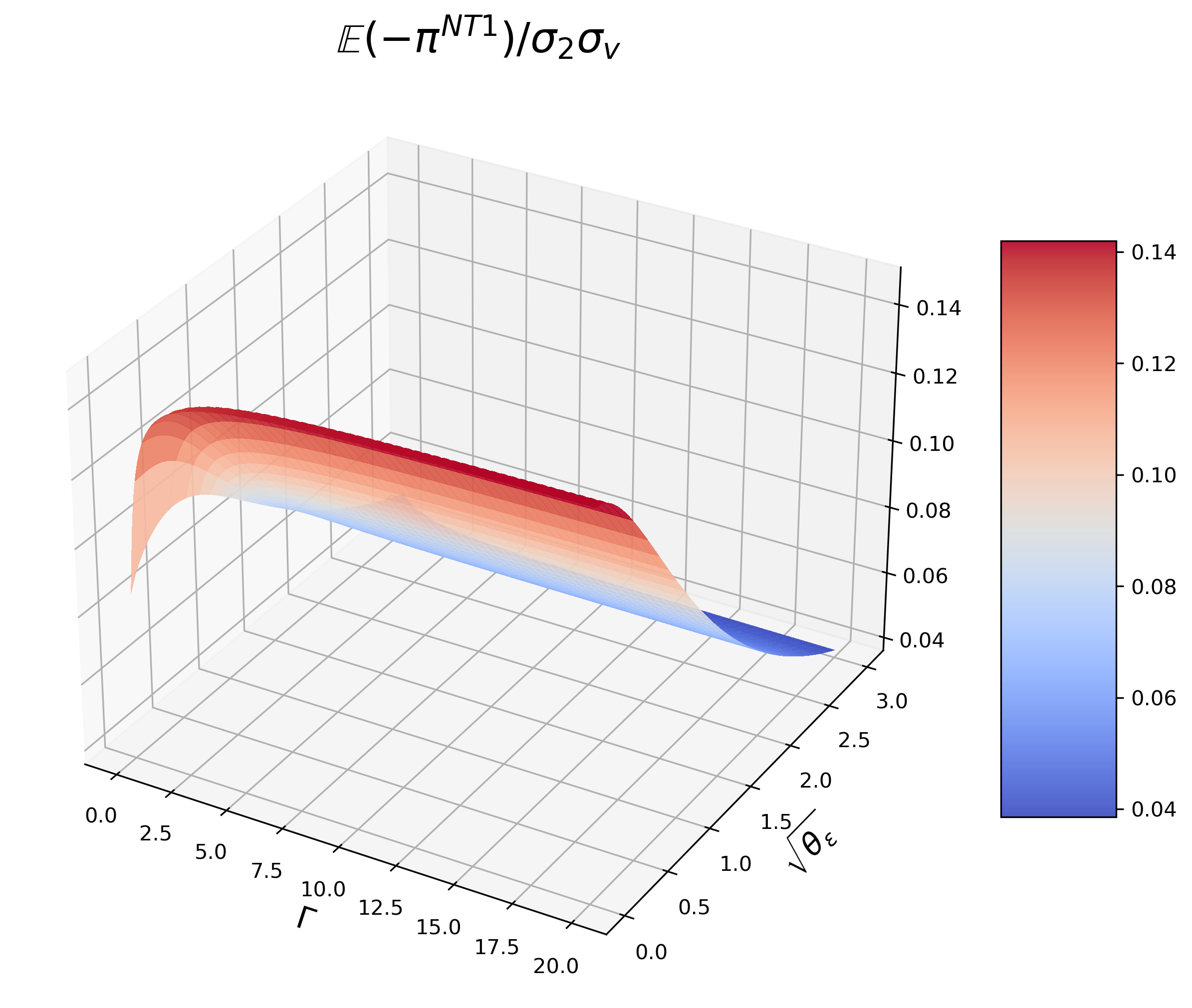}
    }
\subcaptionbox{$\theta_1=1.$}{
    \includegraphics[width = 0.29\textwidth]{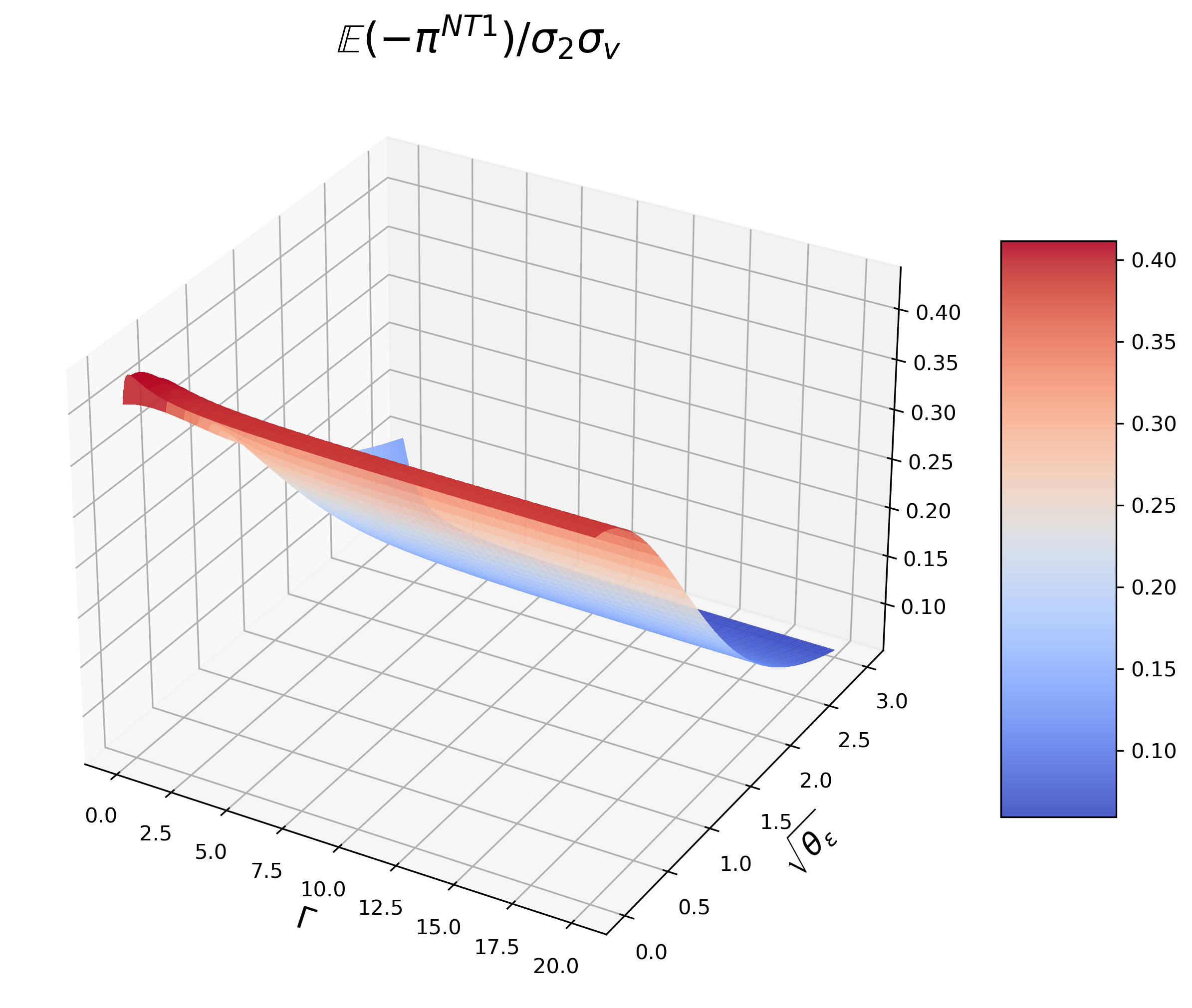}
    }
\caption{Loss of noise traders in period 1.}
\label{figlossNT1}
\end{figure}

As shown in Figure \ref{figlossNT1}, expectedly, the aggregate loss of noise traders in period 1 grows with $\theta_1,$ since a larger $\theta_1$ indicates that more uninformed investors are suffering from trading with informed investors.

\begin{figure}[!htbp]
    \centering
\subcaptionbox{$\theta_1=10^{-4}.$}{
    \includegraphics[width = 0.29\textwidth]{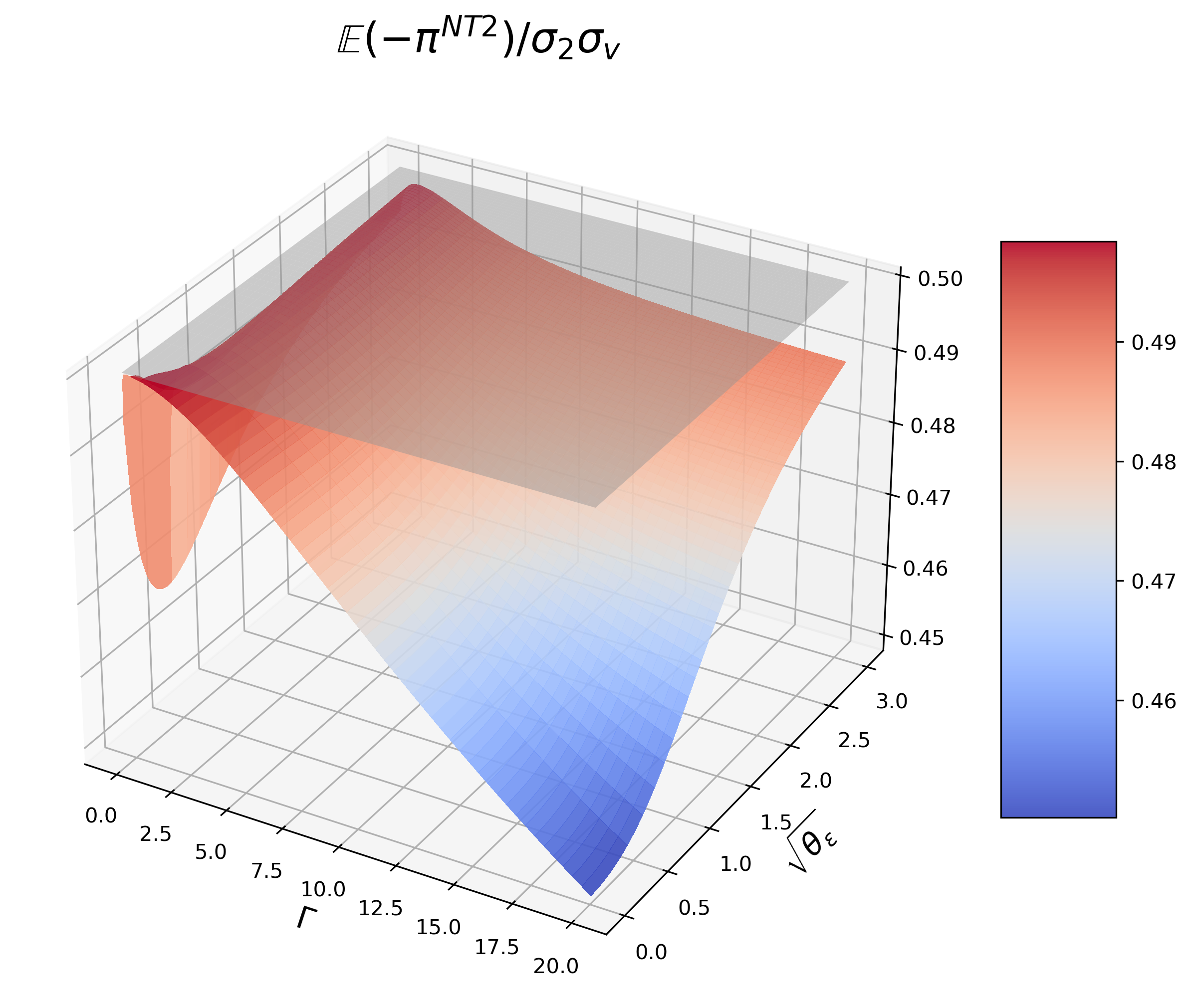}
    }
\subcaptionbox{$\theta_1=0.1.$}{
    \includegraphics[width = 0.29\textwidth]{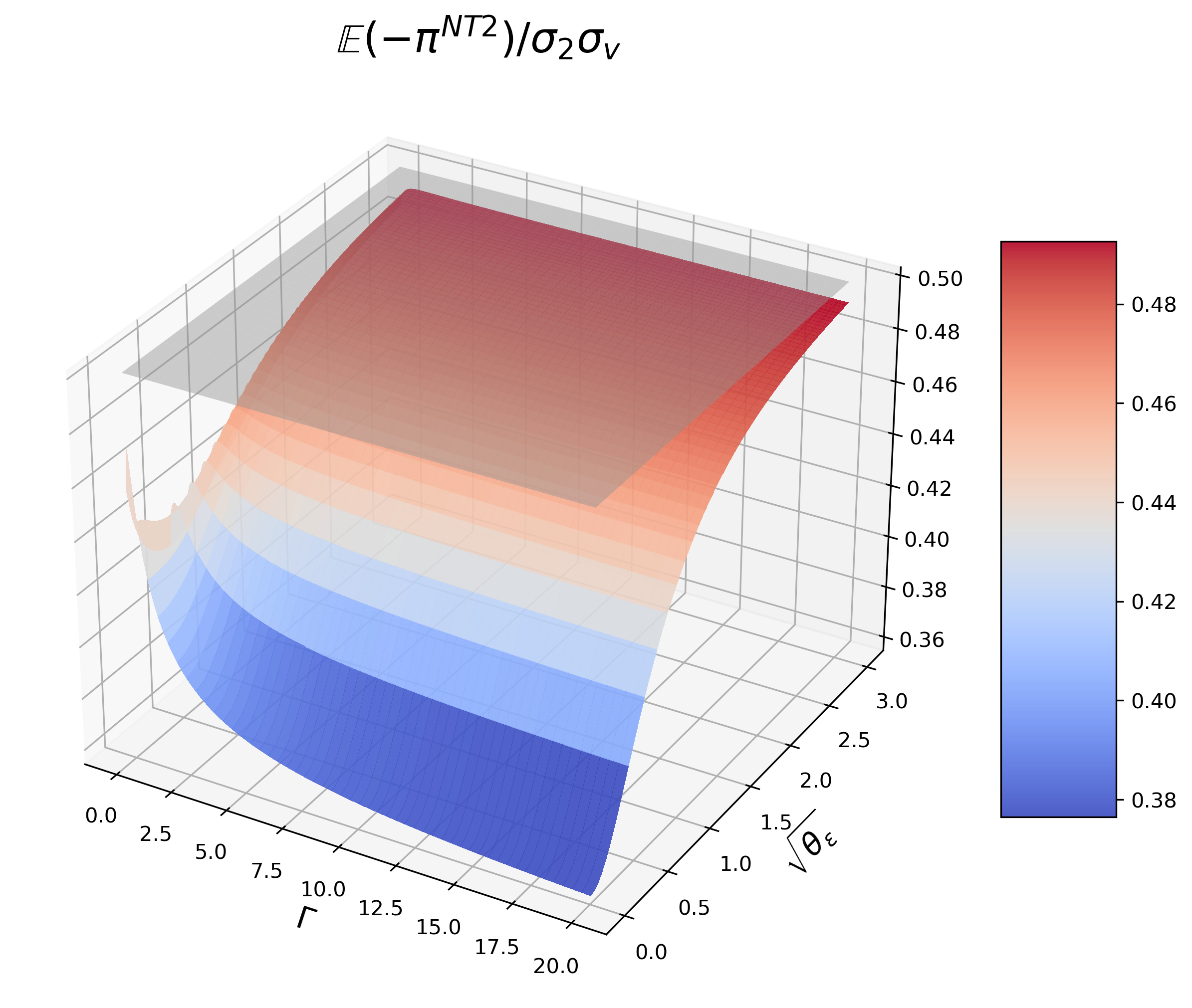}
    }
\subcaptionbox{$\theta_1=1.$}{
    \includegraphics[width = 0.29\textwidth]{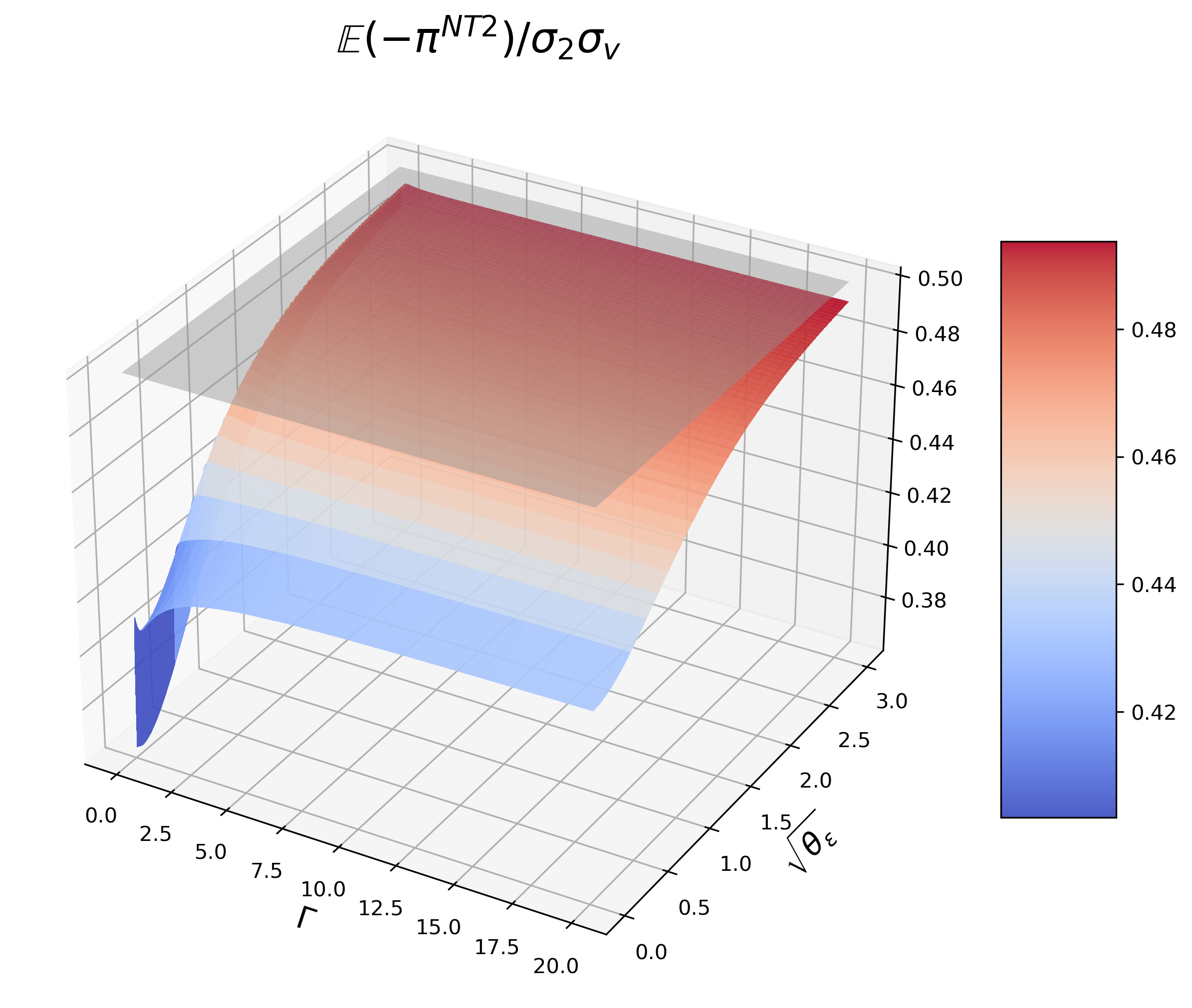}
    }
\caption{Loss of noise traders in period 2. The grey plane refers $\mathbbm{E}(-\pi^{\text{NT2}})=\frac{\sigma_2\sigma_v}{2},$\\
which is the loss of noise traders without HFT.}
\label{figlossNT2}
\end{figure}

Figure \ref{figlossNT2} states that the loss of noise traders in period 2 is always less than that without HFT, i.e., 
HFT benefits normal-speed noise traders.  In Kyle's model, where there are only IT, dealers, and noise traders, the informativeness of order flow $y=i+u_2$ is $\frac{\sigma_v}{2\sigma_2}$. While, with HFT, information about $v$ has been exposed through time-1 trading $y_1$. Thus, the price is less sensitive to the surprise component in $y_2$ (i.e., $y_2-\mathbbm{E}(y_2|y_1)$), which makes the informativeness $\lambda_{22}<\frac{\sigma_v}{2\sigma_2},$ implying less loss of noise traders in period 2.

\subsubsection{Price discovery} The price information is unfolded earlier through HFT's trading. In general, the time-1 pricing error is reduced as $\Gamma$ increases and $\theta_\varepsilon$ decreases. In most cases, IT trades more aggressively with a larger $\Gamma$ and hence leaks more information to HFT. On the other hand,  a smaller $\theta_\varepsilon$ lets HFT predict more precisely. All of the above make $x_1$ contain more information about $v$. However, the time-2 pricing error is always $\frac{\sigma_v^2}{2},$ as proved by Theorem \ref{mainthm}, which is the result of IT's optimization. IT will only release information to this extent, regardless of whether she is facing Small-IT or Round-Tripper.
\begin{figure}[!htbp]
    \centering
\subcaptionbox{$\theta_1=10^{-4}.$}{
    \includegraphics[width = 0.28\textwidth]{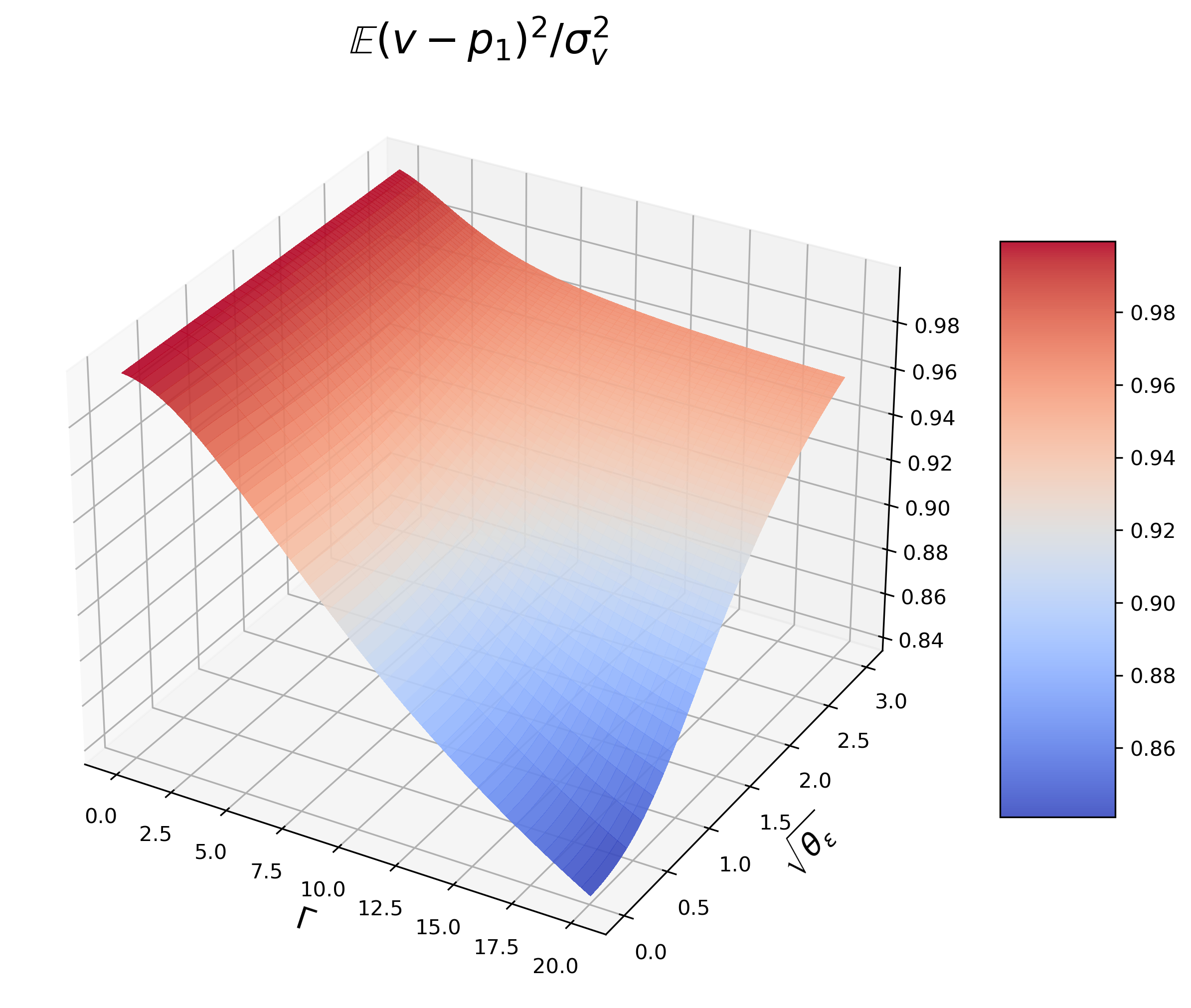}
    }
\subcaptionbox{$\theta_1=0.1.$}{
    \includegraphics[width = 0.28\textwidth]{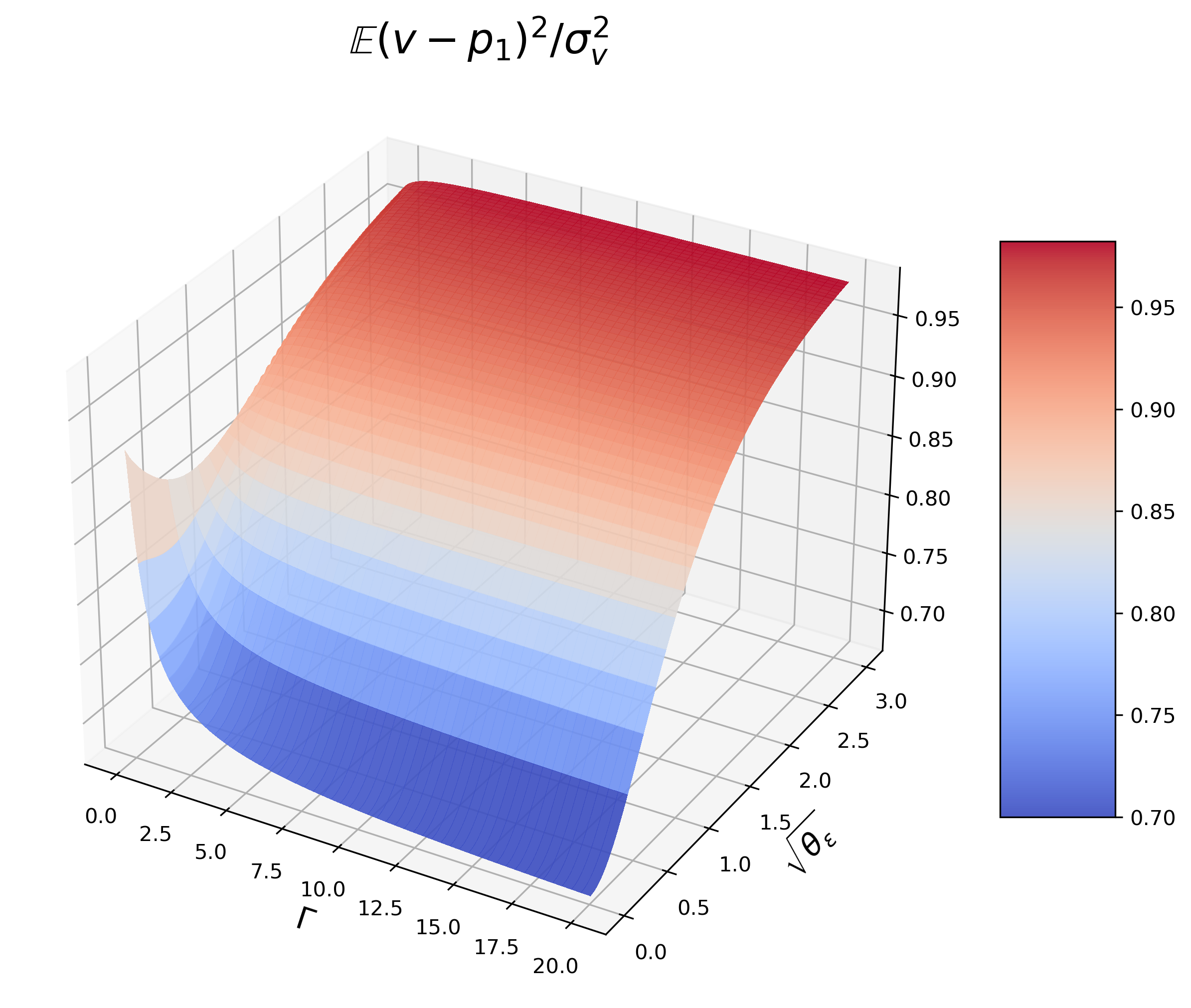}
    }
\subcaptionbox{$\theta_1=1.$}{
    \includegraphics[width = 0.28\textwidth]{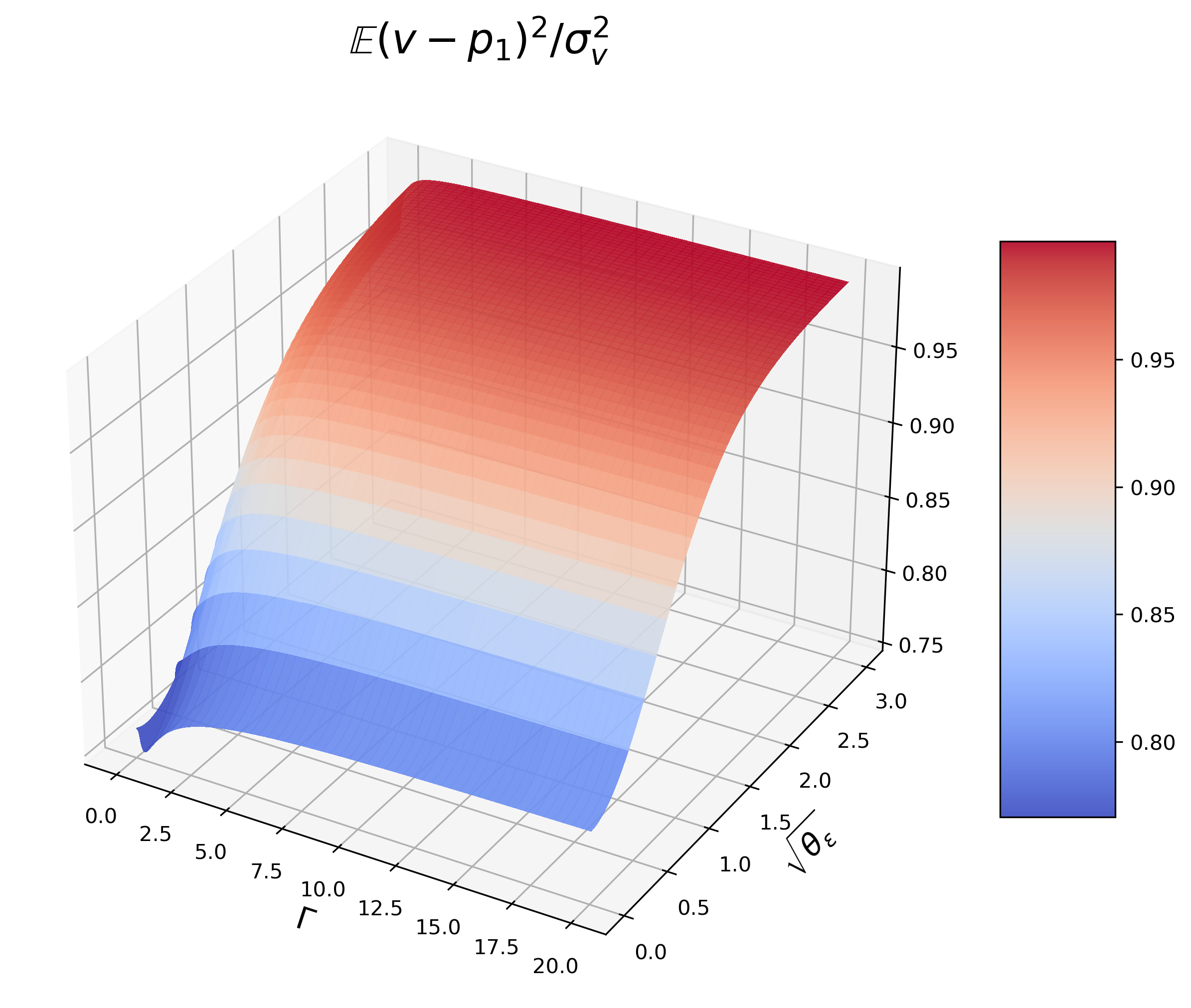}
    }
\caption{Time-1 pricing error.}
\end{figure}

\newpage
\subsection{Equilibrium under extreme conditions}
In this section, we analytically investigate the equilibrium in different limit cases.
\subsubsection{Equilibrium with little high-speed noise trading}
When there is little high-speed noise trading, i.e., $\theta_1\rightarrow0,$ the equilibrium converges to the one where HFT only trades along with IT in period 2, like back-runners in \cite{yang2020back}, and does not trade in front of IT in period 1. That is to say, HFT almost gives up her speed advantage when there are few high-frequency noise traders. It becomes a static game between IT and another (partially) informed trader with the same speed.

\begin{theorem}
\label{theta1rightarrow0thm}
As $\theta_1\rightarrow0,$ in equilibrium, HFT's time-1 order
\begin{equation*}
  x_1\rightarrow0,\ \text{a.s.;}  
\end{equation*}
IT's strategy
\begin{equation*}
i\rightarrow\alpha(v-p_0),\  \text{a.s.;}   
\end{equation*}
the liquidation prices
\begin{equation*}
\begin{aligned}
&p_1\rightarrow p_0,\  \text{a.s.,}\\
&p_2\rightarrow p_0+\lambda_{22}y_2,\  \text{a.s..}
\end{aligned}
\end{equation*}
If $\theta_\varepsilon>0$ or $\Gamma>0,$ HFT's time-2 order
\begin{equation*}
\begin{aligned}
&x_2\rightarrow\beta_{21}\hat{i},\ \text{a.s.,}   
\end{aligned}
\end{equation*}
where $\beta_{21}\geq0$ solves the equation
\begin{equation}
\label{theta1limitbeta21}
0=\beta_{21}-\frac{2 \beta_{21}+1}{2 \sqrt{\frac{1}{\beta_{21}^2 \theta_{\varepsilon}+1}} (2 \beta_{21}^2 \theta_{\varepsilon}+2 \beta_{21} \theta_{\varepsilon}+\theta_{\varepsilon}+1) (\sqrt{\frac{1}{\beta_{21}^2 \theta_{\varepsilon}+1}}+2 \Gamma)},
\end{equation}
and
\begin{equation*}
\begin{aligned}
&\alpha=\frac{\sigma_2}{\sigma_v}\frac{\sqrt{1+\beta_{21}^2\theta_\varepsilon}}{1+\beta_{21}},\\
&\lambda_{22}=\frac{\sigma_v}{2\sigma_2}\frac{1}{\sqrt{1+\beta_{21}^2\theta_\varepsilon}}.
\end{aligned}
\end{equation*}
The expected profit of IT 
\begin{equation*}
\mathbbm{E}(\pi^{\text{IT}})\rightarrow\frac{\sigma_2\sigma_v}{2}\frac{\sqrt{1+\beta_{21}^2\theta_\varepsilon}}{1+\beta_{21}}\leq\frac{\sigma_2\sigma_v}{2};
\end{equation*}
the expected profit and  expected inventory penalty of HFT
\begin{equation*}
\begin{aligned}
&\mathbbm{E}(\pi^{\text{HFT}})\rightarrow\frac{\sigma_2\sigma_v}{2}\frac{\beta_{21}(1-\beta_{21}\theta_\varepsilon)}{\sqrt{1+\beta_{21}^2\theta_\varepsilon}(1+\beta_{21})},\\ 
&\mathbbm{E}(-\gamma x_2^2)\rightarrow-\sigma_2\sigma_v\Gamma\beta_{21}^2[\theta_\varepsilon+\frac{1+\beta_{21}^2\theta_\varepsilon}{(1+\beta_{21})^2}].
\end{aligned}
\end{equation*}
If $\theta_\varepsilon=\Gamma=0,$
HFT's time-2 order
\begin{equation*}
\begin{aligned}
&x_2\rightarrow\beta_{21}(v-p_0),\ \text{a.s.,}   
\end{aligned}
\end{equation*}
and
\begin{equation*}
\begin{aligned}
&\alpha=\beta_{21}=\frac{\sigma_2}{\sigma_v}\frac{\sqrt{2}}{2},\\
&\lambda_{22}=\frac{\sigma_v}{2\sigma_2}\frac{2\sqrt{2}}{3}.
\end{aligned}
\end{equation*}
The expected profit of IT and HFT
\begin{equation*}
\mathbbm{E}(\pi^{\text{IT}}),\mathbbm{E}(\pi^{\text{HFT}})\rightarrow\frac{\sigma_2\sigma_v}{2}\frac{\sqrt{2}}{3}.
\end{equation*}
\end{theorem}

Without noise trading to cover HFT's time-1 trading, it will bring huge transaction costs to her, so she prefers to do nothing.  For IT, the information leakage causes HFT to trade along with her, her profit is lower than that without HFT. Larger the inventory aversion $\gamma$, less aggressive HFT is, and less hurt IT is.

\subsubsection{Equilibrium when HFT must clear all positions}
When $\Gamma\rightarrow\infty,$ it is an extreme case where HFT's inventory aversion is so large that any ending positions will produce a huge cost. It is rigorously proved that HFT clears all positions at $t=2$, i.e., 
\begin{equation*}
\beta_{21}=\beta_{22}=0,\beta_{23}=-1.
\end{equation*}
The $x_1\_x_2$ strategy becomes $x\_(-x)$ strategy and we denote
\begin{equation*}
x=\beta\hat{i}.
\end{equation*}
The limit equilibrium is equivalent with the one where HFT maximizes the short term profit
\begin{equation*}
\mathbbm{E}\left(x(p_2-p_1)|\hat{i}\right).
\end{equation*}
\begin{theorem}
   \label{thmgam=infty} 
As $\Gamma\rightarrow\infty,$ for any $\theta_1>0,\theta_\varepsilon\geq0,$  there exists a unique equilibrium where HFT's trading intensity $\beta\in(0,1)$ solves:
   \begin{equation}
\label{gam=inftybeta}
\begin{aligned}
 0=& \beta^6(4\theta_1\theta_\varepsilon^2+\theta_1\theta_\varepsilon^3+2\theta_1^2\theta_\varepsilon^2+2\theta_\varepsilon^2+\theta_\varepsilon^3)+\beta^5(4\theta_1\theta_\varepsilon+4\theta_1\theta_\varepsilon^2+2\theta_1\theta_\varepsilon^3+8\theta_1^2\theta_\varepsilon+4\theta_1^2\theta_\varepsilon^2+4\theta_1^3\theta_\varepsilon)\\
+ & \beta^4(2\theta_1\theta_\varepsilon+\theta_1\theta_\varepsilon^2-11\theta_1^2\theta_\varepsilon-8\theta_1^2\theta_\varepsilon^2-13\theta_1^3\theta_\varepsilon) 
+\beta^3(2\theta_1^2 + 2\theta_1^3 + 8\theta_1^2\theta_\varepsilon + 4\theta_1^2\theta_\varepsilon^2+ 16\theta_1^3\theta_\varepsilon ) \\
  - & \beta^2( \theta_1^2\theta_\varepsilon +5\theta_1^3 + 9\theta_1^3\theta_\varepsilon) + \beta(4\theta_1^3 + 2\theta_1^3\theta_\varepsilon) -\theta_1^3,
\end{aligned}
\end{equation}
and
\begin{equation*}
\begin{aligned}
&\Lambda_1=\frac{\beta\sqrt{[\beta^2\theta_\varepsilon(\theta_1+1)+\theta_1][\theta_1(1-\beta)^2+\beta^2(\theta_\varepsilon+1)]}}{\beta^2[\beta^2\theta_\varepsilon(\theta_1+1)+\theta_1]+(\beta^2\theta_\varepsilon+\theta_1)[\theta_1(1-\beta)^2+\beta^2(\theta_\varepsilon+1)]},\\
&\Lambda_{21}=\frac{\beta^2\theta_\varepsilon+\beta}{2\sqrt{[\theta_1(1-\beta)^2+\beta^2(\theta_\varepsilon+1)][\theta_1+\beta^2\theta_\varepsilon(\theta_1+1)]}},\\
&\Lambda_{22}=\frac{\beta^2\theta_\varepsilon+(1-\beta)\theta_1}{2\sqrt{[\theta_1(1-\beta)^2+\beta^2(\theta_\varepsilon+1)][\theta_1+\beta^2\theta_\varepsilon(\theta_1+1)]}},\\
&A=\sqrt{\frac{\theta_1+\beta^{2}\theta_\varepsilon(\theta_1+1)}{\theta_1(1-\beta)^2+\beta^{2}(\theta_\varepsilon+1)}}.\\
\end{aligned}
\end{equation*}
Thus, the expected profit of IT is
\begin{equation*}
\mathbbm{E}(\pi^{\text{IT}})=\frac{\sigma_2\sigma_v}{2}A;
\end{equation*}
the expected profit and expected inventory penalty of HFT are
\begin{equation*}
\begin{aligned}
\mathbbm{E}(\pi_1^{\text{HFT}}+\pi_2^{\text{HFT}})=&\frac{\sigma_2\sigma_v}{2}\frac{\beta\theta_1[(1-\beta)\theta_1+\beta^2\theta_\varepsilon]}{\sqrt{[\beta^2\theta_\varepsilon(\theta_1+1)+\theta_1][\theta_1(1-\beta)^2+\beta^2(\theta_\varepsilon+1)]}}\\
&\frac{\beta^2\theta_\varepsilon[(1-2\beta)-\beta\theta_\varepsilon]+(1-\beta)\theta_1[1-\beta(1-2\beta)\theta_\varepsilon]}{(1-\beta)^2\theta_1^2+\beta^{4}\theta_\varepsilon(\theta_\varepsilon+2)+2\beta^2\theta_1[1+(\beta^2-\beta+1)\theta_\varepsilon]},\\
\mathbbm{E}[-\gamma(x_1+x_2)^2]=&0.
\end{aligned}
\end{equation*}
\end{theorem}

The limit equilibrium is consistent with the one in Xu and Cheng (2023) \cite{xu2023large}. Corresponding to the conclusions in Section \ref{secequilibriumgeneral}, the exact forms of $\Tilde{\theta}_\varepsilon(\theta_1,\infty)$ and $\hat{\theta}_\varepsilon(\theta_1,\infty)$ can be derived in this extreme case: IT makes more profits with Round-Tripper when
\begin{equation*}
\label{Gaminfcond-benefit} \theta_\varepsilon>\Tilde{\theta}_\varepsilon(\theta_1,\infty)=\max(\frac{-(\theta_1+5)+2\sqrt{4\theta_1^2+10\theta_1+5}}{-5\theta_1},0);
\end{equation*}
IT's profit increases with $\theta_\varepsilon$ if 
    \begin{equation*}
0\leq\theta_\varepsilon<\hat{\theta}_\varepsilon(\theta_1,\infty)=\max(\frac{1-\theta_1-2\theta_1^2}{3\theta_1},0).
    \end{equation*}
The calculation can be found in Section \ref{secappendix}. 

\begin{figure}[!htbp]
    \centering
    \subcaptionbox{HFT's influences on IT.}{\includegraphics[height=0.15\textheight]{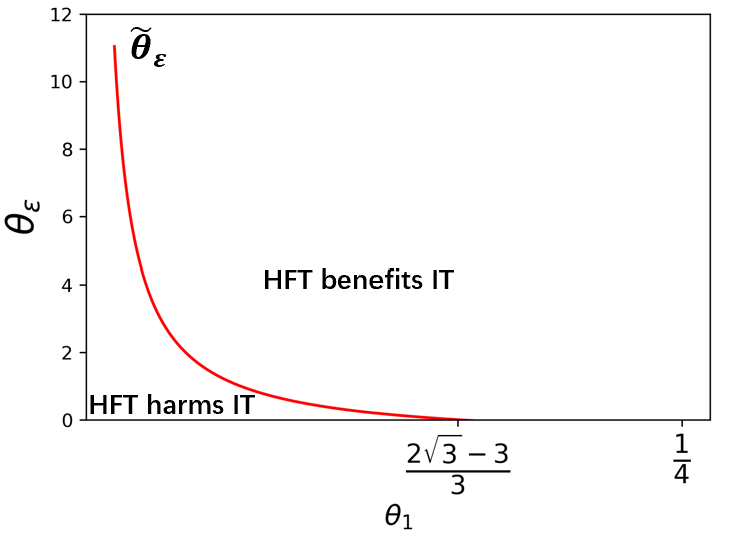}}
\subcaptionbox{$\theta_\varepsilon$'s influences on IT's profit.}{\includegraphics[height=0.15\textheight]{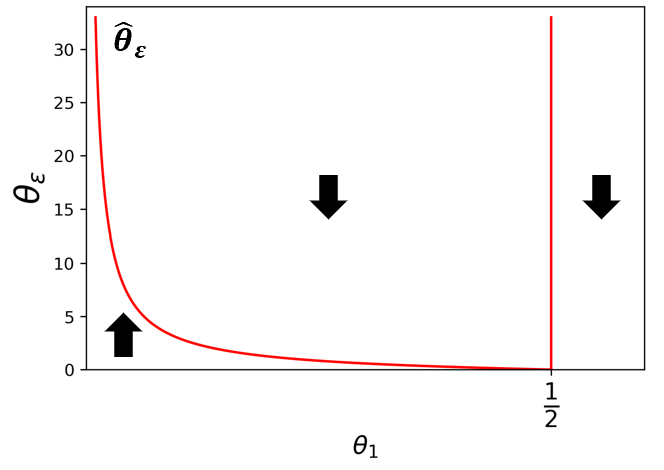}}
\caption{The exact forms of $\Tilde{\theta}_\varepsilon(\theta_1,\infty)$ and $\hat{\theta}_\varepsilon(\theta_1,\infty)$}
\label{figforthmGaminf}
\end{figure}

In Figure \ref{figforthmGaminf}, the above conclusions are visually displayed. From (a), when the size of high-speed noise trading $\theta_1>\frac{2\sqrt{3}-3}{3},$ IT is benefited even if her order is perfectly anticipated by HFT. From (b), when $\theta_1>\frac{1}{2}$, adding any noise into HFT's signal will decrease IT's profit.

\section{An Extension to Multi-Asset Anticipatory Trading}
\label{secmulti}
Following Caballé and Krishnan (1994) and Molino et al. (2020), the current model can be extended to a multi-asset case, where both IT and HFT trade on several assets. IT's incoming orders are predicted by HFT and HFT implements the corresponding anticipatory strategies. In section \ref{subsecmultiasset1}, the model is set up and solved.

Due to HFT's stronger ability to process information and lack of clear investment intentions, she may apply the obtained information to other correlated assets which are not traded by IT. That is to say, HFT may not hold exactly the same assets as IT. Motivated by this, in section \ref{subsecmultiasset2}, another variation of the model is considered: IT trades on one asset and her trading is detected by HFT, but HFT will apply the information to this and other related assets. It gives us more insights into anticipatory trading.

\subsection{An extended multi-asset Kyle model}
\label{subsecmultiasset1}
For simplicity, we assume there are two assets in the market:
\begin{equation*}
\boldsymbol{v}=\begin{pmatrix}
    v_1\\
    v_2
\end{pmatrix}\sim N(\boldsymbol{p_0},\boldsymbol{\Sigma_v}),\ \boldsymbol{p_0}=\begin{pmatrix}
    p_{01}\\
    p_{02}
\end{pmatrix},\  \boldsymbol{\Sigma_v}=\begin{pmatrix}
\sigma_{v_1}^2&\rho\sigma_{v_1}\sigma_{v_2}\\
\rho\sigma_{v_1}\sigma_{v_2}&\sigma_{v_2}^2
\end{pmatrix}.
\end{equation*}

At $t=0,$ IT receives the signal $\boldsymbol{v}$ and sends the order 
$\boldsymbol{i}=\begin{pmatrix}
     i_1\\
     i_2
 \end{pmatrix},
 $
 which is delayed to be executed at $t=2.$ The orders are predicted by HFT: HFT receives a signal
\begin{equation*}
\hat{\boldsymbol{i}}=\boldsymbol{i}+\boldsymbol{\varepsilon},\ \boldsymbol{\varepsilon}\sim N(0,\boldsymbol{\Sigma_\varepsilon}),\ \boldsymbol{\Sigma_\varepsilon}=\begin{pmatrix}
\sigma_{\varepsilon_1}^2&0\\
0&\sigma_{\varepsilon_2}^2
\end{pmatrix}.
\end{equation*}
At $t=1,$ HFT sends preemptive orders
$\boldsymbol{x_1}=\begin{pmatrix}
    x_{11}\\
    x_{12}
\end{pmatrix};$
at $t=2$, HFT sends orders
$\boldsymbol{x_2}=\begin{pmatrix}
    x_{21}\\
    x_{22}
\end{pmatrix},$
where the noise $\boldsymbol{\varepsilon}$ is independent with other random variable. Consequently, the order flows at $t=1$ and $2$ are
\begin{equation*}
\begin{aligned}
&\boldsymbol{y_1}=\boldsymbol{x_1}+\boldsymbol{u_1}=\begin{pmatrix}
    y_{11}\\
    y_{12}
\end{pmatrix},\\
&\boldsymbol{y_2}=\boldsymbol{i}+\boldsymbol{x_2}+\boldsymbol{u_2}=\begin{pmatrix}
    y_{21}\\
    y_{22}
\end{pmatrix},\\
\end{aligned}
\end{equation*}
where noise trading
\begin{equation*}
\begin{aligned}
&\boldsymbol{u_1}=\begin{pmatrix}
    u_{11}\\
    u_{12}
\end{pmatrix}\sim N(\boldsymbol{0},\boldsymbol{\Sigma_{1}}),\ \boldsymbol{\Sigma_{1}}=\begin{pmatrix}
\sigma_{11}^2&0\\
0&\sigma_{12}^2
\end{pmatrix},\\
&\boldsymbol{u_2}=\begin{pmatrix}
    u_{21}\\
    u_{22}
\end{pmatrix}\sim N(\boldsymbol{0},\boldsymbol{\Sigma_{2}}),\ \boldsymbol{\Sigma_{2}}=\begin{pmatrix}
\sigma_{21}^2&0\\
0&\sigma_{22}^2
\end{pmatrix}.
\end{aligned}
\end{equation*}

\begin{figure}[!htbp]
    \centering
    \includegraphics[height=0.13\textheight]{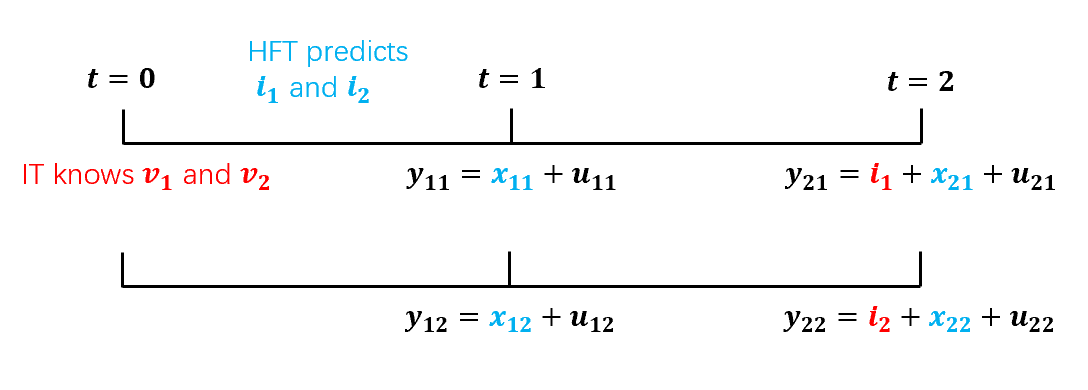}
    \caption{Timeline of the extended multi-asset Kyle model.}
    \label{figtimeline-multi1}
\end{figure}

A linear conjecture is still applied:
\begin{equation*}
\begin{aligned}
&\boldsymbol{p_1}=\boldsymbol{p_0}+\boldsymbol{\lambda_1}\boldsymbol{y_1},\ \boldsymbol{p_{2}}=\boldsymbol{p_0}+\boldsymbol{\lambda_{21}}\boldsymbol{y_1}+\boldsymbol{\lambda_{22}}\boldsymbol{y_2},\\
&\boldsymbol{i}=\boldsymbol{\alpha}(\boldsymbol{v}-\boldsymbol{p_0}),\\
&\boldsymbol{x_1}=\boldsymbol{\beta_1}\hat{\boldsymbol{i}},\ \boldsymbol{x_2}=\boldsymbol{\beta_{21}}\hat{\boldsymbol{i}}+\boldsymbol{\beta_{22}}\boldsymbol{u_1}+\boldsymbol{\beta_{23}}\boldsymbol{x_1},    
\end{aligned}
\end{equation*}
where $\boldsymbol{\lambda_1},\boldsymbol{\lambda_{21}},\boldsymbol{\lambda_{22}},\boldsymbol{\alpha},\boldsymbol{\beta_1},\boldsymbol{\beta_{21}},\boldsymbol{\beta_{22}},\boldsymbol{\beta_{23}}\in\mathbbm{R}^{2\times2}$. In the following, $\boldsymbol{A}_{ij}$ represents the element of $\boldsymbol{A}$ at the $i$-th row and the $j$-th column.

\textbf{Dealers' quotes.} Dealers set the transaction prices based on the order flow:
\begin{equation*}
\begin{aligned}
\boldsymbol{p_1}=\mathbbm{E}(\boldsymbol{v}|\boldsymbol{y_1}),\ \boldsymbol{p_2}=\mathbbm{E}(\boldsymbol{v}|\boldsymbol{y_1},\boldsymbol{y_2}).
\end{aligned}
\end{equation*}
So
{\footnotesize
\begin{equation}
\label{multi1-lams}
\begin{aligned}
\boldsymbol{\lambda_1}=&\boldsymbol{\Sigma_v}\boldsymbol{\alpha}^\textsuperscript{T}\boldsymbol{\beta_1}^\textsuperscript{T}[\boldsymbol{\beta_1}(\boldsymbol{\alpha}\boldsymbol{\Sigma_v}\boldsymbol{\alpha}^\textsuperscript{T}+\boldsymbol{\Sigma_\varepsilon})\boldsymbol{\beta_1}^\textsuperscript{T}+\boldsymbol{\Sigma_1}]^{-1},\\
\begin{pmatrix}
\boldsymbol{\lambda_{21}}&\boldsymbol{\lambda_{22}}
\end{pmatrix}=& \boldsymbol{\Sigma_v}\boldsymbol{\alpha}^\textsuperscript{T}\begin{pmatrix}
\boldsymbol{\beta_1}^\textsuperscript{T}&(\boldsymbol{I}+\boldsymbol{\beta})^\textsuperscript{T}\\
  \end{pmatrix}\\
 &\begin{pmatrix}
\boldsymbol{\beta_1}(\boldsymbol{\alpha}\boldsymbol{\Sigma_v}\boldsymbol{\alpha}^\textsuperscript{T}+\boldsymbol{\Sigma_\varepsilon})\boldsymbol{\beta_1}^\textsuperscript{T}+\boldsymbol{\Sigma_1}&\boldsymbol{\beta_1}\boldsymbol{\alpha}\boldsymbol{\Sigma_v}\boldsymbol{\alpha}^\textsuperscript{T}(\boldsymbol{I}+\boldsymbol{\beta})^\textsuperscript{T}+\boldsymbol{\beta_1}\boldsymbol{\Sigma_\varepsilon}\boldsymbol{\beta}^\textsuperscript{T}+\boldsymbol{\Sigma_1}\boldsymbol{\beta_{22}}^\textsuperscript{T}\\
\boldsymbol{\beta_1}\boldsymbol{\alpha}\boldsymbol{\Sigma_v}\boldsymbol{\alpha}^\textsuperscript{T}(\boldsymbol{I}+\boldsymbol{\beta})^\textsuperscript{T}+\boldsymbol{\beta_1}\boldsymbol{\Sigma_\varepsilon}\boldsymbol{\beta}^\textsuperscript{T}+\boldsymbol{\Sigma_1}\boldsymbol{\beta_{22}}^\textsuperscript{T}&(\boldsymbol{I}+\boldsymbol{\beta})\boldsymbol{\alpha}\boldsymbol{\Sigma_v}\boldsymbol{\alpha}^\textsuperscript{T}(\boldsymbol{I}+\boldsymbol{\beta})^\textsuperscript{T}+\boldsymbol{\beta}\boldsymbol{\Sigma_\varepsilon}\boldsymbol{\beta}^\textsuperscript{T}+\boldsymbol{\beta_{22}}\boldsymbol{\Sigma_1}\boldsymbol{\beta_{22}}^\textsuperscript{T}+\boldsymbol{\Sigma_2}
\end{pmatrix}^{-1},
\end{aligned}
\end{equation}}
where $\boldsymbol{\beta}=\boldsymbol{\beta_{21}}+\boldsymbol{\beta_{23}\beta_1}.$

\textbf{IT's strategies.} IT maximizes her expected profits
\begin{equation*}
\mathbbm{E}(\boldsymbol{i}^\textsuperscript{T}(\boldsymbol{v}-\boldsymbol{p_2})|\boldsymbol{v})=-\boldsymbol{i}^\textsuperscript{T}\boldsymbol{\lambda_2}\boldsymbol{i}+\boldsymbol{i}^\textsuperscript{T}(\boldsymbol{v}-\boldsymbol{p_0}),\\
\end{equation*}
where $\boldsymbol{\lambda_2}=\boldsymbol{\lambda_{21}\beta_1}+\boldsymbol{\lambda_{22}}(\boldsymbol{I}+\boldsymbol{\beta}).$ So
\begin{equation}
\label{multi1-alpha}
\boldsymbol{\alpha}=\frac{1}{2}\boldsymbol{\lambda_{2s}}^{-1},
\end{equation}
where $\boldsymbol{A_s}\triangleq\frac{\boldsymbol{A}+\boldsymbol{A}^\textsuperscript{T}}{2}$ is the symmetric part of $\boldsymbol{A}$. The SOC implies that $\boldsymbol{\lambda_{2s}}$ must be positive-definite.

\textbf{HFT's strategies.} The objective function of HFT is composed by her expected profits together with the following inventory penalty:
\begin{equation*}
-(\boldsymbol{x_1}+\boldsymbol{x_2})^\textsuperscript{T}\boldsymbol{\gamma}(\boldsymbol{x_1}+\boldsymbol{x_2}),\ \boldsymbol{\gamma}=\begin{pmatrix}
\gamma_1&\gamma_3\\
\gamma_3&\gamma_2
\end{pmatrix}.
\end{equation*}
Here $\gamma_3$ measures the cross-asset inventory penalty, which may come from HFT's fund limitation and cross-asset price impact.

In period 2, HFT maximizes
\begin{equation*}
\begin{aligned}
&\mathbbm{E}\left(\boldsymbol{x_2}^\textsuperscript{T}(\boldsymbol{v}-\boldsymbol{p_2})-(\boldsymbol{x_1}+\boldsymbol{x_2})^\textsuperscript{T}\boldsymbol{\gamma}(\boldsymbol{x_1}+\boldsymbol{x_2})|\hat{\boldsymbol{i}},\boldsymbol{y_1}\right)\\
=&-\boldsymbol{x_2}^\textsuperscript{T}(\boldsymbol{\lambda_{22}}+\boldsymbol{\gamma})\boldsymbol{x_2}+\boldsymbol{x_2}^\textsuperscript{T}[\boldsymbol{\eta}\hat{\boldsymbol{i}}-\boldsymbol{\lambda_{21}}\boldsymbol{u_1}-(\boldsymbol{\lambda_{21}}+2\boldsymbol{\gamma})\boldsymbol{x_1}]-\boldsymbol{x_1}^\textsuperscript{T}\boldsymbol{\gamma}\boldsymbol{x_1},\\
\end{aligned}
\end{equation*}
where $\boldsymbol{\eta}=(\boldsymbol{I}-\boldsymbol{\lambda_{22}\alpha})\boldsymbol{\Sigma_v\alpha}^\textsuperscript{T}(\boldsymbol{\alpha}\boldsymbol{\Sigma_v\alpha}^\textsuperscript{T}+\boldsymbol{\Sigma_\varepsilon})^{-1}.$ So
\begin{equation}
\label{multi1-beta2}
\begin{aligned}
\boldsymbol{\beta_{21}}=\frac{1}{2}(\boldsymbol{\lambda_{22}}+\boldsymbol{\gamma})_{\textbf{s}}^{-1}\boldsymbol{\eta},\ \boldsymbol{\beta_{22}}= -\frac{1}{2}(\boldsymbol{\lambda_{22}}+\boldsymbol{\gamma})_{\textbf{s}}^{-1}\boldsymbol{\lambda_{21}},\ \boldsymbol{\beta_{23}}=-\frac{1}{2}(\boldsymbol{\lambda_{22}}+\boldsymbol{\gamma})_{\textbf{s}}^{-1}(\boldsymbol{\lambda_{21}}+2\boldsymbol{\gamma
})
\end{aligned}
\end{equation}
and $(\boldsymbol{\lambda_{22}}+\boldsymbol{\gamma})_{\textbf{s}}$ must be positive-definite.

In period 1, HFT maximizes
\begin{equation*}
\begin{aligned}
&\mathbbm{E}\left(\boldsymbol{x_1}^\textsuperscript{T}(\boldsymbol{v}-\boldsymbol{p_1})+\boldsymbol{x_2}^\textsuperscript{T}(\boldsymbol{v}-\boldsymbol{p_2})-(\boldsymbol{x_1}+\boldsymbol{x_2})^\textsuperscript{T}\boldsymbol{\gamma}(\boldsymbol{x_1}+\boldsymbol{x_2})|\hat{\boldsymbol{i}}\right)\\
=&-\boldsymbol{x_1}^\textsuperscript{T}[\boldsymbol{\lambda_1}+\boldsymbol{\gamma}-(\boldsymbol{\lambda_{21}}+2\boldsymbol{\gamma
})^\textsuperscript{T}\boldsymbol{B}(\boldsymbol{\lambda_{21}}+2\boldsymbol{\gamma
})]\boldsymbol{x_1}+\boldsymbol{x_1}^\textsuperscript{T}[\boldsymbol{\mu}-2(\boldsymbol{\lambda_{21}}+2\boldsymbol{\gamma
})^\textsuperscript{T}\boldsymbol{B}\boldsymbol{\eta}]+\mathbbm{E}(\boldsymbol{u_1}^\textsuperscript{T}\boldsymbol{\lambda_{21}}^\textsuperscript{T}\boldsymbol{B}\boldsymbol{\lambda_{21}}\boldsymbol{u_1}),
\end{aligned}
\end{equation*}
where $\boldsymbol{B}=\frac{1}{2}(\boldsymbol{\lambda_{22}}+\boldsymbol{\gamma})_{\textbf{s}}^{-1}-\frac{1}{4}(\boldsymbol{\lambda_{22}}+\boldsymbol{\gamma})_{\textbf{s}}^{-1}(\boldsymbol{\lambda_{22}}+\boldsymbol{\gamma})(\boldsymbol{\lambda_{22}}+\boldsymbol{\gamma})_{\textbf{s}}^{-1}$ and $\boldsymbol{\mu}=\boldsymbol{\Sigma_v\alpha}^\textsuperscript{T}(\boldsymbol{\alpha}\boldsymbol{\Sigma_v\alpha}^\textsuperscript{T}+\boldsymbol{\Sigma_\varepsilon})^{-1}.$ So 
\begin{equation}
\label{multi1-beta1}
\begin{aligned}
\boldsymbol{\beta_1}=&\frac{1}{2}[\boldsymbol{\lambda_1}+\boldsymbol{\gamma}-(\boldsymbol{\lambda_{21}}+2\boldsymbol{\gamma
})^\textsuperscript{T}\boldsymbol{B}(\boldsymbol{\lambda_{21}}+2\boldsymbol{\gamma
})]_\textbf{s}^{-1}[\boldsymbol{\mu}-2(\boldsymbol{\lambda_{21}}+2\boldsymbol{\gamma
})^\textsuperscript{T}\boldsymbol{B}\boldsymbol{\eta}]  
\end{aligned}
\end{equation}
and $[\boldsymbol{\lambda_1}+\boldsymbol{\gamma}-(\boldsymbol{\lambda_{21}}+2\boldsymbol{\gamma
})^\textsuperscript{T}\boldsymbol{B}(\boldsymbol{\lambda_{21}}+2\boldsymbol{\gamma
})]_\textbf{s}$ must be positive-definite.

\textbf{Equilibrium.} \eqref{multi1-lams}, \eqref{multi1-alpha}, \eqref{multi1-beta2}, \eqref{multi1-beta1} and SOCs form the equilibrium conditions. The equilibrium can be solved out through numerical methods. Conclusions about HFT's anticipatory strategies and their effects on IT and market quality remain the same as in the single-asset case. In other words, the conclusions in Section \ref{secequilibriumgeneral} are robust when multiple assets are considered.

\subsection{When HFT applies the information to correlated assets}
\label{subsecmultiasset2}

In this section, we consider the following case: IT has private information of asset 1 and does informed trading on it, HFT acquires information from the prediction to IT's order and applies it to asset 1 and 2. At $t=0,$ IT only receives the signal $v_1$ and sends the order $i_1$ on asset 1. Thus, $\boldsymbol{i}=\begin{pmatrix}
   i_1\\
    0
\end{pmatrix}.$ After that, HFT receives the signal 
\begin{equation*}
\hat{i}_1=i_1+\varepsilon,\ \varepsilon\sim N(0,\sigma_\varepsilon^2),
\end{equation*}
where the noise $\varepsilon$ is independent with other random variables.

\begin{figure}[!htbp]
    \centering
    \includegraphics[height=0.13\textheight]{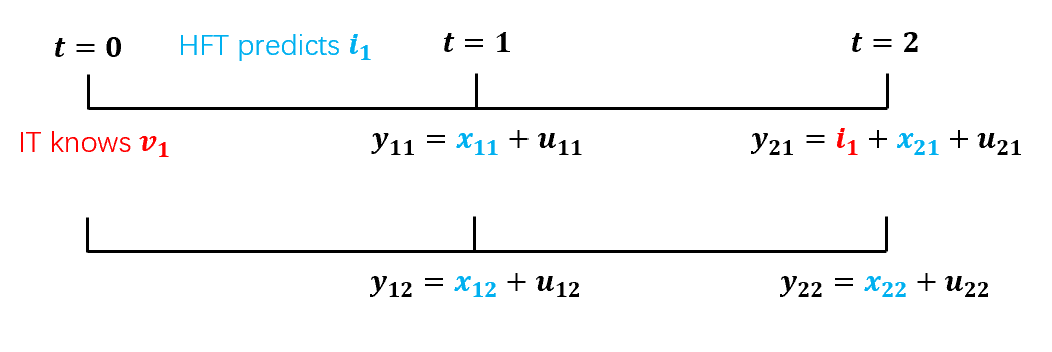}
    \caption{Timeline when HFT applies the information to multiple assets.}
    \label{figtimeline-multi2}
\end{figure}


Dealers of each asset offer the prices based on the order flow of this single asset:
\begin{equation*}
\begin{aligned}
p_{1j}=\mathbbm{E}(v_j|y_{1j}),\  p_{2j}=\mathbbm{E}(v_j|y_{1j},y_{2j}),\ j=1,2.
\end{aligned}
\end{equation*}
IT maximizes her expected profit only on asset 1:
\begin{equation*}
\mathbbm{E}(i_1(v_1-p_{21})|v_1).\\
\end{equation*}
HFT's objective function is still composed by the expected profits with inventory penalty on both assets.

Now the linear conjecture becomes
\begin{equation*}
\begin{aligned}
&\boldsymbol{p_1}=\boldsymbol{p_0}+\boldsymbol{\lambda_1}\boldsymbol{y_1},\ \boldsymbol{p_{2}}=\boldsymbol{p_0}+\boldsymbol{\lambda_{21}}\boldsymbol{y_1}+\boldsymbol{\lambda_{22}}\boldsymbol{y_2},\\
&i_1=\alpha(v_1-p_{01}),\\
&\boldsymbol{x_1}=\boldsymbol{\beta_1}\hat{i}_1,\boldsymbol{x_2}=\boldsymbol{\beta_{21}}\hat{i}_1+\boldsymbol{\beta_{22}}\boldsymbol{u_1}+\boldsymbol{\beta_{23}}\boldsymbol{x_1},\\
\end{aligned}
\end{equation*}
where $\boldsymbol{\lambda_1},\boldsymbol{\lambda_{21}},\boldsymbol{\lambda_{22}},\boldsymbol{\alpha},\boldsymbol{\beta_{22}},\boldsymbol{\beta_{23}}\in\mathbbm{R}^{2\times2}$ and $\boldsymbol{\beta_1},\boldsymbol{\beta_{21}}\in\mathbbm{R}^{2\times1}.$
The equilibrium can be solved out similarly as in Section \ref{subsecmultiasset1}. $\boldsymbol{\beta}_{\textbf{1}11}$ and $\text{sgn}(\rho)\boldsymbol{\beta}_{\textbf{1}21}$ represent the trading directions of $x_{11}$ and $x_{12}$, $\boldsymbol{\beta}_{\textbf{21}11}+\boldsymbol{\beta}_{\textbf{23}11}\boldsymbol{\beta}_{\textbf{1}11}+\boldsymbol{\beta}_{\textbf{23}12}\boldsymbol{\beta}_{\textbf{1}21}$ and $\text{sgn}(\rho)(\boldsymbol{\beta}_{\textbf{21}21}+\boldsymbol{\beta}_{\textbf{23}21}\boldsymbol{\beta}_{\textbf{1}11}+\boldsymbol{\beta}_{\textbf{23}22}\boldsymbol{\beta}_{\textbf{1}21})$ represent the trading directions
of $x_{21}$ and $x_{22}$.

Following the baseline parameters used in \cite{yang2020back}, we set the normalized price $p_{01}=p_{02}=1$; the daily volatility $\sigma_{v_1}=\sigma_{v_2}=\sqrt{0.00036};$ the sizes of noise trading $\sigma_{11},\sigma_{12},\sigma_{21},\sigma_{22}=0.6,0.5,1,1\text{ million shares};$ the signal noise $\sigma_\varepsilon=0.2 \text{ million shares}.$ The correlation $\rho=0.8,$ which implies that the assets are highly positive correlated.\footnote{The results with a negative $\rho$ are actually similar.}
Let
\begin{equation}
\gamma_1=\gamma_2=4\gamma_3\in[0,0.5],
\end{equation}
where the bound $0.5$ is large enough because the impact coefficients of this market are approximately $O(10^{-3})$.
\begin{figure}[!htbp]
    \centering
    \subcaptionbox{Direction of HFT's time-1 trading.}{
    \includegraphics[width = 0.3\textwidth]{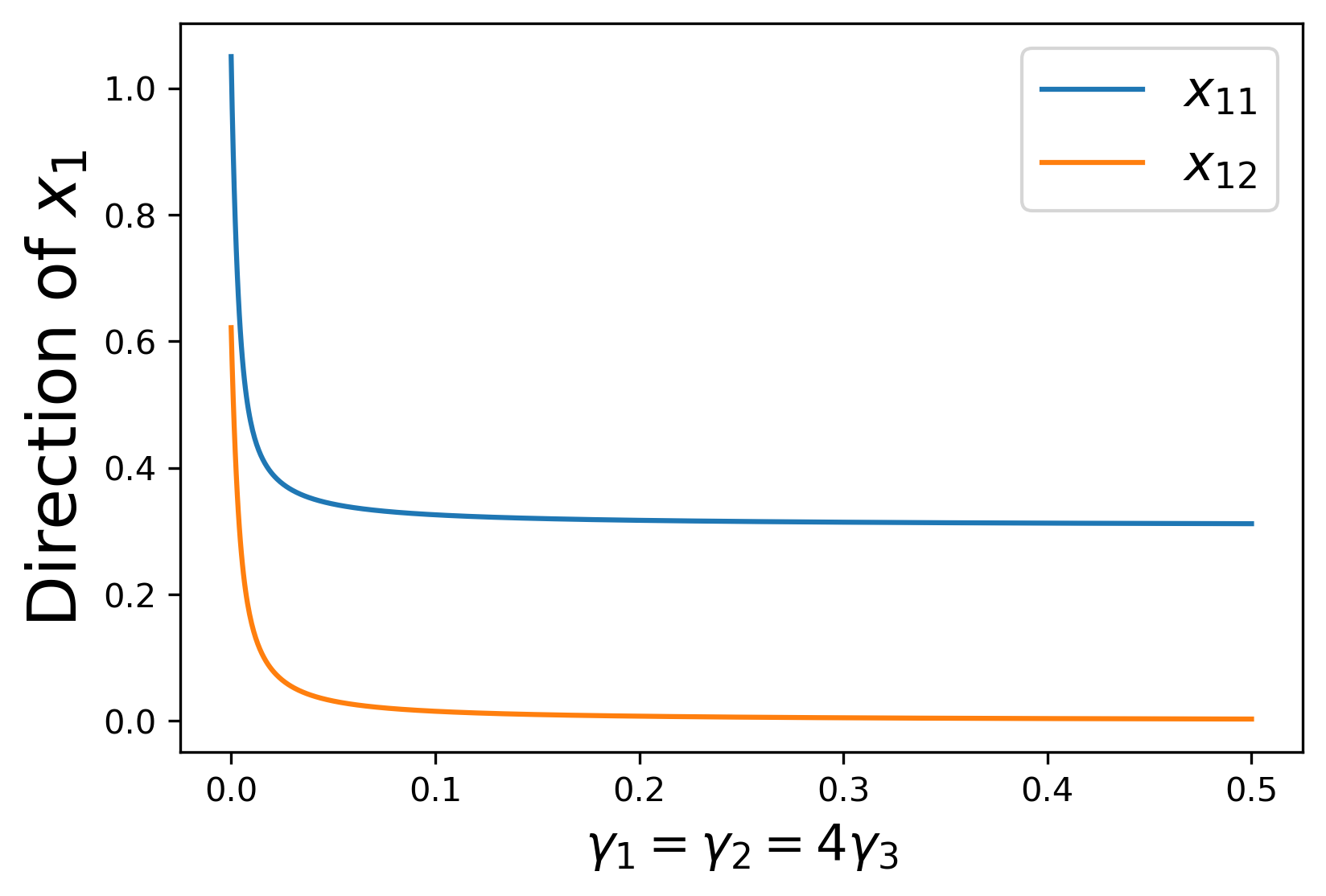}
    }
\subcaptionbox{Direction of HFT's time-2 trading.}{
\includegraphics[width = 0.3\textwidth]{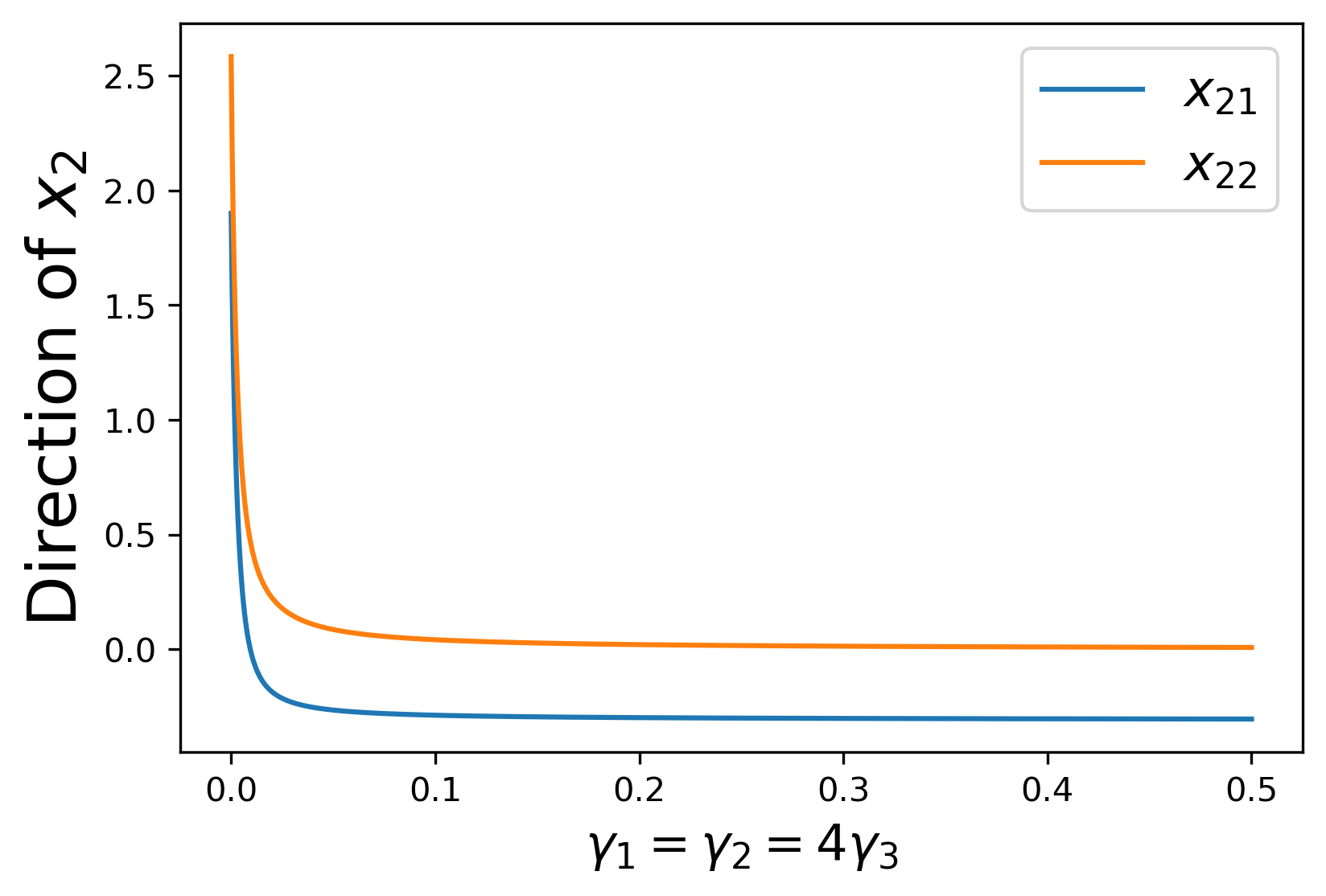}
    }
  \subcaptionbox{IT's intensity.}{
\includegraphics[width = 0.3\textwidth]{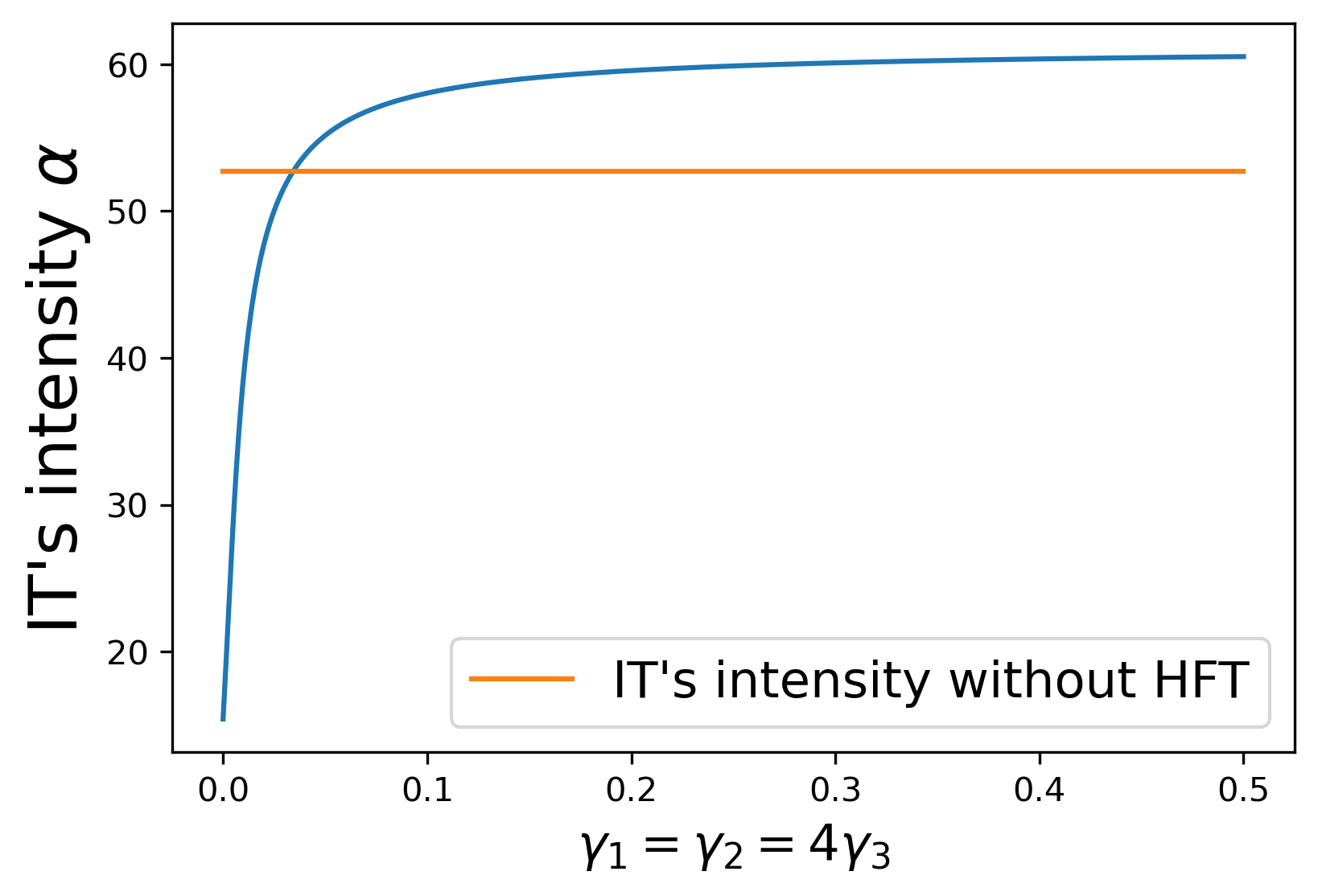}
    }
\caption{Investors' behavior in multi-asset model.}
\label{figmultiasset}
\end{figure}

From Figure \ref{figmultiasset}, we find the following two facts. (1) On asset 1, HFT implements anticipatory trading after predicting IT's order. With the increase of inventory aversion, her role changes from Small-IT to Round-Tripper and may benefit IT, as in the single-asset case. (2) On asset 2, HFT always act as Small-IT. Since $v_2$ is correlated with $v_1$, the signal $\hat{i}_1$ also helps HFT infer the value of $v_2,$ and the large inventory aversion makes her investment more conservative. Without a large trader on asset 2 to bring price impact and act as a counterparty, HFT cannot make any impact profit, so she will not trade in the opposite direction to control inventory as Round-Tripper. Actually, if HFT only trades on asset 2, she will also always play the role of Small-IT.

\section{Conclusion}
\label{secconclusion}
We build up a theoretical model to investigate high-frequency order anticipation strategies with inventory management, and their effects on large trader IT and the market. The way HFT implements anticipatory trading can be quite different when the inventory aversion and signal precision change: she may act as Small-IT or Round-Tripper. When HFT acts as Round-Tripper, she provides IT with liquidity and may benefit her; when HFT acts as Small-IT, she takes liquidity away and always harms IT. 
Some surprising results appear when the size of high-speed noise trading differs: (1) when it is relatively small, Round-Tripper may harm the large trader more than Small-IT and a sufficiently inventory-averse HFT may profit more with a less accurate signal; (2) when it is relatively large and Round-Tripper is beneficial to IT, adding any noise in HFT's signal can decrease IT's profit. In the presence of HFT, the loss of normal-speed noise traders is reduced and price discovery is accelerated. However, in any case, the final level of price discovery remains unchanged. If almost all the noise traders are as slow as IT, HFT gives up her speed advantage and becomes normal-speed, too. If HFT is extremely inventory averse, she becomes a short-term Round-Tripper who makes profit only from the market impact. Conclusions about anticipatory trading's effects are robust when IT and HFT both trade on multiple correlated assets. In the case where HFT may not hold exactly the same assets as IT, it is found that the existence of a large trader who brings price impact is a necessary condition for an anticipary HFT to play the role of Round-Tripper.


\newpage
\section*{Appendix}
\label{secappendix}
\appendix
\noindent
\textbf{Proof of Theorem \ref{mainthm}.} Substitute \eqref{lam1}, \eqref{beta21}, \eqref{beta22}, \eqref{beta23} into \eqref{alpha}, we have
\begin{equation}
\label{lambda21}
\Lambda_{21}=\frac{A^3 \Lambda_{22} [2 (\beta_1-1) \Gamma-\Lambda_{22}]+A^2 \Gamma+2 A \theta_\varepsilon \Lambda_{22} [(\beta_1-1) \Gamma-\Lambda_{22}]+\theta_\varepsilon (\Gamma+\Lambda_{22})}{A\beta_1 (A^2+\theta_\varepsilon) (2 \Gamma+\Lambda_{22})}.
\end{equation}
Then substitute \eqref{lam1}, \eqref{beta21}, \eqref{beta22}, \eqref{beta23}, \eqref{lambda21} into \eqref{lam21}, \eqref{lam22}, \eqref{beta1} and three SOCs, we get the simplified system:
{\footnotesize\begin{equation}
\label{systemmain}
\begin{aligned}
0=&\beta_1-(A^4 (-\Lambda_{22}) (\Lambda_{22} (2 \Gamma+\Lambda_{22})-4 \beta_1 \Gamma (\Gamma+\Lambda_{22}))+A^3 \Lambda_{22} ((2 \beta_1+3) \Gamma+2 \beta_1 \Lambda_{22}+\Lambda_{22})\\
+&A^2 (4 \beta_1 \theta_{\varepsilon} \Gamma^2 \Lambda_{22}+\Gamma (2 (2 \beta_1-1) \theta_{\varepsilon} \Lambda_{22}^2-1)-2 \theta_{\varepsilon} \Lambda_{22}^3)+A (2 \beta_1+3) \theta_{\varepsilon} \Lambda_{22} (\Gamma+\Lambda_{22})-\theta_{\varepsilon} (\Gamma+\Lambda_{22}))/\\
&(\beta_1 (A^2+\theta_{\varepsilon})^2 (\Gamma+\Lambda_{22}) (2 \Gamma+\Lambda_{22}) (\frac{4 A \beta_1}{A^2 \beta_1^2+\beta_1^2 \theta_{\varepsilon}+\theta_1}\\
-&\frac{(A^3 (4 \beta_1 \Gamma (\Gamma+\Lambda_{22})-\Lambda_{22} (2 \Gamma+\Lambda_{22}))+A^2 \Gamma+2 A \theta_{\varepsilon} (\Gamma+\Lambda_{22}) (2 \beta_1 \Gamma-\Lambda_{22})+\theta_{\varepsilon} (\Gamma+\Lambda_{22}))^2}{A^2 \beta_1^2 (A^2+\theta_{\varepsilon})^2 (\Gamma+\Lambda_{22}) (2 \Gamma+\Lambda_{22})^2}+4 \Gamma)),\\
0=&(2 (\beta_1-1) \Gamma-\Lambda_{22}) \Lambda_{22} (4 ((\theta_{\varepsilon}+\theta_1+1) \Gamma^2+2 (\theta_{\varepsilon}+1) \Lambda_{22} \Gamma+(\theta_{\varepsilon}+1) \Lambda_{22}^2) \beta_1^2-4 \Gamma (2 \Gamma+\Lambda_{22}) \theta_1 \beta_1\\
+&(2 \Gamma+\Lambda_{22})^2 \theta_1) A^7-(8 \theta_{\varepsilon} \Gamma (\Gamma+\Lambda_{22})^2 \beta_1^3-4 ((\theta_{\varepsilon}+\theta_1-1) \Gamma^3+\Lambda_{22} (2 \theta_{\varepsilon}-\theta_1-3) \Gamma^2\\
+&(\theta_{\varepsilon}-3) \Lambda_{22}^2 \Gamma-\Lambda_{22}^3) \beta_1^2+4 \Gamma (2 \Gamma^2-\Lambda_{22} \Gamma-\Lambda_{22}^2) \theta_1 \beta_1-(\Gamma-\Lambda_{22}) (2 \Gamma+\Lambda_{22})^2 \theta_1) A^6\\
+&(8 \theta_{\varepsilon} \Gamma \Lambda_{22} ((2 \theta_{\varepsilon}+3 \theta_1+3) \Gamma^2+2 (2 \theta_{\varepsilon}+3) \Lambda_{22} \Gamma
+(2 \theta_{\varepsilon}+3) \Lambda_{22}^2) \beta_1^3-2 \theta_{\varepsilon} (4 \Lambda_{22} (2 \theta_{\varepsilon}+7 \theta_1+3) \Gamma^3\\
+&(2 \Lambda_{22}^2 (11 \theta_{\varepsilon}+8 \theta_1+16)-1) \Gamma^2+(4 (5 \theta_{\varepsilon}+7) \Lambda_{22}^3-2 \Lambda_{22}) \Gamma
+\Lambda_{22}^2 ((6 \theta_{\varepsilon}+8) \Lambda_{22}^2-1)) \beta_1^2\\
+&2 \Gamma ((11 \theta_{\varepsilon}+4) \Lambda_{22}^3+4 \Gamma^2 (6 \theta_{\varepsilon} \Lambda_{22}+\Lambda_{22})+\Gamma (8 (4 \theta_{\varepsilon}+1) \Lambda_{22}^2-1)) \theta_1 \beta_1
-(2 \Gamma+\Lambda_{22}) (2 (3 \theta_{\varepsilon}+2) \Lambda_{22}^3\\
+&4 \Gamma^2 (2 \theta_{\varepsilon} \Lambda_{22}+\Lambda_{22})+\Gamma (2 (7 \theta_{\varepsilon}+4) \Lambda_{22}^2-1)) \theta_1) A^5+(-16 \theta_{\varepsilon}^2 \Gamma (\Gamma+\Lambda_{22})^2 \beta_1^3\\
+&4 \theta_{\varepsilon} ((2 \theta_{\varepsilon}+3 \theta_1-1) \Gamma^3+\Lambda_{22} (6 \theta_{\varepsilon}-\theta_1-3) \Gamma^2+3 (2 \theta_{\varepsilon}-1) \Lambda_{22}^2 \Gamma+(2 \theta_{\varepsilon}-1) \Lambda_{22}^3) \beta_1^2\\
-&8 \theta_{\varepsilon} \Gamma^2 (2 \Gamma+\Lambda_{22}) \theta_1 \beta_1
+(\Gamma+\Lambda_{22}) ((8 \theta_{\varepsilon}+4) \Gamma^2+4 (2 \theta_{\varepsilon} \Lambda_{22}+\Lambda_{22}) \Gamma+\theta_{\varepsilon} \Lambda_{22}^2) \theta_1) A^4\\
+&\theta_{\varepsilon} (8 \theta_{\varepsilon} \Gamma \Lambda_{22} ((\theta_{\varepsilon}+3 \theta_1+3) \Gamma^2+2 (\theta_{\varepsilon}+3) \Lambda_{22} \Gamma+(\theta_{\varepsilon}+3) \Lambda_{22}^2) \beta_1^3-4 \theta_{\varepsilon} \Lambda_{22} (2 (\theta_{\varepsilon}+5 \theta_1+3) \Gamma^3\\
+&\Lambda_{22} (6 \theta_{\varepsilon}+7 \theta_1+17) \Gamma^2+2 (3 \theta_{\varepsilon}+8) \Lambda_{22}^2 \Gamma+(2 \theta_{\varepsilon}+5) \Lambda_{22}^3) \beta_1^2+2 \Gamma (8 (\theta_{\varepsilon}+1) \Lambda_{22}^3\\
+&4 (5 \theta_{\varepsilon}+4) \Gamma \Lambda_{22}^2+4 (3 \theta_{\varepsilon}+2) \Gamma^2 \Lambda_{22}-\Lambda_{22}-2 \Gamma) \theta_1 \beta_1-\Lambda_{22} (\Gamma+\Lambda_{22}) (8 (\theta_{\varepsilon}+2) \Gamma^2\\
+&4 (4 \theta_{\varepsilon}+7) \Lambda_{22} \Gamma
+4 (2 \theta_{\varepsilon}+3) \Lambda_{22}^2-1) \theta_1) A^3+4 \theta_{\varepsilon} (-2 \theta_{\varepsilon}^2 \Gamma (\Gamma+\Lambda_{22})^2 \beta_1^3+\theta_{\varepsilon} ((\theta_{\varepsilon}+3 \theta_1+1) \Gamma^3\\
+&\Lambda_{22} (4 \theta_{\varepsilon}+\theta_1+3) \Gamma^2+(5 \theta_{\varepsilon}+3) \Lambda_{22}^2 \Gamma+(2 \theta_{\varepsilon}+1) \Lambda_{22}^3) \beta_1^2-\theta_{\varepsilon} \Gamma (2 \Gamma^2+3 \Lambda_{22} \Gamma+\Lambda_{22}^2) \theta_1 \beta_1\\
+&(\Gamma+\Lambda_{22})^2 ((\theta_{\varepsilon}+2) \Gamma+2 \theta_{\varepsilon} \Lambda_{22}+\Lambda_{22}) \theta_1) A^2+2 \theta_{\varepsilon}^2 (4 \theta_{\varepsilon} \Gamma \Lambda_{22} ((\theta_1+1) \Gamma^2+2 \Lambda_{22} \Gamma+\Lambda_{22}^2) \beta_1^3\\
-&\theta_{\varepsilon} (\Gamma+\Lambda_{22}) (4 \Lambda_{22}^3+8 \Gamma \Lambda_{22}^2+4 \Gamma^2 (\theta_1+1) \Lambda_{22}+\Lambda_{22}+\Gamma) \beta_1^2\\
+&\Gamma (4 \Lambda_{22}^3+8 \Gamma \Lambda_{22}^2+4 \Gamma^2 \Lambda_{22}-\Lambda_{22}-\Gamma) \theta_1 \beta_1-(\Gamma+\Lambda_{22})^2 (4 \Lambda_{22}^2+4 \Gamma \Lambda_{22}+1) \theta_1) A\\
+&4 \theta_{\varepsilon}^2 (\Gamma+\Lambda_{22}) (\theta_{\varepsilon} ((\theta_1+1) \Gamma^2+2 \Lambda_{22} \Gamma+\Lambda_{22}^2) \beta_1^2+(\Gamma+\Lambda_{22})^2 \theta_1),\\
0=&A^4 \Lambda_{22} (4 \beta_1^2 (\Gamma^2 (\theta_{\varepsilon}+\theta_1+1)+2 (\theta_{\varepsilon}+1) \Gamma \Lambda_{22}+(\theta_{\varepsilon}+1) \Lambda_{22}^2)-4 \beta_1 \Gamma \theta_1 (2 \Gamma+\Lambda_{22})+\theta_1 (2 \Gamma+\Lambda_{22})^2)\\
-&2 A^3 (2 \beta_1^2 \theta_{\varepsilon} (\Gamma+\Lambda_{22})^2-2 \beta_1 \Gamma^2 \theta_1+\Gamma \theta_1 (2 \Gamma+\Lambda_{22}))+A^2 (4 \beta_1^2 \theta_{\varepsilon} \Lambda_{22} (\Gamma^2 (\theta_{\varepsilon}+2 \theta_1+2)\\
+&2 (\theta_{\varepsilon}+2) \Gamma \Lambda_{22}
+(\theta_{\varepsilon}+2) \Lambda_{22}^2)-4 \beta_1 \theta_{\varepsilon} \Gamma \Lambda_{22} \theta_1 (2 \Gamma+\Lambda_{22})+\theta_1 (4 (\theta_{\varepsilon}+1) \Gamma^2 \Lambda_{22}\\
+&8 (\theta_{\varepsilon}+1) \Gamma \Lambda_{22}^2+4 (\theta_{\varepsilon}+1) \Lambda_{22}^3-2 \Gamma-\Lambda_{22}))-4 A \theta_{\varepsilon} (\beta_1^2 \theta_{\varepsilon} (\Gamma+\Lambda_{22})^2-\beta_1 \Gamma^2 \theta_1+\theta_1 (\Gamma+\Lambda_{22})^2)\\
+&4 \theta_{\varepsilon} \Lambda_{22} (\beta_1^2 \theta_{\varepsilon} (\Gamma^2 (\theta_1+1)+2 \Gamma \Lambda_{22}+\Lambda_{22}^2)+\theta_1 (\Gamma+\Lambda_{22})^2),\\
0<&\Lambda_{22}+\Gamma,\\
0<&A,\\
0<&4\Gamma+\frac{4 A \beta_1}{A^2 \beta_1^2+\beta_1^2 \theta_{\varepsilon}+\theta_1}\\
-&\frac{(A^3 (4 \beta_1 \Gamma (\Gamma+\Lambda_{22})-\Lambda_{22} (2 \Gamma+\Lambda_{22}))+A^2 \Gamma+2 A \theta_{\varepsilon} (\Gamma+\Lambda_{22}) (2 \beta_1 \Gamma-\Lambda_{22})+\theta_{\varepsilon} (\Gamma+\Lambda_{22}))^2}{A^2 \beta_1^2 (A^2+\theta_{\varepsilon})^2 (\Gamma+\Lambda_{22}) (2 \Gamma+\Lambda_{22})^2}.
\end{aligned}
\end{equation}}

For investors' profits and market variables, we only prove the time-2 pricing error is always $\frac{\sigma_v^2}{2}.$ Substitute \eqref{lam21}, \eqref{lam22} into \eqref{alpha}, we have
\begin{equation}
\label{Asolve}
A=\sqrt{\frac{\theta_1\theta_\varepsilon(\beta_{21}+\beta_1(\beta_{23}-\beta_{22}))^2+\beta_1^2\theta_\varepsilon+\theta_1}{\theta_1(1+\beta_{21}+\beta_1(\beta_{23}-\beta_{22}))^2+(1+\theta_\varepsilon)\beta_1^2}}.
\end{equation}
Calculate $\mathbbm{E}(v-p_2)^2$ and substitute \eqref{lam21}, \eqref{lam22}, \eqref{Asolve} into it, we prove it.

\noindent\textbf{Proof of Proposition \ref{propexistence}.} For the non-existence of equilibrium under \eqref{conjecture}, we only prove the case $\theta_\varepsilon=0$ and $\Gamma=0.$
Substitute \eqref{beta21},\eqref{beta22},\eqref{beta23},\eqref{beta1} into \eqref{alpha},
\begin{equation*}
A(-\Lambda_{21}^2+2\Lambda_{21}\Lambda_{22}+4A\Lambda_1\Lambda_{21})=0.
\end{equation*}
Since $A>0,$ we have
\begin{equation}
\label{A>01}
A=\frac{\Lambda_{21}(\Lambda_{21}-2\Lambda_{22})}{4\Lambda_1\Lambda_{22}^2}.
\end{equation}
By SOC \eqref{SOC1} and \eqref{SOC2}, we have
\begin{equation}
\label{A>02}
\Lambda_{22}>0,\Lambda_1>\Lambda_{22}\beta_{23}^2>0,
\end{equation}
Combining \eqref{A>01}, \eqref{A>02} and $A>0$, if $\Lambda_{21}>0$, we have $\Lambda_{21}>2\Lambda_{22}.$ Substitute \eqref{A>01} into \eqref{lam1}, it becomes
\begin{equation*}
\Lambda_1(\Lambda_{21}^2+4\Lambda_{22}^2(4\Lambda_1^2\theta_1-1))=0.
\end{equation*}
Since $\Lambda_1>0,$
\begin{equation*}
\Lambda_{22}=\frac{\Lambda_{21}}{2\sqrt{1-4\Lambda_1^2\theta_1}}>\frac{\Lambda_{21}}{2}.
\end{equation*}
We got two contradictory conclusions. If $\Lambda_{21}<0,$ substitute \eqref{A>01} into \eqref{lam21}, we have
\begin{equation*}
\begin{aligned}
0=&64 \Lambda_1^2 \Lambda_{21} \Lambda_{22}^4 \theta_1+\Lambda_1 (4 \Lambda_{21}^3 \Lambda_{22} \theta_1-8 \Lambda_{21}^2 \Lambda_{22}^2 \theta_1+16 \Lambda_{21} \Lambda_{22}^3-32 \Lambda_{22}^4)+\Lambda_{21}^5 \theta_1\\
-&4 \Lambda_{21}^4 \Lambda_{22} \theta_1+4 \Lambda_{21}^3 \Lambda_{22}^2 \theta_1+4 \Lambda_{21}^3 \Lambda_{22}^2-16 \Lambda_{21}^2 \Lambda_{22}^3+16 \Lambda_{21} \Lambda_{22}^4,
\end{aligned}
\end{equation*}
however the RHS is always strictly negative, so the equilibrium does not exist.

\begin{figure}[!htbp]
    \centering
\includegraphics[width = 0.27\textwidth]{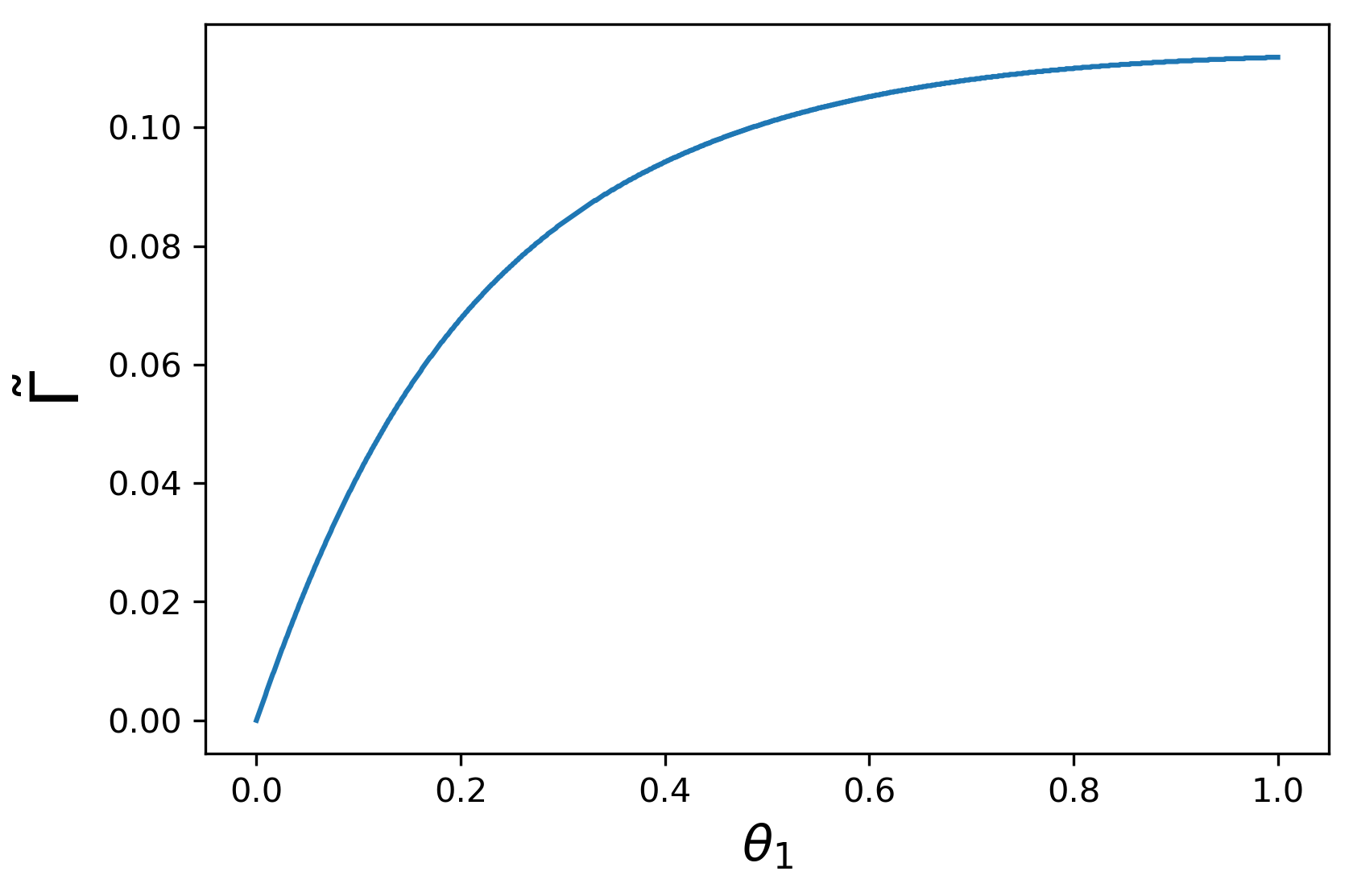}
\caption{Given $\theta_1\in(0,1]$ and $\theta_\varepsilon=0$, the critical $\Tilde{\Gamma}(\theta_1)$ for equilibrium existence.}
    \label{figtildeGam}
\end{figure}

The exact form of $\Tilde{\Gamma}$ is hard to find. So we numerically find $\Tilde{\Gamma}$ for $\theta\in(0,1]$, as shown in Figure \ref{figtildeGam}.

When considering the equilibrium with duopolistic informed investors, the equilibrium can be similarly derived as in the beginning of Section \ref{secresults}. The equilibrium is characterized by a simplified system of $\{\Lambda
_1,\Lambda_{21},\Lambda_{22}\}$:
{\footnotesize
\begin{equation}
\label{systemduopolistic}
\begin{aligned}
0&=4 \Gamma^2 (16 \Lambda_1^3 \theta_1-24 \Lambda_1^2 \theta_1 (\Lambda_{21}-\Lambda_{22})+3 \Lambda_1 (3 \Lambda_{21}^2 \theta_1-6 \Lambda_{21} \Lambda_{22} \theta_1+3 \Lambda_{22}^2 \theta_1-1)+3 (\Lambda_{21}-\Lambda_{22}))\\
&+2 \Gamma (48 \Lambda_1^3 \Lambda_{22} \theta_1-4 \Lambda_1^2 \theta_1 (2 \Lambda_{21}^2+9 \Lambda_{21} \Lambda_{22}-9 \Lambda_{22}^2)+2 \Lambda_1 (3 \Lambda_{21}^3 \theta_1-3 \Lambda_{21}^2 \Lambda_{22} \theta_1+\Lambda_{21}-6 \Lambda_{22})-2 \Lambda_{21}^2\\
&+12 \Lambda_{21} \Lambda_{22}-9 \Lambda_{22}^2)+36 \Lambda_1^3 \Lambda_{22}^2 \theta_1-12 \Lambda_1^2 \Lambda_{21}^2 \Lambda_{22} \theta_1+\Lambda_1 (\Lambda_{21}^4 \theta_1+\Lambda_{21}^2-9 \Lambda_{22}^2)-\Lambda_{21}^2 (\Lambda_{21}-3 \Lambda_{22}),\\
0&=6 \Gamma^4 (4 \Lambda_1^2 \Lambda_{21} (4 \Lambda_{22}^2+1) \theta_1+4 \Lambda_1 (-\Lambda_{22}^2 (6 \Lambda_{21}^2 \theta_1+1)-2 \Lambda_{21}^2 \theta_1+6 \Lambda_{21} \Lambda_{22}^3 \theta_1+\Lambda_{21} \Lambda_{22} \theta_1)\\
&+\Lambda_{21}^3 (9 \Lambda_{22}^2+4) \theta_1-2 \Lambda_{21}^2 \Lambda_{22} (9 \Lambda_{22}^2+2) \theta_1+\Lambda_{21} \Lambda_{22}^2 (9 \Lambda_{22}^2 \theta_1+\theta_1+4)-3 \Lambda_{22}^3)+8 \Gamma^3 \Lambda_{22} (8 \Lambda_1^2 \Lambda_{21} (14 \Lambda_{22}^2+3) \theta_1\\
&-2 \Lambda_1 (4 \Lambda_{21}^3 \Lambda_{22} \theta_1+\Lambda_{21}^2 (66 \Lambda_{22}^2+19) \theta_1-\Lambda_{21} \Lambda_{22} (66 \Lambda_{22}^2 \theta_1+7 \theta_1+2)+17 \Lambda_{22}^2)+6 \Lambda_{21}^4 \Lambda_{22} \theta_1+2 \Lambda_{21}^3 (15 \Lambda_{22}^2+7) \theta_1\\
&-\Lambda_{21}^2 \Lambda_{22} (72 \Lambda_{22}^2 \theta_1+9 \theta_1+4)+\Lambda_{21} \Lambda_{22}^2 (36 \Lambda_{22}^2 \theta_1+\theta_1+34)-21 \Lambda_{22}^3)+4 \Gamma^2 \Lambda_{22} (4 \Lambda_1^2 \Lambda_{21} \Lambda_{22} (73 \Lambda_{22}^2+13) \theta_1\\
&-2 \Lambda_1 (\Lambda_{21}^3 (22 \Lambda_{22}^2 \theta_1+\theta_1)+30 \Lambda_{21}^2 (4 \Lambda_{22}^3+\Lambda_{22}) \theta_1-\Lambda_{21} \Lambda_{22}^2 (120 \Lambda_{22}^2 \theta_1+5 \theta_1+11)+53 \Lambda_{22}^3)+\Lambda_{21}^5 \Lambda_{22} \theta_1\\
&+2 \Lambda_{21}^4 (12 \Lambda_{22}^2 \theta_1+\theta_1)+\Lambda_{21}^3 \Lambda_{22} (12 \Lambda_{22}^2+13) \theta_1-\Lambda_{21}^2 \Lambda_{22}^2 (72 \Lambda_{22}^2 \theta_1+\theta_1+19)+\Lambda_{21} \Lambda_{22}^3 (36 \Lambda_{22}^2 \theta_1-2 \theta_1+97)\\
&-48 \Lambda_{22}^4)+2 \Gamma \Lambda_{22}^2 (48 \Lambda_1^2 \Lambda_{21} \Lambda_{22} (7 \Lambda_{22}^2+1) \theta_1-2 \Lambda_1 (2 \Lambda_{21}^3 (20 \Lambda_{22}^2 \theta_1+\theta_1)+\Lambda_{21}^2 \Lambda_{22} (72 \Lambda_{22}^2+19) \theta_1\\
&+2 \Lambda_{21} \Lambda_{22}^2 (-36 \Lambda_{22}^2 \theta_1+\theta_1-10)+72 \Lambda_{22}^3)+4 \Lambda_{21}^5 \Lambda_{22} \theta_1+2 \Lambda_{21}^4 (12 \Lambda_{22}^2 \theta_1+\theta_1)+8 \Lambda_{21}^3 \Lambda_{22} (1-3 \Lambda_{22}^2) \theta_1\\
&-\Lambda_{21}^2 \Lambda_{22}^2 (\theta_1+28)+108 \Lambda_{21} \Lambda_{22}^3-36 \Lambda_{22}^4)+\Lambda_{22}^3 (16 \Lambda_1^2 \Lambda_{21} \Lambda_{22} (9 \Lambda_{22}^2+1) \theta_1-2 \Lambda_1 (\Lambda_{21}^3 (24 \Lambda_{22}^2 \theta_1+\theta_1)+5 \Lambda_{21}^2 \Lambda_{22} \theta_1\\
&-12 \Lambda_{21} \Lambda_{22}^2+36 \Lambda_{22}^3)+\Lambda_{21} (4 \Lambda_{21}^4 \Lambda_{22} \theta_1+\Lambda_{21}^3 \theta_1+\Lambda_{21}^2 \Lambda_{22} \theta_1-12 \Lambda_{21} \Lambda_{22}^2+36 \Lambda_{22}^3)),\\
0&=16 \Gamma^4 (\Lambda_{22}^2 (\theta_1 (16 \Lambda_1^2-24 \Lambda_1 \Lambda_{21}+9 \Lambda_{21}^2-2)+1)-2 \theta_1 (2 \Lambda_1^2-3 \Lambda_1 \Lambda_{21}+\Lambda_{21}^2)+6 \Lambda_{22}^3 \theta_1 (4 \Lambda_1-3 \Lambda_{21})\\
&+\Lambda_{22} (5 \Lambda_{21} \theta_1-6 \Lambda_1 \theta_1)+9 \Lambda_{22}^4 \theta_1)+8 \Gamma^3 (2 \Lambda_{22} (\Lambda_{22}^2 (56 \Lambda_1^2 \theta_1-4 \theta_1+5)-14 \Lambda_1^2 \theta_1+66 \Lambda_1 \Lambda_{22}^3 \theta_1-17 \Lambda_1 \Lambda_{22} \theta_1\\
&+18 \Lambda_{22}^4 \theta_1)+2 \Lambda_{21}^2 \theta_1 (-4 \Lambda_1 \Lambda_{22}^2+\Lambda_1+15 \Lambda_{22}^3-3 \Lambda_{22})+\Lambda_{21} \Lambda_{22} (-132 \Lambda_1 \Lambda_{22}^2 \theta_1+32 \Lambda_1 \theta_1-72 \Lambda_{22}^3 \theta_1+\Lambda_{22} (21 \theta_1-2))\\
&+2 \Lambda_{21}^3 (3 \Lambda_{22}^2-1) \theta_1)+4 \Gamma^2 \Lambda_{22} (\Lambda_{22} (\Lambda_{22}^2 (292 \Lambda_1^2 \theta_1-8 \theta_1+37)-72 \Lambda_1^2 \theta_1+240 \Lambda_1 \Lambda_{22}^3 \theta_1-64 \Lambda_1 \Lambda_{22} \theta_1+36 \Lambda_{22}^4 \theta_1)\\
&+\Lambda_{21}^2 (-44 \Lambda_1 \Lambda_{22}^2 \theta_1+12 \Lambda_1 \theta_1+12 \Lambda_{22}^3 \theta_1+\Lambda_{22} \theta_1+\Lambda_{22})+\Lambda_{21} \Lambda_{22} (-240 \Lambda_1 \Lambda_{22}^2 \theta_1+54 \Lambda_1 \theta_1-72 \Lambda_{22}^3 \theta_1\\
&+\Lambda_{22} (23 \theta_1-14))+\Lambda_{21}^4 \Lambda_{22} \theta_1+8 \Lambda_{21}^3 (3 \Lambda_{22}^2-1) \theta_1)+2 \Gamma \Lambda_{22} (4 \Lambda_{22}^2 (3 \Lambda_{22}^2 (28 \Lambda_1^2 \theta_1+5)-20 \Lambda_1^2 \theta_1+36 \Lambda_1 \Lambda_{22}^3 \theta_1\\
&-10 \Lambda_1 \Lambda_{22} \theta_1)+2 \Lambda_{21}^2 \Lambda_{22} (-40 \Lambda_1 \Lambda_{22}^2 \theta_1+11 \Lambda_1 \theta_1-12 \Lambda_{22}^3 \theta_1+\Lambda_{22} (5 \theta_1+2))+2 \Lambda_{21} \Lambda_{22}^2 (-72 \Lambda_1 \Lambda_{22}^2 \theta_1+13 \Lambda_1 \theta_1\\
&+\Lambda_{22} (\theta_1-16))+\Lambda_{21}^4 (4 \Lambda_{22}^2-1) \theta_1+\Lambda_{21}^3 \Lambda_{22} (24 \Lambda_{22}^2-7) \theta_1)+\Lambda_{22}^2 (4 (9 \Lambda_{22}^4 (4 \Lambda_1^2 \theta_1+1)-8 \Lambda_1^2 \Lambda_{22}^2 \theta_1)\\
&+\Lambda_{21}^2 \Lambda_{22} (-48 \Lambda_1 \Lambda_{22}^2 \theta_1+12 \Lambda_1 \theta_1+\Lambda_{22} (\theta_1+4))-4 \Lambda_{21} \Lambda_{22}^2 (\Lambda_1 \theta_1+6 \Lambda_{22})+\Lambda_{21}^4 (4 \Lambda_{22}^2-1) \theta_1),\\
0&<\Lambda_{22},\\
0&<4 \Gamma (\Lambda_1-\Lambda_{21}+\Lambda_{22})+4 \Lambda_1 \Lambda_{22}-\Lambda_{21}^2.
\end{aligned}
\end{equation}}
After solving it,
\begin{equation*}
\begin{aligned}
&A=\frac{4 \Gamma (\Lambda_1-\Lambda_{21}+\Lambda_{22})+\Lambda_{22} (2 \Lambda_1-\Lambda_{21})}{\Lambda_{22} [\Gamma (8 \Lambda_1-6 \Lambda_{21}+6 \Lambda_{22})+6 \Lambda_1 \Lambda_{22}-\Lambda_{21}^2]},\\
&\beta_1=\frac{\sigma_2}{\sigma_v}\frac{-2 \Gamma+\Lambda_{21}-3 \Lambda_{22}}{\Gamma (-8 \Lambda_1+6 \Lambda_{21}-6 \Lambda_{22})-6 \Lambda_1 \Lambda_{22}+\Lambda_{21}^2},\\
&\beta_{21}=\frac{\sigma_2}{\sigma_v}\frac{1-\Lambda_{22}A}{2(\Lambda_{22}+\Gamma)},\\
&\beta_{22}=-\frac{\Lambda_{21}}{2 (\Lambda_{22}+\Gamma)},\\
&\beta_{23}=-\frac{2 \Gamma+\Lambda_{21}}{2 (\Lambda_{22}+\Gamma)}.
\end{aligned}
\end{equation*}

\noindent\textbf{Proof of Theorem \ref{theta1rightarrow0thm}.} When $\theta_\varepsilon>0,0\leq\Gamma<\infty$ or  $\theta_\varepsilon\geq0,0<\Gamma<\infty,$ the limit equilibrium under \eqref{conjecture} always exists. Through numerical experiments, we find that $\beta_1=O(\theta_1^a).$ Then we derive $a$ through theoretical analyses. Given IT's intensity $0<\alpha<\infty,$ if $a>1,$
\begin{equation*}
\begin{aligned}
&\lambda_1\sim0,\\
&\Sigma\sim\alpha^2\sigma_v^2\sigma_1^2(1+\beta_{21})^2+\sigma_1^2\sigma_\varepsilon^2\beta_{21}^2+\sigma_2^2\sigma_1^2\sim O(\sigma_1^2),\\
&\lambda_{22}\sim\frac{\alpha\sigma_v^2}{\Sigma}\sigma_1^2(1+\beta_{21})\sim O(1),\\
&\lambda_{21}\sim-\frac{\alpha\sigma_v^2}{\Sigma}\sigma_1^2\beta_{22}(1+\beta_{21}).\\
\end{aligned}
\end{equation*}
When $\Gamma=0,$ SOC \eqref{SOC2} fails. When $\Gamma>0,$ in limit, we have
\begin{equation*}
\beta_{22}=-\frac{\lambda_{21}}{\lambda_{22}}
\end{equation*}
and
\begin{equation*}
\beta_{22}=-\frac{\lambda_{21}}{2(\lambda_{22}+\gamma)},
\end{equation*}
which hold simultaneously only if 
\begin{equation*}
\lambda_{21}\rightarrow\infty,
\end{equation*}
so SOC \eqref{SOC2} can fail.

If $\frac{1}{2}<a<1,$ 
\begin{equation*}
\begin{aligned}
&\lambda_1\sim O(\frac{\beta_1}{\sigma_1^2}),\\
&\Sigma\sim\alpha^2\sigma_v^2\sigma_1^2(1+\beta_{21})^2+\sigma_1^2\sigma_\varepsilon^2\beta_{21}^2+\sigma_2^2\sigma_1^2\sim O(\sigma_1^2),\\
&\lambda_{22}\sim\frac{\alpha\sigma_v^2}{\Sigma} \sigma_1^2(1+\beta_{21}) \sim O(1),\\
&\lambda_{21}\sim\frac{\alpha\sigma_v^2}{\Sigma} (-\sigma_1^2\beta_{22}(1+\beta_{21})+\sigma_2^2\beta_1-\sigma_\varepsilon^2\beta_1\beta_{21})\geq O(\frac{\beta_1}{\sigma_1^2}).
\end{aligned}
\end{equation*}
Then the SOC \eqref{SOC2} is negative if $\theta_1$ is small enough.

If $a=\frac{1}{2}$,
\begin{equation*}
\begin{aligned}
&\lambda_1\sim O(\frac{1}{\sigma_1}),\\
&\Sigma\sim\alpha^2\sigma_v^2(\sigma_1^2(1+\beta_{21})^2+\beta_1^2(\sigma_2^2+\sigma_\varepsilon^2))+\sigma_1^2\sigma_\varepsilon^2\beta_{21}^2+\sigma_2^2(\beta_1^2\sigma_\varepsilon^2+\sigma_1^2)\sim O(\sigma_1^2),\\
&\lambda_{22}\sim\frac{\alpha\sigma_v^2}{\Sigma} (\sigma_1^2(1+\beta_{21})+\sigma_\varepsilon^2\beta_1^2)\sim O(1),\\
&\lambda_{21}\sim\frac{\alpha\sigma_v^2}{\Sigma} (-\sigma_1^2\beta_{22}(1+\beta_{21})+\sigma_2^2\beta_1-\sigma_\varepsilon^2\beta_1(\beta_{21}+\beta_{23}\beta_1))\geq O(\frac{1}{\sigma_1}),
\end{aligned}
\end{equation*}
The SOC \eqref{SOC2} can be negative either.

If $0<a<\frac{1}{2}$,
\begin{equation*}
\begin{aligned}
&\lambda_1\sim O(\frac{1}{\beta_1}),\\
&\Sigma\sim\alpha^2\sigma_v^2\beta_1^2(\sigma_2^2+\sigma_\varepsilon^2)+\beta_1^2\sigma_2^2\sigma_\varepsilon^2\sim O(\beta_1^2),\\
&\lambda_{22}\sim\frac{\alpha\sigma_v^2}{\Sigma} \sigma_\varepsilon^2\beta_1^2\sim O(1),\\
&\lambda_{21}\sim\frac{\alpha\sigma_v^2}{\Sigma} (\sigma_2^2\beta_1-\sigma_\varepsilon^2\beta_1(\beta_{21}+\beta_{23}\beta_1))\geq O(\frac{1}{\beta_1}),
\end{aligned}
\end{equation*}
The SOC \eqref{SOC2} can be negative either.

Consequently, we must have $\beta_1\sim O(\theta_1)$. We assume 
\begin{equation*}
\lim_{\theta_1\rightarrow0}\frac{\beta_1}{\theta_1}=\zeta(\theta_\varepsilon,\Gamma),
\end{equation*}
which can be found through numerical methods:
\begin{figure}[!htbp]
    \centering
\subcaptionbox{$\Gamma\in[0,1].$}{
    \includegraphics[width = 0.27\textwidth]{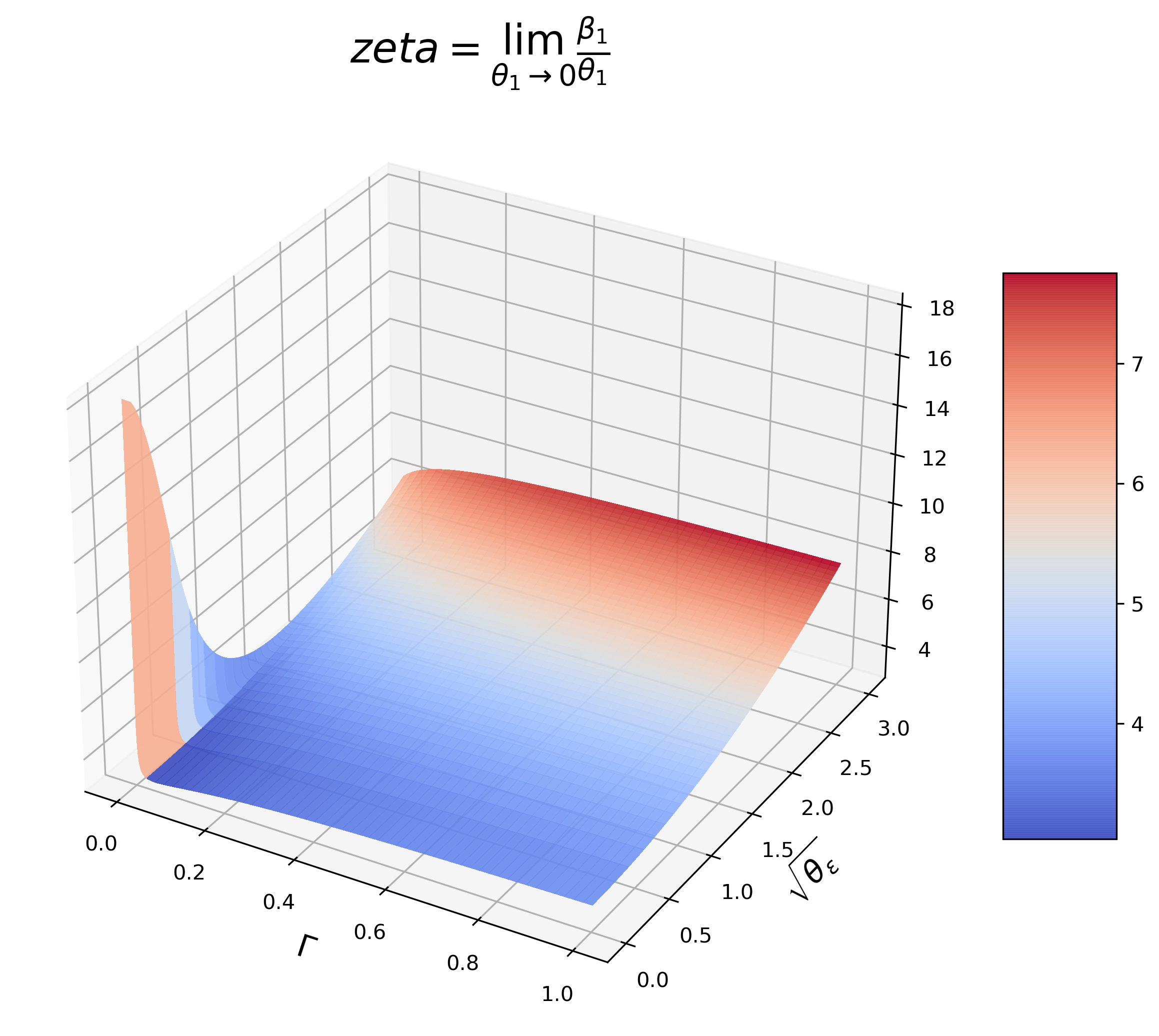}
    }
\subcaptionbox{$\Gamma\in[1,20].$}{
    \includegraphics[width = 0.27\textwidth]{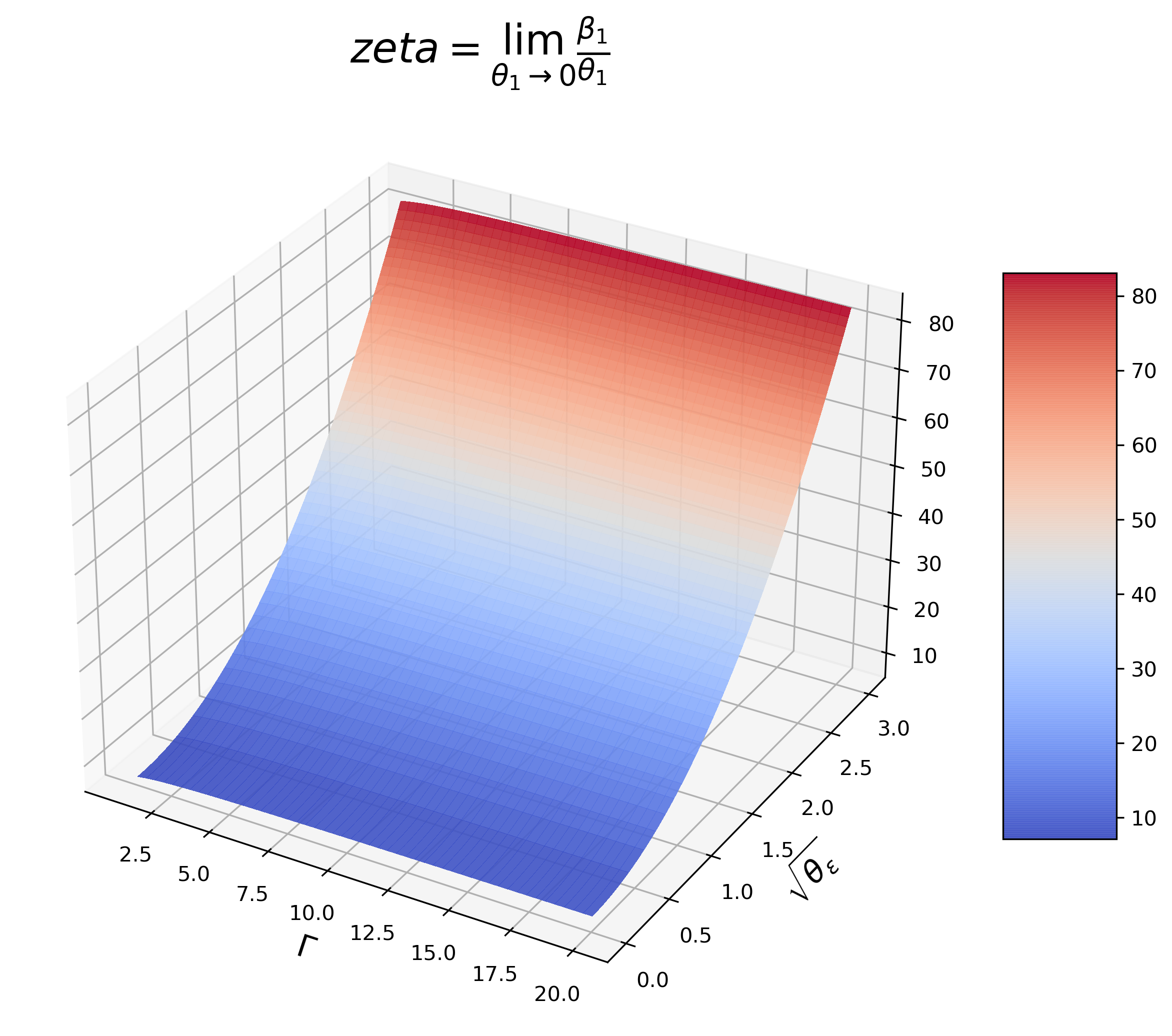}
    }
\caption{$\zeta(\theta_\varepsilon,\Gamma),\theta_\varepsilon\in[10^{-4},9]$.}
\end{figure}

Actually, SOC \eqref{SOC2} holds only if $|\beta_{23}|<\infty$, i.e., $\lambda_{21}<\infty.$ So when $\theta_1\rightarrow0,$ 
the equilibrium conditions as $\theta_1\rightarrow0$ are given by
\begin{equation}
\label{theta1limitequilibrium}
\begin{aligned}
&\lambda_{22}=\frac{\alpha\sigma_v^2(1+\beta_{21})}{\alpha^2\sigma_v^2(1+\beta_{21})^2+\sigma_\varepsilon^2\beta_{21}^2+\sigma_2^2},\\
&\alpha=\frac{1}{2(\lambda_{22}(1+\beta_{21}))},\\
&\beta_{21}=\frac{1-\lambda_{22}\alpha}{2(\lambda_{22}+\gamma)}\frac{\alpha\sigma_v^2}{\alpha\sigma_v^2+\sigma_\varepsilon^2},\\
&\alpha>0,\\
&\lambda_{22}+\gamma>0.
\end{aligned}
\end{equation}
If it is solved, $\{\lambda_1,\lambda_{21},\beta_{22},\beta_{23}\}$ are given by
\begin{equation*}
\begin{aligned}
&\lambda_1=\alpha\sigma_v^2\zeta,\\
&\lambda_{21}=\frac{\alpha\sigma_v^2(-\beta_{22}(1+\beta_{21})+\sigma_2^2\zeta-\sigma_\varepsilon^2\zeta\beta_{21})}{\alpha^2\sigma_v^2(1+\beta_{21})^2+\sigma_\varepsilon^2\beta_{21}^2+\sigma_2^2},\\
&\beta_{22}=-\frac{\lambda_{21}}{2(\lambda_{22}+\gamma)},\\
&\beta_{23}=-\frac{\lambda_{21}+2\gamma}{2(\lambda_{22}+\gamma)}.\\
\end{aligned}
\end{equation*}

Further simplify three equations in \eqref{theta1limitequilibrium}, we find that $\beta_{21}$ satisfies \eqref{theta1limitbeta21}.
We denote the RHS of Equation \eqref{theta1limitbeta21} $f(\beta_{21}),$ when $\theta_\varepsilon>0$ or $0<\Gamma<\infty,$
we have 
\begin{equation*}
\begin{aligned}
&f(0)<0,f(\infty)>0, \\
&f'(\beta_{21})>0,\forall\beta_{21}>0.
\end{aligned}
\end{equation*}
so there must exist a unique $\beta_{21}>0$, such that $f(\beta_{21})=0$. 

In the limit, $\mathbbm{E}(\pi^{\text{IT}})<\frac{\sigma_2\sigma_v}{2}$ if and only if
\begin{equation}
\label{theta1=0harm}
\beta_{21}^2(\theta_\varepsilon-1)<2\beta_{21},
\end{equation}
which holds for $\theta_\varepsilon\leq1.$ When $\theta_\varepsilon>1,$
\begin{equation*}
f(\frac{2}{\theta_\varepsilon-1})=\frac{3\theta_\varepsilon^2-2\theta_\varepsilon-1+8\Gamma(\theta_\varepsilon^2+2\theta_\varepsilon+8)}{2(\theta_\varepsilon-1)(\theta_\varepsilon+1)(\theta_\varepsilon-1+2\Gamma(\theta_\varepsilon+1))}>0,
\end{equation*}
\eqref{theta1=0harm} holds.

For $\theta_\varepsilon=\Gamma=0,$ by similar arguments, we can prove that $\beta_1\sim O(\theta_1).$ And the conditions of limit equilibrium is given by:
\begin{equation*}
\begin{aligned}
&\lambda_{22}=\frac{\sigma_v^2(\alpha+\beta_{21})}{\sigma_v^2(\alpha+\beta_{21})\sigma_v^2+\sigma_2^2},\\
&\alpha=\frac{1-\lambda_{22}\beta_{21}}{2\lambda_{22}},\\
&\beta_{21}=\frac{1-\lambda_{22}\alpha}{2\lambda_{22}},
\end{aligned}
\end{equation*}
which can be solved out explicitly:
\begin{equation*}
\lambda_{22}=\frac{2\sqrt{2}}{3}\frac{\sigma_v}{2\sigma_2},\ \alpha=\beta_{21}=\frac{\sqrt{2}}{2}\frac{\sigma_2}{\sigma_v}.
\end{equation*}


\noindent\textbf{Proof of Theorem \ref{thmgam=infty}}. Substitute \eqref{beta21},\eqref{beta23} into \eqref{beta1} and \eqref{SOC2}, we have
\begin{equation}
\label{gam=infty1}
\begin{aligned}
&\beta_1=\frac{\alpha\sigma_v^2}{\alpha^2\sigma_v^2+\sigma_\varepsilon^2}\frac{\lambda_{21}(\alpha\lambda_{22}-1)+2\lambda_{22}(\alpha\gamma+1)}{4\lambda_1\lambda_{22}-\lambda_1^2+4\gamma(\lambda_{22}+\lambda_1-\lambda_{21})},
\end{aligned}
\end{equation}
and 
\begin{equation}
\label{gam=infty2}
4\gamma(\lambda_{22}+\lambda_1-\lambda_{21})+4\lambda_1\lambda_{22}-\lambda_{21}^2>0
\end{equation}
When $\gamma=\Gamma\rightarrow\infty,$
\begin{equation*}
\beta_{21}=\beta_{22}=0,\beta_{23}=-1,
\end{equation*}
So $x_2=-x_1$. Denote $\beta_1=\beta$, the market impact coefficients follow
\begin{equation}
\label{gam=infty3}
\begin{aligned}
&\lambda_1=\frac{\alpha\beta\sigma_v^2}{\beta_1^2(\alpha^2\sigma_v^2+\sigma_\varepsilon^2)+\sigma_1^2},\\
&\lambda_{21}=\frac{\alpha\sigma_v^2(\beta^2\sigma_\varepsilon^2+\beta\sigma_2^2)}{\alpha^2\sigma_v^2(\sigma_1^2(1-\beta)^2+\sigma_2^2\beta^2+\sigma_\varepsilon^2\beta^2)+\beta^2\sigma_\varepsilon^2(\sigma_1^2+\sigma_2^2)+\sigma_1^2\sigma_2^2},\\
&\lambda_{22}=\frac{\alpha\sigma_v^2(\beta^2\sigma_\varepsilon^2+(1-\beta)\sigma_1^2)}{\alpha^2\sigma_v^2(\sigma_1^2(1-\beta)^2+\sigma_2^2\beta^2+\sigma_\varepsilon^2\beta^2)+\beta^2\sigma_\varepsilon^2(\sigma_1^2+\sigma_2^2)+\sigma_1^2\sigma_2^2}.\\
\end{aligned}
\end{equation}
Now $\lambda_2=\lambda_{21}\beta+\lambda_{22}(1-\beta)$, IT's optimal intensity is
\begin{equation}
\label{gam=infty4}
\alpha=\frac{1}{2\lambda_2}.
\end{equation}
Substitute \eqref{gam=infty3} into \eqref{gam=infty4}, we get
\begin{equation}
\label{gam=infty5}
\alpha=\frac{\sigma_2}{\sigma_v}\sqrt{\frac{\theta_1+\beta^{2}\theta_\varepsilon(\theta_1+1)}{\theta_1(1-\beta)^2+\beta^{2}(\theta_\varepsilon+1)}}.
\end{equation}
HFT's optimal intensity is
\begin{equation}
\label{gam=infty6}
\beta=\frac{\alpha^2\sigma_v^2}{\alpha^2\sigma_v^2+\sigma_\varepsilon^2}\frac{\lambda_{22}}{2(\lambda_{22}+\lambda_1-\lambda_{21})}.
\end{equation}
Substitute \eqref{gam=infty3} and \eqref{gam=infty5} into \eqref{gam=infty6}, we get \eqref{gam=inftybeta}. The only SOC \eqref{gam=infty2} becomes
\begin{equation}
\label{gam=inftysoc}
\lambda_{22}+\lambda_1-\lambda_{21}>0.
\end{equation}

Now we prove the existence and uniqueness of equilibrium. By \eqref{gam=infty6}, $\beta$ has the same sign with $\lambda_{22}.$ If $\beta\leq0,$ by \eqref{gam=infty3}, $\lambda_{22}>0,$ which contradicts. So in equilibrium we must have $\beta>0.$

Next we need only to prove that the root of Equation \eqref{gam=inftybeta} must be in $(0,1)$ and unique, satisfies the SOC. If we see the right side of Equation \eqref{gam=inftybeta} the function of $\beta$: $f(\beta;\theta_1,\theta_\varepsilon),$ we have
\begin{equation*}
\begin{aligned}
&f(0;\theta_1,\theta_\varepsilon)=-\theta_1^3<0,\\
&f(1;\theta_1,\theta_\varepsilon)=2 \theta_1^2 + 6 \theta_1 \theta_\varepsilon + 4 \theta_1^2\theta_\varepsilon + 2 \theta_\varepsilon^2 + 9 \theta_1 \theta_\varepsilon^2 + 2 \theta_1^2 \theta_\varepsilon^2 + \theta_\varepsilon^3 + 3 \theta_1 \theta_\varepsilon^3>0,
\end{aligned}
\end{equation*}
then there must exist $\beta\in(0,1)$, satisfying Equation \eqref{gam=inftybeta}. For $\beta\geq1,$ it is easy to verify $f'(\beta;\theta_1,\theta_\varepsilon)$ is also positive. Thus there is no root of Equation \eqref{gam=inftybeta} on $[1,+\infty)$.

Now we denote $\beta^*$ the smallest root of $f$ on $(0,1)$. The SOC \eqref{gam=inftysoc} is equivalent to
\begin{equation*}
\alpha\sigma_v^2(\sigma_1^2+\alpha^2\beta\sigma_v^2)((1-\beta)\sigma_1^2+\beta^2\sigma_\varepsilon^2)>0,
\end{equation*}
which holds for $\beta\in(0,1).$

\begin{equation*}
\begin{aligned}
f^{'}(\beta;\theta_1,\theta_\varepsilon)=&\theta_1^2\theta_\varepsilon(40\beta^4-44\beta^3+24\beta^2-2\beta)+\theta_1^2\theta_\varepsilon^2(12\beta^5+20\beta^4-32\beta^3+12\beta^2)\\
+&\theta_1^3\theta_\varepsilon(20\beta^4-42\beta^3+48\beta^2-18\beta+2)+\theta_1^3(6\beta^2-10\beta+4)\\
+&\theta_1\theta_\varepsilon(20\beta^4+8\beta^3)+\theta_1\theta_\varepsilon^2(24\beta^5+20\beta^4+4\beta^3)\\
+&\theta_1\theta_\varepsilon^3(6\beta^5+10\beta^4)+6\beta^5(2\theta_\varepsilon^2+\theta_\varepsilon^3)+6\beta^2\theta_1^2.
\end{aligned}
\end{equation*}

For $\beta\in(0,0.099),40\beta^4-44\beta^3+24\beta^2-2\beta<0.$ If this leads to $f'(\beta)<0$ for $\beta\in(0,0.099),$ we must have $\beta^*>0.099,$ since $f(0)<0.$ So the negativity would not cause $f'(\beta)<0$ for $\beta\in[\beta^*,1).$

For $\beta\in(0.205,0.325),20\beta^4-42\beta^3+48\beta^2-18\beta+2<0.$ If $\theta_1\leq\theta_\varepsilon,$
\begin{equation*}
\begin{aligned}
&\theta_1^3\theta_\varepsilon(20\beta^4-42\beta^3+48\beta^2-18\beta+2)+\theta_1^2\theta_\varepsilon^2(12\beta^5+20\beta^4-32\beta^3+12\beta^2)\\
\geq&\theta_1^3\theta_\varepsilon(12\beta^5+40\beta^4-74\beta^3+60\beta^2-18\beta+2)\\
\geq&0.
\end{aligned}
\end{equation*}
If $\theta_1>\theta_\varepsilon$ and we want to have
\begin{equation*}
\theta_1^3\theta_\varepsilon(20\beta^4-42\beta^3+48\beta^2-18\beta+2)+\theta_1^3(6\beta^2-10\beta+1)\geq0,
\end{equation*}
$\theta_\varepsilon\leq21.$

For $\beta\in(\frac{2}{3},1),6\beta^2-10\beta+4<0.$ If $\theta_1\leq\theta_\varepsilon,$
\begin{equation*}
\begin{aligned}
&\theta_1^3(6\beta^2-10\beta+4)+\theta_1^2\theta_\varepsilon(40\beta^4-44\beta^3+24\beta^2-2\beta)\\
\geq&\theta_1^3(40\beta^4-44\beta^3+30\beta^2-12\beta+4)\\
\geq&0.
\end{aligned}
\end{equation*}
If $\theta_1>\theta_\varepsilon$ and we want to have
\begin{equation*}
\theta_1^3(6\beta^2-10\beta+4)+6\beta^2\theta_1^2\geq0,
\end{equation*}
$\theta_1\leq24.$

Now we have proved that if $\theta_1\leq\theta_\varepsilon$ or $\theta_\varepsilon<\theta_1\leq21,f^{'}(\beta)\geq0$ for $\beta\in[\beta^*,1)$. So $f(\beta)>0$ for $\beta\in(\beta^*,1)$, $\beta^*$ is the unique root of $f$ on $(0,1).$ We only need to prove the uniqueness for $\theta_1>21>\theta_\varepsilon$ and $\theta_1>\theta_\varepsilon>21$, which is equivalent to prove $f(\beta)>0$ for $\beta\in(\beta^*,1]$.
\begin{equation*}
\begin{aligned}
f(\beta;\theta_1,\theta_\varepsilon)=&\theta_1^2\theta_\varepsilon(8\beta^5-11\beta^4+8\beta^3-\beta^2)+\theta_1^2\theta_\varepsilon^2(2\beta^6+4\beta^5-8\beta^4+4\beta^3)\\
+&\theta_1^3\theta_\varepsilon(4\beta^5-13\beta^4+16\beta^3-9\beta^2+2\beta)+\theta_1^3(2\beta^3-5\beta^5+4\beta-1)\\
+&\theta_1\theta_\varepsilon(4\beta^5+2\beta^4)+\theta_1\theta_\varepsilon^2(4\beta^6+4\beta^5+\beta^4)\\
+&\theta_1\theta_\varepsilon^3(\beta^6+2\beta^5)+\beta^6(2\theta_\varepsilon^2+\theta_\varepsilon^3)+2\beta^3\theta_1^2.
\end{aligned}
\end{equation*}

When $\theta_1>21,$ and for $\beta\in(0,0.154),8\beta^5-11\beta^4+8\beta^3-\beta^2<0,$ but 
\begin{equation*}
\begin{aligned}
&\theta_1^2\theta_\varepsilon(8\beta^5-11\beta^4+8\beta^3-\beta^2)+\theta_1^3\theta_\varepsilon(4\beta^5-13\beta^4+16\beta^3-9\beta^2+2\beta)\\
\geq&\theta_1^2\theta_\varepsilon(8\beta^5-11\beta^4+8\beta^3-\beta^2+21(4\beta^5-13\beta^4+16\beta^3-9\beta^2+2\beta))\\
\geq&0.
\end{aligned}
\end{equation*}
Then for $\beta\in(0,1),$
\begin{equation*}
f(\beta;\theta_1,\theta_\varepsilon)\geq\theta_1^3(2\beta^3-5\beta^5+4\beta-1)+2\beta^3\theta_1^2=f(\beta;\theta_1,\theta_\varepsilon=0),
\end{equation*}
we would analyze the special case $\theta_\varepsilon=0$ in subsection 4.1.1 , and we employ some conclusions in advance. 

We now prove $f(\beta;\theta_1,\theta_\varepsilon=0)>0$ for $\beta\in(\beta^*(\theta_\varepsilon=0),1)$. Since
\begin{equation*}
\begin{aligned}
&f(\beta;\theta_1,\theta_\varepsilon=0)=\theta_1^2g(\beta),\\
&g(\beta)=2\beta^3(\theta_1+1)-5\theta_1\beta^2+4\theta_1\beta-\theta_1,
\end{aligned}
\end{equation*}
it is equivalent to investigate $g(\beta).$

$g(\beta)$ is (1) increasing when $21<\theta_1<24$ or (2) has a unique minimum point  $\Tilde{\beta}=\frac{5\theta_1+\sqrt{\theta_1^2-24\theta_1}}{6(\theta_1+1)}$ on $(\beta^*(\theta_\varepsilon=0),1]$, when $\theta_1\geq24.$ Correspondingly, for $\beta\in(\beta^*(\theta_\varepsilon=0),1]$, (1) $g(\beta)>g(\beta^*(\theta_\varepsilon=0))=0$, (2) $g(\beta)\geq g(\Tilde{\beta})>0.$ Since $\beta^*(\theta_\varepsilon=0)\leq\frac{1}{2},$ we have $f(\beta;\theta_1,\theta_\varepsilon)>0$ for $\beta\in(\frac{1}{2},1)$ when $\theta_1>21$.

When $\theta_\varepsilon<21,$ $f(\beta;\theta_1,\theta_\varepsilon)$ is increasing for $\beta\in[\beta^*,\frac{2}{3}],$ then it is also increasing for $\beta\in[\beta^*,\frac{1}{2}],$ which implies $f(\beta;\theta_1,\theta_\varepsilon)>0$ for $\beta\in(\beta^*,1)$.

When $\theta_\varepsilon>21,f(\beta;\theta_1,\theta_\varepsilon)$ may decrease for $\beta\in(0.206,0.324).$ The only possible negative part in $f$ is
$\theta_1^3(2\beta^3-5\beta^5+4\beta-1)$. However,
\begin{equation*}
\begin{aligned}
&\theta_1^3\theta_\varepsilon(4\beta^5-13\beta^4+16\beta^3-9\beta^2+2\beta)+\theta_1^3(2\beta^3-5\beta^5+4\beta-1)\\
\geq&\theta_1^3(21(4\beta^5-13\beta^4+16\beta^3-9\beta^2+2\beta)+2\beta^3-5\beta^5+4\beta-1)\\
>&0,
\end{aligned}
\end{equation*}
so we must have $f(\beta;\theta_1,\theta_\varepsilon)>0$ for $\beta\in(\beta^*,1]$ no matter $\beta^*<0.206,$ $\beta^*>0.324$ or $\beta^*\in(0.206,0.324).$

The expected profit and market quality variables are from direct calculations.


\noindent\textbf{Derivation of $\Tilde{\theta}_\varepsilon(\theta_1,\infty)$. }For the sake of accuracy, we use $\beta^*$ to represent the equilibrium value of $\beta$. $A<1$ is equivalent to
$\beta^*(1+\theta_1-\theta_1\theta_\varepsilon)-2\theta_1<0,$ which holds for $\theta_1\theta_\varepsilon\geq1.$
If $\theta_1\theta_\varepsilon<1,$ we need $\beta^*<\frac{2\theta_1}{1+\theta_1-\theta_\varepsilon},$ i.e.,
\begin{equation*}
f(\frac{2\theta_1}{1+\theta_1-\theta_\varepsilon})>0\Longleftrightarrow-5\theta_1^2\theta_\varepsilon^2+2\theta_1^2\theta_\varepsilon+3\theta_1^2+10\theta_1\theta_\varepsilon+6\theta_1-1>0.
\end{equation*}
We could rearrange it as
\begin{equation*}
h(\theta_\varepsilon)=-5\theta_1^2\theta_\varepsilon^2+\theta_\varepsilon(2\theta_1^2+10\theta_1)+3\theta_1^2+6\theta_1-1>0.
\end{equation*}
If $3\theta_1^2+6\theta_1-1\geq0,\theta_1\geq\frac{2\sqrt{3}-3}{3},h(\theta_\varepsilon)>0$ when $0\leq\theta_\varepsilon<x_2,$ where
\begin{equation*}
x_2=\frac{(\theta_1+5)+2\sqrt{4\theta_1^2+10\theta_1+5}}{5\theta_1}.
\end{equation*}
 Since $x_2>\frac{1}{\theta_1},$ $A<1$ holds.
 
 If $3\theta_1^2+6\theta_1-1<0,0<\theta_1<\frac{2\sqrt{3}-3}{3},h(\theta_\varepsilon)>0$ when $x_1<\theta_\varepsilon<x_2$, where 
 $x_1=\Tilde{\theta}_\varepsilon.$ Since $x_1<\frac{1}{\theta_1}<x_2,$ we have $A<1$ when $\theta_\varepsilon>\Tilde{\theta}_\varepsilon.$
 


\noindent\textbf{Derivation of $\Hat{\theta}_\varepsilon(\theta_1,\infty)$. }
 For the sake of accuracy, we still use $\beta^*$ to represent the equilibrium value of $\beta$. 
\begin{equation*}
\begin{aligned}
&A(\theta_1,\theta_\varepsilon)=\sqrt{\frac{\theta_1+\beta^{*2}\theta_\varepsilon(\theta_\varepsilon+1)}{\theta_1(1-\beta^*)^2+(1+\theta_\varepsilon)\beta^{*2}}},\\
&\frac{\partial A(\theta_1,\theta_\varepsilon)}{\partial\theta_\varepsilon}=\frac{\partial A (\beta^*,\theta_1,\theta_\varepsilon)}{\partial\beta^*}\frac{\partial\beta^*}{\partial\theta_\varepsilon}+\frac{\partial A(\beta^*,\theta_1,\theta_\varepsilon)}{\partial\theta_\varepsilon}.
\end{aligned}
\end{equation*}

$\frac{\partial A(\theta_1,\theta_\varepsilon)}{\partial\theta_\varepsilon}$ has the same sign as 
\begin{equation*}
2\theta_1(\theta_1-(1+\theta_1)\beta^*)(1+\beta^*\theta_\varepsilon)\frac{\partial\beta^*}{\partial\theta_\varepsilon}+\beta^{*2}(\theta_1-(1+\theta_1)\beta^*)^2.
\end{equation*}
If $\theta_1-(1+\theta_1)\beta^*>0,\frac{\partial A(\theta_1,\theta_\varepsilon)}{\partial\theta_\varepsilon}\leq0$ is equivalent to
\begin{equation*}
g(\beta^*;\theta_1,\theta_\varepsilon)=\beta^{*2}(\theta_1-(1+\theta_1)\beta^*)\frac{\partial f}{\partial\beta^*}-2\theta_1(1+\beta^*\theta_\varepsilon)\frac{\partial f}{\partial\theta_\varepsilon}\leq0.
\end{equation*}
For simplicity, we use $\beta$ to represent $\beta^*$.
\begin{equation*}
\begin{aligned}
g&=\theta_1^2(-4\beta^4-14\beta^5)+\theta_1^3y_1+\theta_1^4(-4\beta+22\beta^2-46\beta^3+42\beta^4-14\beta^5)\\
&+\theta_1\theta_\varepsilon(-16\beta^6-20\beta^7)+\theta_1^2\theta_\varepsilon(-2\beta^4-36\beta^5+32\beta^6-60\beta^3)\\
&+\theta_1^3\theta_\varepsilon(-18\beta^3+60\beta^4-110\beta^5+112\beta^6-60\beta^7)+\theta_1^4\theta_\varepsilon(-2\beta^2-2\beta^3+34\beta^4-74\beta^5+64\beta^6-20\beta^7)\\
&-12\beta^8\theta_\varepsilon^2+\theta_1\theta_\varepsilon^2(-10\beta^6-16\beta^7-36\beta^8)+\theta_1^2\theta_\varepsilon^2(-24\beta^5+26\beta^6-32\beta^7-36\beta^8)\\
&+\theta_1^3\theta_\varepsilon^2(-4\beta^4-12\beta^5+36\beta^6-16\beta^7-12\beta^8)-6\beta^8\theta_\varepsilon^3\\
&+\theta_1\theta_\varepsilon^3(-10\beta^7-12\beta^8)+\theta_1^2\theta_\varepsilon^3(-2\beta^6-10\beta^7-6\beta^8),\\
y_1&=2\beta^2-20\beta^3+38\beta^4-28\beta^5.
\end{aligned}
\end{equation*}
Except $y_1,$ other coefficients are both non-positive for $\beta\in[0,\frac{1}{2}].$ (Actually, $\beta^*\leq\beta^*(\theta_1=\infty,\theta_\varepsilon=0)=\frac{1}{2}$.) $y_1\geq0$ when $\beta\in[0,0.1286].$ Let
\begin{equation*}
\begin{aligned}
p&=-4\beta^4-14\beta^5,\\
q&=-4\beta+22\beta^2-46\beta^3+42\beta^4-14\beta^5.
\end{aligned}
\end{equation*}
$\theta_1^2p+\theta_1^3y_1+\theta_1^4q\leq0$ is equivalent to $p+\theta_1y_1+\theta_1^2q\leq0,\Delta=y_1^2-4pq\geq0$ when $\beta\in[0,0.0328].$ Thus if 
\begin{equation*}
\theta_1\geq\frac{-y_1-\sqrt{\Delta}}{2q}|_{max}=0.0089,
\end{equation*}
and $\theta_\varepsilon>\frac{1-\theta_1-2\theta_1^2}{3\theta_1},$ we have $\frac{\partial A(\theta_1,\theta_\varepsilon)}{\partial\theta_\varepsilon}\leq0$. In a word, we specify two conditions, (1) $\theta_1\geq\frac{1}{2},$ (2) $0.0089\leq\theta_1<\frac{1}{2},\theta_\varepsilon>\frac{1-\theta_1-2\theta_1^2}{3\theta_1}.$

 When $\theta_1-(1+\theta_1)\beta^*\leq0,\frac{\partial A(\theta_1,\theta_\varepsilon)}{\partial\theta_\varepsilon}\geq0,$ which implies $\theta_\varepsilon<\frac{1-\theta_1-2\theta_1^2}{3\theta_1}.$ Thus when $0<\theta_1<\frac{1}{2}$ and $\theta_\varepsilon<\frac{1-\theta_1-2\theta_1^2}{3\theta_1},\frac{\partial A(\theta_1,\theta_\varepsilon)}{\partial\theta_\varepsilon}\geq0.$

\newpage
\addcontentsline{toc}{section}{Reference} 
\bibliographystyle{unsrt}

\begin{thebibliography}{10}

\bibitem{breckenfelder2020does}
Johannes Breckenfelder.
\newblock How does competition among high-frequency traders affect market liquidity?
\newblock {\em European Central Bank Research bulletin}, 78, 2020.

\bibitem{sec2010}
Securities and Exchange Commission.
\newblock Concept release on equity market structure.
\newblock Technical report, Securities and Exchange Commission, 2010.

\bibitem{sec2014}
Staff of~the Division~of Trading and Markets.
\newblock Staff report on equity market structure literature review part {II}: High frequency trading.
\newblock Technical report, Securities and Exchange Commission, 2014.

\bibitem{sec2020}
Staff of~the US~Securities and Exchange Commission.
\newblock Staff report on algorithmic trading in {US} capital markets.
\newblock Technical report, Securities and Exchange Commission, 2020.

\bibitem{brunnermeier2005predatory}
Markus~K Brunnermeier and Lasse~Heje Pedersen.
\newblock Predatory trading.
\newblock {\em The Journal of Finance}, 60(4):1825--1863, 2005.

\bibitem{korajczyk2019high}
Robert~A Korajczyk and Dermot Murphy.
\newblock High-frequency market making to large institutional trades.
\newblock {\em The Review of Financial Studies}, 32(3):1034--1067, 2019.

\bibitem{hirschey2021high}
Nicholas Hirschey.
\newblock Do high-frequency traders anticipate buying and selling pressure?
\newblock {\em Management Science}, 67(6):3321--3345, 2021.

\bibitem{bessembinder2016liquidity}
Hendrik Bessembinder, Allen Carrion, Laura Tuttle, and Kumar Venkataraman.
\newblock Liquidity, resiliency and market quality around predictable trades: Theory and evidence.
\newblock {\em Journal of Financial Economics}, 121(1):142--166, 2016.

\bibitem{murphy2017short}
Dermot~P Murphy and Ramabhadran~S Thirumalai.
\newblock Short-term return predictability and repetitive institutional net order activity.
\newblock {\em Journal of Financial Research}, 40(4):455--477, 2017.

\bibitem{brogger2021market}
S{\o}ren~Bundgaard Br{\o}gger.
\newblock The market impact of predictable flows: Evidence from leveraged {VIX} products.
\newblock {\em Journal of Banking \& Finance}, 133:106280, 2021.

\bibitem{yan2022sunshine}
Lei Yan, Scott~H Irwin, and Dwight~R Sanders.
\newblock Sunshine vs. predatory trading effects in commodity futures markets: New evidence from index rebalancing.
\newblock {\em Journal of Commodity Markets}, 26:100195, 2022.

\bibitem{kyle1985continuous}
Albert~S Kyle.
\newblock Continuous auctions and insider trading.
\newblock {\em Econometrica: Journal of the Econometric Society}, pages 1315--1335, 1985.

\bibitem{2017The}
Andrei Kirilenko, Albert~S. Kyle, Mehrdad Samadi, and Tugkan Tuzun.
\newblock The flash crash: High frequency trading in an electronic market.
\newblock {\em The Journal of Finance}, 72(3), 2017.

\bibitem{sauglam2020order}
Mehmet Sa{\u{g}}lam.
\newblock Order anticipation around predictable trades.
\newblock {\em Financial Management}, 49(1):33--67, 2020.

\bibitem{brogaard2014discovery}
Jonathan Brogaard, Terrence Hendershott, and Ryan Riordan.
\newblock High-frequency trading and price discovery.
\newblock {\em The Review of Financial Studies}, 27(8):2267--2306, 2014.

\bibitem{yang2020back}
Liyan Yang and Haoxiang Zhu.
\newblock Back-running: Seeking and hiding fundamental information in order flows.
\newblock {\em The Review of Financial Studies}, 33(4):1484--1533, 2020.

\bibitem{li2018high}
Wei Li.
\newblock High frequency trading with speed hierarchies.
\newblock {\em Available at SSRN 2365121}, 2018.

\bibitem{rocsu2019fast}
Ioanid Ro{\c{s}}u.
\newblock Fast and slow informed trading.
\newblock {\em Journal of Financial Markets}, 43:1--30, 2019.

\bibitem{hoffmann2014dynamic}
Peter Hoffmann.
\newblock A dynamic limit order market with fast and slow traders.
\newblock {\em Journal of Financial Economics}, 113(1):156--169, 2014.

\bibitem{foucault2016news}
Thierry Foucault, Johan Hombert, and Ioanid Ro{\c{s}}u.
\newblock News trading and speed.
\newblock {\em The Journal of Finance}, 71(1):335--382, 2016.

\bibitem{baldauf2020high}
Markus Baldauf and Joshua Mollner.
\newblock High-frequency trading and market performance.
\newblock {\em The Journal of Finance}, 75(3):1495--1526, 2020.

\bibitem{menkveld2016economics}
Albert~J Menkveld.
\newblock The economics of high-frequency trading: Taking stock.
\newblock {\em Annual Review of Financial Economics}, 8:1--24, 2016.

\bibitem{carlin2007episodic}
Bruce~Ian Carlin, Miguel~Sousa Lobo, and S~Viswanathan.
\newblock Episodic liquidity crises: Cooperative and predatory trading.
\newblock {\em The Journal of Finance}, 62(5):2235--2274, 2007.

\bibitem{schoneborn2009liquidation}
Torsten Sch{\"o}neborn and Alexander Schied.
\newblock Liquidation in the face of adversity: stealth vs. sunshine trading.
\newblock In {\em EFA 2008 Athens Meetings Paper}, 2009.

\bibitem{herrmann2020inventory}
Sebastian Herrmann, Johannes Muhle-Karbe, Dapeng Shang, and Chen Yang.
\newblock Inventory management for high-frequency trading with imperfect competition.
\newblock {\em SIAM Journal on Financial Mathematics}, 11(1):1--26, 2020.

\bibitem{cardaliaguet2018mean}
Pierre Cardaliaguet and Charles-Albert Lehalle.
\newblock Mean field game of controls and an application to trade crowding.
\newblock {\em Mathematics and Financial Economics}, 12:335--363, 2018.

\bibitem{huang2019mean}
Xuancheng Huang, Sebastian Jaimungal, and Mojtaba Nourian.
\newblock Mean-field game strategies for optimal execution.
\newblock {\em Applied Mathematical Finance}, 26(2):153--185, 2019.

\bibitem{bernhardt2004informed}
Dan Bernhardt and Jianjun Miao.
\newblock Informed trading when information becomes stale.
\newblock {\em The Journal of Finance}, 59(1):339--390, 2004.

\bibitem{bernhardt2008front}
Dan Bernhardt and Bart Taub.
\newblock Front-running dynamics.
\newblock {\em Journal of Economic Theory}, 138(1):288--296, 2008.

\bibitem{2001Public}
S.~Huddart, J.~S. Hughes, and C.~B. Levine.
\newblock Public disclosure and dissimulation of insider trades.
\newblock {\em Econometrica}, 69(3):665--681, 2001.

\bibitem{cox2006using}
David~A Cox, John Little, and Donal O'shea.
\newblock {\em Using algebraic geometry}, volume 185.
\newblock Springer Science \& Business Media, 2006.

\bibitem{xu2023large}
Ziyi Xu and Xue Cheng.
\newblock Are large traders harmed by front-running {HFT}s?
\newblock {\em Procedia Computer Science}, 221:501--508, 2023.

\end{thebibliography}

\end{document}